# Surface codes: Towards practical large-scale quantum computation

Austin G. Fowler

*Centre for Quantum Computation and Communication Technology,*
*School of Physics, The University of Melbourne, Victoria 3010, Australia*

Matteo Mariantoni

*Department of Physics, University of California, Santa Barbara, CA 93106-9530, USA and*
*California Nanosystems Institute, University of California, Santa Barbara, CA 93106-9530, USA*

John M. Martinis and Andrew N. Cleland

*California Nanosystems Institute, University of California, Santa Barbara, CA 93106-9530, USA*

(Dated: October 26, 2012)

This article provides an introduction to surface code quantum computing. We first estimate the size and speed of a surface code quantum computer. We then introduce the concept of the stabilizer, using two qubits, and extend this concept to stabilizers acting on a two-dimensional array of physical qubits, on which we implement the surface code. We next describe how logical qubits are formed in the surface code array and give numerical estimates of their fault-tolerance. We outline how logical qubits are physically moved on the array, how qubit braid transformations are constructed, and how a braid between two logical qubits is equivalent to a controlled-NOT. We then describe the single-qubit Hadamard, $\hat{S}$ and $\hat{T}$ operators, completing the set of required gates for a universal quantum computer. We conclude by briefly discussing physical implementations of the surface code. We include a number of appendices in which we provide supplementary information to the main text.

In this article we describe the surface code approach to quantum computing. We have attempted to maximize clarity and simplicity, while perhaps sacrificing some rigor. The article is targeted at an audience with a good grounding in basic quantum mechanics, but assumes no additional knowledge regarding surface codes, error correction, or topological information processing. We do however assume some prior knowledge of the basics of qubits and quantum computing, including familiarity with single qubit operations such as $\hat{X}$ bit-flips and $\hat{Z}$ phase-flips, the Hadamard gate, and the two-qubit controlled-NOT (CNOT) gate as well as the multi-qubit Toffoli gate. We also assume a working understanding of quantum circuits and their terminology. We refer the uninitiated reader to one of the excellent texts on this topic, e.g. Ref. [1] or [2].

## I. BACKGROUND

Quantum computers provide a means to solve certain problems that cannot be solved in a reasonable period of time using a conventional, classical computer. These problems include factoring very large numbers into their primes, which on a quantum computer can be accomplished relatively quickly using Shor's algorithm [3], and searching large, unstructured data sets, which can be done on a quantum computer using Grover's search algorithm [4, 5]. A number of physical systems are being explored for their use in quantum computing, including ions, spins in semiconductors, and superconducting circuits [6, 7]. However, none of these systems perform sufficiently well to serve directly as computational qubits. It is however possible to construct a *logical* qubit from a collection of physical qubits, such that the logical qubit performs much better than the individual physical qubits.

One approach to building a quantum computer is based on *surface codes* [8, 9], operated as *stabilizer codes* [10]. The surface codes evolved from an invention of Alexei Kitaev known as *toric codes* [11–14], which arose from his efforts to develop simple models for topological order, using qubits distributed on the surface of a toroid. The toroidal geometry employed by Kitaev turned out to be unnecessary, and planar versions (thus "surface codes") were developed by Bravyi and Kitaev as well as Freedman and Meyer [8, 15].

One of the significant advantages of surface codes is their relative tolerance to local errors, as was first described by Preskill and co-workers [16]. In this publication, the critical logical CNOT operation was implemented using stacked layers of surfaces, a three-dimensional structure that significantly complicates potential physical implementations, but the ability of the surface codes to withstand large error rates was apparent: These authors showed that the surface codes could handle error rates of almost 3% per surface code clock cycle, assuming the ability to measure a four-qubit operator.

Raussendorf and co-workers then discovered that the logical CNOT operation could be implemented by braid transformations on a single surface, a highly significant simplification [17–19]. These authors also evaluated error tolerances for a fully planar implementation using only one- and two-qubit nearest-neighbor gates, arriving at an error threshold of 0.75% per operation.

The literature on surface codes is somewhat opaque. Increasingly accessible descriptions have appeared in more recent publications [20, 21], including thorough analyses of errors and their propagation [21, 22] and the ongoing development of efficient classical control software [23]. A number of authors are working on improving the classical processing associated with the surface code [24–28], as well as other two-dimensional topological codes [29–32].

The tolerance of surface codes to errors, with a per-operation error rate as high as about 1% [22, 23], is far less stringent than that of other quantum computational approaches. For example, calculations of error tolerances of the Steane and Bacon-Shor codes, implemented on two-dimensional lattices with nearest-neighbor coupling, find per-step thresholds of about $2 \times 10^{-5}$ [33, 34], thus requiring three orders of magnitude lower error rate than the surface code. The error tolerance of the surface code, along with a simple two-dimensional physical layout with only nearest-neighbor coupling, makes a surface code architecture one of the most realistic approaches to building a solid-state quantum computer.

As the reader will discover, the price paid for the high error tolerance is that implementations of the surface code involve large numbers of physical qubits — as do most other approaches to quantum computing [33, 34]. It takes a minimum of thirteen physical qubits to implement a single logical qubit. A reasonably fault-tolerant logical qubit that can be used effectively in a surface code takes of order $10^3$ to $10^4$ physical qubits.[1] The number of physical qubits needed to find a number's prime factors, using Shor's algorithm [3], depends on a tradeoff between physical size and time of computation. These trade-offs are apparent in Table I, where we display the approximate number of computational logical qubits, the number of sequential Toffoli gates, and the number of total Toffoli gates needed to factor an $N$-bit number, for different types of factoring circuits.[2] In general, while the computational time gets smaller as one moves down the table, the spatial extent of the circuit becomes larger, as the number of logical qubits (the circuit "footprint") scales with the ratio of the total number of Toffoli gates to the number of sequential Toffoli gates.

We can make a rough estimate of the time and circuit size needed to factor a number with $N = 2,000$ bits (600

| Number of computational logical qubits | Sequential Toffoli gates | Total Toffoli gates | References |
|---|---|---|---|
| $2N$ | $40N^3$ | $40N^3$ | [35–37] |
| $5N$ | $600N^2$ | $\mathcal{O}(N^3 \log N)$ | [38] |
| $2N^2$ | $15N \log^2 N$ | $\mathcal{O}(N^3 \log^2 N)$ | [39] |
| $\mathcal{O}(N^3)$ | $\mathcal{O}(\log^3 N)$ | $\mathcal{O}(N^3 \log^3 N)$ | [40] |

TABLE I. Trade-off between number of computational logical qubits and number of sequential and total Toffoli gate operations for factoring an $N$-bit number into its primes using Shor's algorithm. Each line in the table corresponds to a different quantum circuit implementing the algorithm. The physical size of a circuit scales with the ratio of the total number of Toffoli gates to the number of sequential Toffoli gates.

decimal digits), using a circuit size scaling as in the first line of Table I, and making assumptions about the physical qubit error rates and gate times; more details on this estimate are provided in Appendix M.[3] This Shor's algorithm implementation is constructed from ideas in Ref. [35–37], and involves a resource-intensive modular exponentiation that requires approximately $40N^3 \approx 3 \times 10^{11}$ sequential Toffoli gates. The modular exponentiation thus determines the total execution time for the factoring algorithm. A highly optimized version of this circuit [41] can complete each Toffoli gate in approximately three physical qubit measurement cycles. If we assume a physical qubit measurement time of 100 ns, it will take about 26.7 hours to complete the exponentiation.

The spatial extent of the circuit is determined in part by the number of computational logical qubits, which for this circuit is about $2N = 4,000$. A much larger part of the surface code is however needed to generate and purify the special ancilla $|A_L\rangle$ states that are used in the Toffoli gates. Each Toffoli gate consumes seven $|A_L\rangle$ states [37]. In total, the exponentiation circuit therefore requires approximately $280N^3 \approx 2.2 \times 10^{12}$ $|A_L\rangle$ states. The surface code must be able to generate these states at a rate sufficient to keep pace with the exponentiation circuit.

The number of physical qubits needed to define a logical qubit is strongly dependent on the error rate in the physical qubits. Error rates just below the threshold require larger numbers of physical qubits per logical qubit, while error rates substantially smaller than the threshold allow smaller numbers of physical qubits. Here we assume an error rate approximately one-tenth the threshold rate, which implies that we need about 14,500 physical qubits per logical qubit to give a sufficiently low logical error rate to successfully execute the algorithm.

We can now estimate the number of physical qubits needed for our factoring problem. The 4,000 computational logical qubits require a total of about $4000 \times 14,500 = 58$ million physical qubits. As discussed in Appendix M, generating and purifying one $|A_L\rangle$ state takes a surface code area of about 800,000 physical qubits, and takes about 100 $\mu$s. Generating the full set of $2.2 \times 10^{12}$ $|A_L\rangle$ states over the 26.7 hours of execution takes about a billion physical qubits. The computational qubits, which are separate from those used to generate the $|A_L\rangle$ states, therefore comprise less than 6% of the quantum computer footprint, so in total we see that with the assumed error rate, the full quantum computer needs about a billion physical qubits, operating for about one day.

---

[1] This number depends strongly on the rate that errors occur on the physical qubits.

[2] A Toffoli gate is a three-qubit gate, with two controls determining the result on one target. The Toffoli can be described as a "controlled-controlled NOT": If both controls are "true", the target is flipped, "false"→"true" or "true"→"false", while otherwise the target is unchanged. This reversible gate can be implemented in quantum logic using a combination of two-qubit controlled-NOTs and single qubit gates, and can be used to construct any arithmetical operation. The Toffoli is needed to implement Shor's algorithm; see Ref. [1, 2].

[3] We look at this circuit in detail, as it has the smallest footprint, albeit with the longest factoring time.

The size of the quantum computer is quite sensitive to the error rate in the physical qubits. For example, improving the overall error rate by about a factor of ten, as detailed in Appendix M, can reduce the number of physical qubits by about an order of magnitude, to about 130 million, although leaving the execution time unchanged.

Further reduction of the qubit overhead can in principle be achieved by speeding up the surface code operation. Faster operation means fewer qubits are needed to generate $|A_L\rangle$ states at the necessary rate. Note however that the algorithm execution time is set by the modular exponentiation, and can only be reduced by reducing the measurement time. However, as will be clear later, operating a surface-code-based quantum computer requires intimate classical monitoring and control of the physical qubits, with the control circuitry driven by classical logic. Reducing the logical gate time will then put severe demands on this classical control hardware. Given the current speed of digital logic, logical gate times for the quantum computer in the range of 0.1-10 microseconds are probably realistic. Longer logical gate times can be used, however this requires more qubits one keeps the overall execution time constant. The number of qubits can be reduced by slowing down the modular exponentiation, as then the rate of $|A_L\rangle$ consumption is reduced, resulting in a smaller footprint for the $|A_L\rangle$ production. This may, however, lead to unacceptably long factoring times. We return to this question, and the issue of physical implementations, in a brief discussion at the end of this article.

## II. INTRODUCTION

Quantum computation relies on the use of qubits, which are two-level quantum systems. The prototypical two-level system is an electron spin in a magnetic field, from which much of the terminology originates; all of the quantum properties of an electron spin are captured by the algebra of the Pauli operators $\hat{\sigma}_x$, $\hat{\sigma}_y$, $\hat{\sigma}_z$ and the identity $\hat{I}$. In keeping with the literature on quantum computation, we will be using the qubit operators $\hat{X}$, $\hat{Y}$ and $\hat{Z}$, defined in terms of the Pauli operators by $\hat{X} = \hat{\sigma}_x$, $\hat{Y} = -i\hat{\sigma}_y$, and $\hat{Z} = \hat{\sigma}_z$ (see Appendix A). The algebra of the qubit operators is almost identical to that of the

Pauli operators:

$$
\begin{aligned}
\hat{X}^2 &= -\hat{Y}^2 = \hat{Z}^2 = \hat{I}, \\
\hat{X}\hat{Z} &= -\hat{Z}\hat{X}, \\
\left[\hat{X}, \hat{Y}\right] &\equiv \hat{X}\hat{Y} - \hat{Y}\hat{X} = -2\hat{Z},
\end{aligned} \tag{1}
$$

where the last relation holds for cyclic permutations of $\hat{X}$, $\hat{Y}$ and $\hat{Z}$, except for $[\hat{Z}, \hat{X}] = +2\hat{Y}$.

Any two-level quantum system that satisfies the relations (1) can in principle be used as a qubit. In fact, any system in which one can define $\hat{X}$ and $\hat{Z}$ operators that satisfy the relations (1) can be used as a qubit, even if the system has more than two degrees of freedom.[4] Just having $\hat{X}$ and $\hat{Z}$ operators on a qubit is not sufficient, as a quantum computer needs a few more single-qubit gates as well as an entangling two-qubit gate. The Solovay-Kitaev theorem [2, 12] implies that one set of operators sufficient to implement an arbitrary quantum algorithm comprises the single-qubit operators $\hat{X}$, $\hat{Z}$, the Hadamard $\hat{H}$, the $\hat{S}$ and $\hat{S}^\dagger$ phase gates, and the $\hat{T}$ and $\hat{T}^\dagger$ gates, as well as a two-qubit controlled-NOT (CNOT) gate (any two-qubit gate from which a CNOT can be constructed is of course acceptable).[5] The matrix representations of these operators are given in Appendix A.

In the surface code, physical qubits are entangled together using a sequence of physical qubit CNOT operations, with subsequent measurements of the entangled states providing a means for error correction and error detection. A set of physical qubits entangled in this way is used to define a *logical* qubit, which due to the entanglement and measurement has far better performance than the underlying physical qubits. We will describe how logical qubits are constructed in the surface code, and also show how the complete set of single logical qubit gates and the two-qubit logical CNOT are constructed, allowing us to implement quantum algorithms based on these logical qubits.

The qubit $\hat{Z}$ eigenstates are called the ground state $|g\rangle$ and the excited state $|e\rangle$. The ground state is the $+1$ eigenstate of $\hat{Z}$, with $\hat{Z}|g\rangle = +|g\rangle$, and the excited state is the $-1$ eigenstate, with $\hat{Z}|e\rangle = -|e\rangle$.[6] It is tempting to think of the qubit as a kind of quantum transistor, with the ground state corresponding to "off" and the excited state to "on". However, in distinct contrast to a classical logic element, a qubit can exist in a superposition of its eigenstates, $|\psi\rangle = \alpha|g\rangle + \beta|e\rangle$, so a qubit can be both "off" and "on" at the same time. A measurement $M_Z$ of the qubit will however return only one of two possible measurement outcomes, $+1$ with the qubit state projected to $|g\rangle$, or $-1$ with the qubit state projected to $|e\rangle$. A quantum state is furthermore relatively delicate, easily perturbed by interactions with the outside world. These interactions can be intentional, or can arise from errors, energy decay from $|e\rangle$ to $|g\rangle$, or from fluctuations in the energy difference between the qubit eigenstates, i.e. fluctuations in the qubit transition frequency. When unintended, these changes to the qubit state comprise quantum errors, and present one of the largest challenges in quantum computing.

---

[4] Note that we are defining a qubit to have two real degrees of freedom.

[5] Note this is not a minimal set, which would be $\hat{T}$, the Hadamard $\hat{H}$ and the CNOT, as we have the identities $\hat{T}^2 = \hat{S}$, $\hat{T}^4 = \hat{Z}$, $\hat{H}\hat{Z}\hat{H} = \hat{X}$, $\hat{Z}\hat{S} = \hat{S}^\dagger$, and $\hat{T}^7 = \hat{T}^\dagger$.

[6] Typically $\hat{Z}$ serves as the energy quantization axis, from which the names originate; the larger eigenvalue for the ground with respect to the excited state is because the Hamiltonian is proportional to $-\hat{Z}$.

Qubit errors can be modeled by introducing random $\hat{X}$ bit-flip and $\hat{Z}$ phase-flip operators in the evolution of the qubit state (we note that the $\hat{Y}$ operator is the combination $\hat{Y} = \hat{Z}\hat{X}$, and is thus a combined bit- and phase-flip). If these errors are rare, the amplitudes for these operators will be correspondingly small. This modeling of errors can describe a quite wide range of single-qubit errors; see e.g. Ref. [42].

The operator model for single-qubit errors implies that these errors can, in principle, be undone by quantum correction gates: An erroneous $\hat{Z}$ can be canceled by subsequently applying an intentional $\hat{Z}$, since $\hat{Z}^2 = \hat{I}$. If we detect all the errors, we can correct them by repeatedly applying quantum correction gates. However, one feature of the surface code is that errors only need to be corrected when they affect measurement outcomes, and thus one merely needs to identify errors, and then correct any measurements that are affected by these errors. This can be done entirely in the classical system used to control the surface code, as we describe in Section IX. For example, a $\hat{Z}$ phase-flip error that is detected immediately can be corrected by changing the sign of any subsequent $\hat{X}$ measurements, whereas an $\hat{X}$ error will have no effect on the same $\hat{X}$ measurement; any subsequent $\hat{Z}$ measurements would similarly have to be corrected for an $\hat{X}$ error but not for a $\hat{Z}$ error. This means that as long as errors can be detected promptly, they can be undone in classical software. One of the important aspects of the surface code is therefore a focus on error *detection* rather than error *correction*.

As errors occur from the random and unpredictable appearance of $\hat{X}$ and $\hat{Z}$ operations, they must be detected by repeatedly measuring each qubit, which can be done with combined $\hat{X}$ and $\hat{Z}$ measurements. However, because $[\hat{X}, \hat{Z}] \neq 0$, sequential measurements of $\hat{X}$ and $\hat{Z}$ on the same qubit conflict with one another, causing random projections of the qubit state onto these operators' respective eigenstates, completely destroying the quantum state. Specifically, a $\hat{Z}$ measurement $M_Z$ probabilistically projects the qubit state onto $|g\rangle$ or $|e\rangle$, yielding the corresponding $+1$ or $-1$ eigenvalues respectively, with complete loss of the initial state's amplitude and phase (unless it was originally in a $\hat{Z}$ eigenstate). A subsequent measurement $M_X$ of $\hat{X}$ will project the qubit state onto the $\hat{X}$ eigenstates $|+\rangle = (|g\rangle + |e\rangle)/\sqrt{2}$ or $|-\rangle = (|g\rangle - |e\rangle)/\sqrt{2}$, with $+1$ and $-1$ measurement outcomes, respectively, again destroying the state amplitude and phase. This again is in contrast with classical logic, in which a logic element has only bit-flip errors between "off" and "on", and measurement of the logic state does not perturb that state.

The projective measurement problem can however be avoided by measuring more than one qubit at a time, making non-destructive quantum error detection possible. Consider a two-qubit system, with qubits $a$ and $b$, which we measure using the two-qubit operators $\hat{X}_a \hat{X}_b$ and $\hat{Z}_a \hat{Z}_b$. These operators, even though they represent separate $\hat{X}$ and $\hat{Z}$ measurements, actually commute:

$$
\begin{aligned}
\left[\hat{X}_a \hat{X}_b, \hat{Z}_a \hat{Z}_b\right] &= \left(\hat{X}_a \hat{X}_b\right)\left(\hat{Z}_a \hat{Z}_b\right) - \left(\hat{Z}_a \hat{Z}_b\right)\left(\hat{X}_a \hat{X}_b\right) \\
&= \hat{X}_a \hat{Z}_a \hat{X}_b \hat{Z}_b - \hat{Z}_a \hat{X}_a \hat{Z}_b \hat{X}_b \\
&= \left(-\hat{Z}_a \hat{X}_a\right)\left(-\hat{Z}_b \hat{X}_b\right) - \hat{Z}_a \hat{X}_a \hat{Z}_b \hat{X}_b \\
&= 0.
\end{aligned}
\tag{2}
$$

We have used the fact that operators on different qubits always commute, so e.g. $[\hat{X}_a, \hat{Z}_b] = 0$. Measurements of these two-qubit operators are thus compatible, so a two-qubit state can actually be a simultaneous eigenstate of both $\hat{X}_a \hat{X}_b$ and $\hat{Z}_a \hat{Z}_b$.[7]

In Table II we display the four simultaneous eigenstates of $\hat{X}_a \hat{X}_b$ and $\hat{Z}_a \hat{Z}_b$, with their corresponding eigenvalues. We see that these states are actually the four Bell states. We note that each qubit has two degrees of freedom (ignoring the overall phase), so the pair has four degrees of freedom. Specifying the two-operator eigenvalues, which are by definition real and thus impose two constraints on the qubits, restricts the system to only one quantum state. The operators $\hat{X}_a \hat{X}_b$ and $\hat{Z}_a \hat{Z}_b$ therefore form a complete set for this two-qubit system.

Any of the eigenstates in Table II can be repeatedly measured by $\hat{X}_a \hat{X}_b$ and $\hat{Z}_a \hat{Z}_b$, and these measurements will not change the state. If there is an externally-induced $\hat{X}$ or $\hat{Z}$ error on one of the qubits, then the two-qubit measurements will project the qubit state onto one of the other two-qubit eigenstates, and the measurement eigenvalues will change, signaling that an error has occurred. For example, consider the state $(|gg\rangle + |ee\rangle)/\sqrt{2}$, with eigenvalues $(+1, +1)$, in the first line of Table II. An $\hat{X}_b$ error on the second qubit will transform this state to $(|ge\rangle + |eg\rangle)/\sqrt{2}$, which is also an eigenstate but with eigenvalues $(-1, +1)$; the change in the first eigenvalue will signal this error. Note however that this error cannot be distinguished from an $\hat{X}_a$ error, which yields the same final state and final eigenvalues. By contrast, a $\hat{Z}_a$ error on the first qubit will transform $(|gg\rangle + |ee\rangle)/\sqrt{2}$ to $(|gg\rangle - |ee\rangle)/\sqrt{2}$, which is an eigenstate with eigenvalues $(+1, -1)$, changing the second eigenvalue (a $\hat{Z}_b$ error will give the same outcome as $\hat{Z}_a$). Hence, while errors can always be detected, they cannot be uniquely identified: A more complex system is needed to achieve that, the surface code being one example.

---

[7] The careful reader may be confused by our notation: When we use a qubit operator product such as $\hat{X}\hat{Z}$ without subscripts on the operators, the two operators act on the same qubit; the $2 \times 2$ matrix representations of each operator are multiplied together to give a $2 \times 2$ matrix representation for the product. When we instead have subscripts, such as $\hat{X}_a \hat{X}_b$, the operators act on *different* qubits, so formally we should write this as an outer product $\hat{X}_a \otimes \hat{X}_b$; this outer product operator acts on a 4-dimensional Hilbert space, so its matrix representation is a $4 \times 4$ matrix formed by the outer product of the two $2 \times 2$ representations of $\hat{X}_a$ and $\hat{X}_b$.

| $\hat{Z}_a\hat{Z}_b$ | $\hat{X}_a\hat{X}_b$ | $|\psi\rangle$ |
|:---:|:---:|:---:|
| $+1$ | $+1$ | $(|gg\rangle + |ee\rangle)/\sqrt{2}$ |
| $+1$ | $-1$ | $(|gg\rangle - |ee\rangle)/\sqrt{2}$ |
| $-1$ | $+1$ | $(|ge\rangle + |eg\rangle)/\sqrt{2}$ |
| $-1$ | $-1$ | $(|ge\rangle - |eg\rangle)/\sqrt{2}$ |

TABLE II. Eigenstates of the two-qubit operators $\hat{Z}_a\hat{Z}_b$ and $\hat{X}_a\hat{X}_b$. The four eigenstates are the Bell states for this system.

The operator products $\hat{X}_a\hat{X}_b$ and $\hat{Z}_a\hat{Z}_b$ are called *stabilizers*. Stabilizers are very important in preserving quantum states: By repeatedly measuring a quantum system using a complete set of commuting stabilizers, the system is forced into a simultaneous and unique eigenstate of all the stabilizers. One can measure the stabilizers without perturbing the system; when the measurement outcomes change, this corresponds to one or more qubit errors, and the quantum state is projected by the measurements onto a different stabilizer eigenstate.

## III. THE SURFACE CODE

The two-qubit stabilizer example demonstrates a simple form of error detection. More complex circuits can detect and precisely identify errors in much larger assemblies of qubits; the surface code is such an example. We implement the surface code on a two-dimensional array of physical qubits, as shown in Fig. 1. The qubits are either *data qubits*, represented by open circles in Fig. 1a, in which the computational quantum states are stored, or *measurement qubits*, represented by filled circles. All of the data and measurement qubits must meet the basic requirements for quantum computation: Initialization, single-qubit rotations, and a two-qubit controlled-NOT (CNOT) between nearest neighbors. In addition, in order to perform a topological version of the Hadamard transformation, the data qubits and measurement qubits must be able to exchange their quantum states (a SWAP operation). A method to measure $\hat{Z}$ for each qubit is also required.

The measurement qubits are used to stabilize and manipulate the quantum state of the data qubits. There are two types of measurement qubits, "measure-Z" qubits, colored green (dark) in Fig. 1a, and "measure-X" qubits, colored orange (light); these are called *Z syndrome* and *X syndrome* qubits in the surface code literature; see e.g. Ref. [20]. Each data qubit is coupled to two measure-Z and to two measure-X qubits, and each measurement qubit is coupled to four data qubits. A measure-Z qubit is used to force its neighboring data qubits $a$, $b$, $c$ and $d$ into an eigenstate of the operator product $\hat{Z}_a\hat{Z}_b\hat{Z}_c\hat{Z}_d$: Each measure-Z qubit therefore measures a $\hat{Z}$ stabilizer. A measure-X qubit likewise forces its neighboring data qubits into an eigenstate of $\hat{X}_a\hat{X}_b\hat{X}_c\hat{X}_d$, and therefore measures an $\hat{X}$ stabilizer.

## IV. QUIESCENT STATE OF THE SURFACE CODE

The measure-Z and measure-X qubits that stabilize the surface code are operated in a very particular sequence, with one complete cycle shown in Fig. 1b and c, for a single measure-Z and measure-X qubit, respectively. After initializing each measure qubit in its ground state $|g\rangle$, the heart of the sequence comprises four CNOT operations followed by a projective measurement. For the measure-Z qubit, the CNOTs target the measure qubit with the four nearest-neighbor data qubits as the controls, with the projective measurement yielding an eigenstate of $\hat{Z}_a\hat{Z}_b\hat{Z}_c\hat{Z}_d$ (see Appendix B, as well as [2]; eigenstates are listed in Table III). For the measure-X qubit, the four CNOTs target the nearest-neighbor data qubits using the measure qubit as the control, and the sequence also includes a Hadamard applied to the measure qubit before and after the CNOTs; the projective measurement yields an eigenstate of $\hat{X}_a\hat{X}_b\hat{X}_c\hat{X}_d$. Hence, after the projective measurement of all the measure qubits in the array, the state $|\psi\rangle$ of all the data qubits simultaneously satisfies $\hat{Z}_a\hat{Z}_b\hat{Z}_c\hat{Z}_d|\psi\rangle = Z_{abcd}|\psi\rangle$, with eigenvalues $Z_{abcd} = \pm 1$, and $\hat{X}_a\hat{X}_b\hat{X}_c\hat{X}_d|\psi\rangle = X_{abcd}|\psi\rangle$ with eigenvalues $X_{abcd} = \pm 1$. Following measurement, the cycle is repeated.[8] The measure qubits in Fig. 1b and c all operate in lock-step, so that every step in the cycle shown in the figure is completed over the entire 2D array before the next step begins. We note that the zig-zag sequence $abcd$ followed by each of the measure qubits is quite particular, and cannot be easily modified while preserving the stabilizer property (see Appendix B).

---

[8] A capital roman letter with a "hat", e.g. $\hat{X}$, designates an operator, while a capital roman letter by itself, $X$, represents the outcome of a measurement of that operator, which must be an eigenvalue of the operator. A stabilizer $\hat{X}_a\hat{X}_b\hat{X}_c\hat{X}_d$ is the outer product of four physical qubit $\hat{X}_j$ operators, so would be represented by a $2^4 \times 2^4 = 16 \times 16$ matrix; its measurement outcome $X_{abcd}$ is an eigenvalue of this matrix. Note measuring the product $\hat{X}_a\hat{X}_b\hat{X}_c\hat{X}_d$ does not yield the same result as measuring each individual $\hat{X}_a$, $\hat{X}_b$, $\hat{X}_c$ and $\hat{X}_d$, as the qubits are in general not in a product eigenstate of the individual $\hat{X}_j$ operators, so measuring the individual $\hat{X}_j$ would cause undesirable projections.

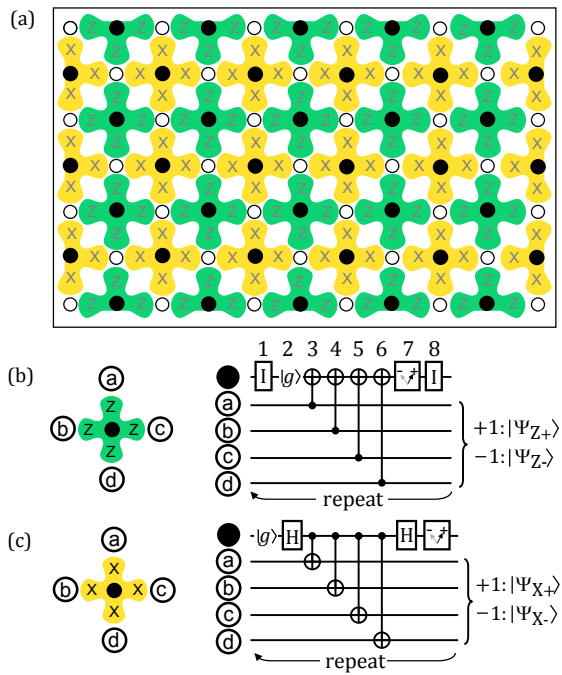

FIG. 1. (Color online)(a) A two-dimensional array implementation of the surface code. Data qubits are open circles (○), measurement qubits are filled circles (●), with measure-Z qubits colored green (dark) and measure-X qubits colored orange (light). Away from the boundaries, each data qubit contacts four measure qubits, and each measure qubit contacts four data qubits; the measure qubits perform four-terminal measurements. On the boundaries, the measure qubits contact only three data qubits and perform three-terminal measurements, and the data qubits contact either two or three measure qubits. The solid line surrounding the array indicates the array boundary. (b) Geometric sequence of operations (left), and quantum circuit (right) for one surface code cycle for a measure-Z qubit, which stabilizes $\hat{Z}_a\hat{Z}_b\hat{Z}_c\hat{Z}_d$. (c) Geometry and quantum circuit for a measure-X qubit, which stabilizes $\hat{X}_a\hat{X}_b\hat{X}_c\hat{X}_d$. The two identity $\hat{I}$ operators for the measure-Z process, which are performed by simply waiting, ensure that the timing on the measure-X qubit matches that of the measure-Z qubit, the former undergoing two Hadamard $\hat{H}$ operations. The identity operators come at the beginning and end of the sequence, reducing the impact of any errors during these steps.

Stabilizer codes have the remarkable property that they do not operate from the system ground state, but instead from the state $|\psi\rangle$ that results from the concurrent measurement of all the stabilizers; we call this the *quiescent state*. The quiescent state $|\psi\rangle$ is randomly selected by completing one full surface code cycle, which is the sequence shown in Fig. 1b and c, starting for exam-ple with all data and measurement qubits in their ground states $|g\rangle$.

| Eigenvalue | $\hat{Z}_a\hat{Z}_b\hat{Z}_c\hat{Z}_d$ | $\hat{X}_a\hat{X}_b\hat{X}_c\hat{X}_d$ |
|---|---|---|
| +1 | $|gggg\rangle$ | $|{+}{+}{+}{+}\rangle$ |
|  | $|ggee\rangle$ | $|{+}{+}{-}{-}\rangle$ |
|  | $|geeg\rangle$ | $|{+}{-}{-}{+}\rangle$ |
|  | $|eegg\rangle$ | $|{-}{-}{+}{+}\rangle$ |
|  | $|egge\rangle$ | $|{-}{+}{+}{-}\rangle$ |
|  | $|gege\rangle$ | $|{+}{-}{+}{-}\rangle$ |
|  | $|egeg\rangle$ | $|{-}{+}{-}{+}\rangle$ |
|  | $|eeee\rangle$ | $|{-}{-}{-}{-}\rangle$ |
| −1 | $|ggge\rangle$ | $|{+}{+}{+}{-}\rangle$ |
|  | $|ggeg\rangle$ | $|{+}{+}{-}{+}\rangle$ |
|  | $|gegg\rangle$ | $|{+}{-}{+}{+}\rangle$ |
|  | $|eggg\rangle$ | $|{-}{+}{+}{+}\rangle$ |
|  | $|geee\rangle$ | $|{+}{-}{-}{-}\rangle$ |
|  | $|egee\rangle$ | $|{-}{+}{-}{-}\rangle$ |
|  | $|eege\rangle$ | $|{-}{-}{+}{-}\rangle$ |
|  | $|eeeg\rangle$ | $|{-}{-}{-}{+}\rangle$ |

TABLE III. Eigenstates for the four-qubit stabilizers $\hat{Z}_a\hat{Z}_b\hat{Z}_c\hat{Z}_d$ and $\hat{X}_a\hat{X}_b\hat{X}_c\hat{X}_d$.

Once selected, in the absence of errors, the same state $|\psi\rangle$ will be maintained by each subsequent cycle of the sequence, with each measure qubit yielding a measurement outcome $X_{abcd}$ or $Z_{abcd}$ equal to that of the previous cycle. This occurs because all $\hat{X}$ and $\hat{Z}$ stabilizers commute with one another. This is trivial for stabilizers that do not have any qubits in common, as $\hat{X}$ and $\hat{Z}$ operators on different qubits always commute. Stabilizers that have qubits in common will always share two such qubits, so every $\hat{X}$ stabilizer shares two data qubits with each neighboring $\hat{Z}$ stabilizer and vice versa.[9] Hence we have, for an $\hat{X}$ and $\hat{Z}$ stabilizer that measure data qubits $a$ and $b$ in common,

$$
\begin{aligned}
[\hat{X}_a\hat{X}_b\hat{X}_c\hat{X}_d &\ , \ \hat{Z}_a\hat{Z}_b\hat{Z}_e\hat{Z}_f] \\
&= (\hat{X}_a\hat{Z}_a)(\hat{X}_b\hat{Z}_b)\hat{X}_c\hat{X}_d\hat{Z}_e\hat{Z}_f \\
&\quad - (\hat{Z}_a\hat{X}_a)(\hat{Z}_b\hat{X}_b)\hat{X}_c\hat{X}_d\hat{Z}_e\hat{Z}_f \\
&= 0,
\end{aligned}
\tag{3}
$$

as we get a minus sign from commuting $\hat{X}_a$ through $\hat{Z}_a$ as well as one from commuting $\hat{X}_b$ through $\hat{Z}_b$. Note the similarity of this four-qubit stabilizer commutator with the two-qubit stabilizer example in Eq. (2).

There are an enormous number of quiescent states that can be selected by the stabilizer measurements: If there are $N$ measure qubits in the array, there are $2^N$ measurement outcomes. The measurements at the end of each surface code cycle randomly project the data qubits onto one of these quiescent states. For the array in Fig. 1, with $N = 38$ measure qubits, there are $2^{38} \approx 3 \times 10^{11}$ possible quiescent states. We also note that the rapid, single-cycle projection onto a quiescent state implies that the quiescent state is not a fully entangled state of the entire array, but instead comprises local collections of highly entangled data qubits, with a smaller degree of entanglement between more distant groups of qubits. This is in some way analogous to the ground state of a superconductor, with Cooper-paired electrons strongly correlated over size scales small compared to the superconducting correlation length, less strongly correlated over larger length scales, yet still described by a single macroscopic order parameter.

---

[9] Note this is true both for qubits in the interior of the array as well as for qubits on the array boundaries.

## V. SINGLE QUBIT ERRORS

Once selected, a quiescent state remains unchanged except when disturbed by errors, for example erroneous single-qubit $\hat{X}$ bit-flip or $\hat{Z}$ phase-flip operations. These errors will be indicated by changes in the measurement outcomes. Consider for example a single data qubit error, represented by the erroneous $\hat{I}_a + \epsilon \hat{Z}_a$ operating on data qubit $a$; $|\epsilon| \ll 1$ is a small number equal to the probability amplitude for the $\hat{Z}$ phase-flip, and we neglect normalization. This error transforms the wavefunction $|\psi\rangle \rightarrow |\psi'\rangle = (\hat{I}_a + \epsilon \hat{Z}_a)|\psi\rangle$. Following this error, when the next surface code measurement cycle is completed, the state $|\psi'\rangle$ will be projected to an eigenstate of all the $\hat{X}_a \hat{X}_b \hat{X}_c \hat{X}_d$ and $\hat{Z}_a \hat{Z}_b \hat{Z}_c \hat{Z}_d$ operator products. This projects $|\psi'\rangle$ either back to the original state $|\psi\rangle$, which occurs with near unit probability and erases the error, or projects it to $\hat{Z}_a|\psi\rangle$, with probability $|\epsilon|^2$; the latter state is an eigenstate of all the stabilizers, as we will see in a moment. In the first case the measurement outcomes are the same as before the error, while in the second case the signs of the two measure-X qubits adjacent to data qubit $a$ will change. This is apparent from the following:

$$\hat{X}_a \hat{X}_b \hat{X}_c \hat{X}_d \left( \hat{Z}_a|\psi\rangle \right) = -\hat{Z}_a \left( \hat{X}_a \hat{X}_b \hat{X}_c \hat{X}_d|\psi\rangle \right)$$
$$= -X_{abcd} \left( \hat{Z}_a|\psi\rangle \right) \quad (4)$$

(recall that $X_{abcd}$ is the measurement outcome of the measure-X qubit). This shows that $\hat{Z}_a|\psi\rangle$ is an eigenstate of this $\hat{X}$ stabilizer, but with the opposite sign from $|\psi\rangle$. The same result applies to the other measure-X qubit adjacent to $a$: $\hat{Z}_a|\psi\rangle$ is an eigenstate of that stabilizer, with opposite sign compared to $|\psi\rangle$. The data qubit error will not change the outcomes of the two neighboring measure-Z qubits, as the $\hat{Z}$ operators all commute: $[\hat{Z}_a \hat{Z}_b \hat{Z}_c \hat{Z}_d, \hat{Z}_a] = 0$. In terms of the quiescent state,

$$\hat{Z}_a \hat{Z}_b \hat{Z}_c \hat{Z}_d \left( \hat{Z}_a|\psi\rangle \right) = \hat{Z}_a \left( \hat{Z}_a \hat{Z}_b \hat{Z}_c \hat{Z}_d|\psi\rangle \right)$$
$$= Z_{abcd} \left( \hat{Z}_a|\psi\rangle \right), \quad (5)$$

clearly again an eigenstate but with unchanged sign. Hence, in this second case, the state $|\psi'\rangle = (\hat{I}_a + \epsilon \hat{Z}_a)|\psi\rangle$ is projected to $\hat{Z}_a|\psi\rangle$, which is a different eigenstate of all the stabilizers in the array.

The erroneous phase-flip $\hat{Z}_a$ causes sign changes in the measurement outcomes of the two measure-X qubits adjacent to qubit $a$, allowing us to detect and localize the phase-flip error. We could then apply a second $\hat{Z}_a$ operator to this qubit, correcting the error (using the fact that $\hat{Z}_a^2 = \hat{I}_a$); however this phase-flip cannot be applied with 100% fidelity, so this could introduce more errors into the surface code state. It is safer to instead handle this phase-flip error in software, by recording on which qubit the phase-flip error occurred. The classical control software simply changes the sign of every subsequent measurement of that data qubit's two adjacent measure-X qubits, thus correcting for the effects of the error. The two neighboring measure-Z qubits are not affected by the error, so their measurement outcomes do not need to be corrected. Both bit- and phase-flip errors are handled in this way, with bit-flip errors corrected by changing the sign of the affected qubit's neighboring measure-Z qubits, leaving the measure-X outcomes unchanged. A second bit-flip error would cancel the first bit-flip error, and a second phase-flip error would cancel the first phase-flip, and any measurement corrections would then end.

Error detection requires that we locate and identify errors. A single data qubit $\hat{Z}_a$ error is signaled by *changes* in the measurement outcomes of the two measure-X qubits adjacent to the affected data qubit, with the changes occurring in one surface code cycle. Single qubit $\hat{X}_a$ errors, projecting to the state $\hat{X}_a|\psi\rangle$, will generate sign changes in the two measure-Z outcomes neighboring the data qubit, while the two neighboring measure-X outcomes will be unchanged. A $\hat{Y}_a = \hat{Z}_a \hat{X}_a$ error, projecting to the eigenstate $\hat{Z}_a \hat{X}_a|\psi\rangle$, will be signalled by sign changes in both of the neighboring measure-X and measure-Z qubits. These types of errors with their correlated signals are shown in Fig. 2. Once such an error has occurred, the state resulting from the projective measurements is again a quiescent state, with the difference from the original $|\psi\rangle$ reported by the measurement outcomes.

Errors occurring in the measurement process itself must also be considered; one such error will yield a sign change for that measure qubit only. On the next cycle, this measurement error will likely vanish, so this error will typically be signaled by a pair of sequential measurement changes occurring on a single measure qubit. Note that a measurement error could of course recur on the subsequent measurement, with a lower probability, and on the measurement following that, with an even lower probability, and so on. Establishing the value of a particular measurement therefore requires several surface code cycles, in order to catch single as well as sequential measurement errors.

The signal from a measurement error, as it is typically isolated on a single measure qubit, is distinct from that of a data qubit error, which is reported by two or more measurements separated in space. Other types of errors, such as CNOT errors, are discussed in [22], and also generate distinct patterns of sign changes in the measure-X and measure-Z qubits.

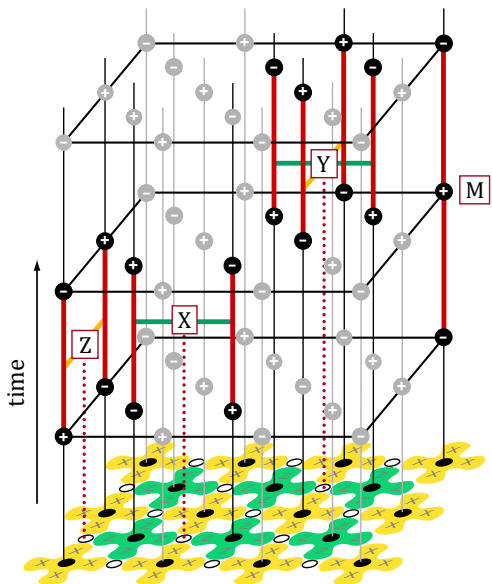

FIG. 2. (Color online) Schematic evolution of measurement outcomes (filled circles with ± signs), over a segment of the 2D array. Time progresses moving up from the array at the bottom of the figure, with measurement steps occurring in each horizontal plane. Vertical heavy red (gray) lines connect time steps in which a measurement outcome has changed, with the spatial correlation indicating an $\hat{X}$ bit-flip error, a $\hat{Z}$ phase-flip error, a $\hat{Y} = \hat{Z}\hat{X}$ error, and temporal correlation a measurement ($M$) error, which is sequential in time.

If the errors are sufficiently rare, the error signals will be well-isolated on the 2D array, i.e. in space as well as in time. The error signals can then be matched up to deduce which specific qubit error occurred, with very high probability of correctly identifying the error. Given the linearity of a quantum computer, if all the qubit errors that occur are correctly identified, it is possible to correct for all these errors in the classical control software, by applying corrective phase- and bit-flips to the qubit measurements, as discussed above. However, if the errors in the array are not so sparse, error identification becomes less straightforward. The *inverse problem*, determining which qubit errors actually occurred to generate a given set of error signals, does not have a unique solution, and alternative sets of qubit errors become likely as the error density increases. If mistakes are made in backing out the qubit errors, these mistakes will result in erroneous conclusions about the computational result. Ultimately this limits the surface code's ability to handle errors.

We will return to this discussion after introducing the surface code logical operators.

## VI. LOGICAL OPERATORS

How does one perform quantum logic in the surface code? It may appear that the surface code completely stabilizes the 2D array, and that it therefore locks the quantum system in a particular state, as in our earlier two-qubit example. However, the set of surface code stabilizers is actually not always complete, so the array can have additional degrees of freedom. These additional degrees of freedom can be used to define logical operators, the first step in defining a logical qubit. We can see this by considering the small 2D array shown in Fig. 3. This array has been drawn with two types of boundaries, terminating with measure-X qubits on the right and left, which we call X boundaries, and terminating with measure-Z qubits on the top and bottom, which we call Z boundaries. The X boundaries are called *smooth boundaries* in the surface code literature, while Z boundaries are called *rough boundaries* [20].

If we count the number of data and measure qubits in the array, we find there are 41 data qubits and 40 measure qubits, so there are $2 \times 41$ degrees of freedom in the data qubits with $2 \times 40$ constraints from the stabilizer measurements (each measurement is real-valued, thus the factor of two). The stabilizers in this array are all linearly independent, and no stabilizer can be written as a product of the other stabilizers, so these $2 \times 40$ constraints are all linearly independent.[10] The two unconstrained degrees of freedom indicate that this small array might serve as a (single) logical qubit; we still need to define the logical operators that manipulate these degrees of freedom.

How do we find the operators that will allow us to manipulate the additional degrees of freedom in the array, without affecting the stabilizers? A clue is provided by the fact that the stabilizer measurements commute because the stabilizers share pairs of data qubits. If we pair up operations on pairs of data qubits, we can create multi-qubit operator products that commute with the stabilizers; this is how we will build the logical operators, which we will term $\hat{X}_L$ and $\hat{Z}_L$.

We remind the reader that a single data qubit $\hat{X}$ operation changes the measure-Z outcomes on either side of the data qubit. Consider however two simultaneous $\hat{X}$ operations on two data qubits $a$ and $b$ that both neighbor

---

one measure-Z qubit. Then we have

$$
\begin{aligned}
\hat{Z}_a \hat{Z}_b \hat{Z}_c \hat{Z}_d \left( \hat{X}_a \hat{X}_b |\psi\rangle \right) &= (-1)^2 \hat{X}_a \hat{X}_b \left( \hat{Z}_a \hat{Z}_b \hat{Z}_c \hat{Z}_d |\psi\rangle \right) \\
&= Z_{abcd} \left( \hat{X}_a \hat{X}_b |\psi\rangle \right).
\end{aligned}
\tag{6}
$$

In other words, the product of two $\hat{X}$ operations commutes with a single $\hat{Z}$ stabilizer: $[\hat{Z}_a \hat{Z}_b \hat{Z}_c \hat{Z}_d, \hat{X}_a \hat{X}_b] = 0$.

Consider now Fig. 3, in particular the operation $\hat{X}_1$ on the left boundary. This operation is detected by the measure-Z qubit just to the right of the flipped data qubit, but we can at the same time perform an $\hat{X}_2$ operation on the data qubit on the other side of that measure-Z. The first $\hat{Z}$ stabilizer now commutes with the product of these two operations, but the measure-Z qubit to the right of $\hat{X}_2$ still reports a change; hence we add a third operation $\hat{X}_3$ to the set, and so on, until we have created a product of concurrent $\hat{X}$ operations $\hat{X}_L = \hat{X}_1 \hat{X}_2 \hat{X}_3 \hat{X}_4 \hat{X}_5$, which connects the two X boundaries on the left and right sides of the array. Every measure-Z qubit now interacts with pairs of data qubits appearing in the $\hat{X}_L$ operator, so by construction the product chain $\hat{X}_L$ commutes with all the $\hat{Z}$ stabilizers in the array (note $\hat{X}_L$ trivially commutes with the $\hat{X}$ stabilizers). Hence, if the $\hat{X}_L$ operator is applied to a quiescent state $|\psi\rangle$, generating the state $|\psi_X\rangle = \hat{X}_L |\psi\rangle$, the new state $|\psi_X\rangle$ will be a quiescent state with identical measurement outcomes to $|\psi\rangle$. Note that $|\psi_X\rangle$ is not equal to $|\psi\rangle$, as we have bit-flipped five data qubits in going from $|\psi\rangle$ to $|\psi_X\rangle$, and as $\hat{X}_L$ cannot be written as a product of stabilizers, $|\psi_X\rangle$ is not trivially related to $|\psi\rangle$ (unless $|\psi\rangle$ is an eigenstate of $X_L$, namely $|+_L\rangle$ or $|-_L\rangle$). The $\hat{X}_L$ operator thus manipulates one of the two degrees of freedom of the array in Fig. 3.

The other degree of freedom in the array can be manipulated by constructing a $\hat{Z}_L$ operator using a completely analogous product of $\hat{Z}$ data qubit operations: We again build a chain of operations using $\hat{Z}$ operators that are paired across each measure-X qubit that would otherwise report a measurement change. The chain of paired $\hat{Z}$ operators must cross the entire array, and further must start and end on a Z-boundary (rather than the X-boundaries on either end of the $\hat{X}_L$ chain). An example of such a chain is $\hat{Z}_L = \hat{Z}_6 \hat{Z}_7 \hat{Z}_3 \hat{Z}_8 \hat{Z}_9$ in Fig. 3. This $\hat{Z}_L$ operator chain commutes with all the stabilizers in the array, and thus when operating on a quiescent state, generates a new state $|\psi_Z\rangle = \hat{Z}_L |\psi\rangle$ with the same measurement outcomes as $|\psi\rangle$. The state $|\psi_Z\rangle$ clearly differs from $|\psi\rangle$, as we have phase-flipped five data qubits, and as $\hat{Z}_L$ cannot be written as a product of stabilizers, $|\psi_Z\rangle$ is not trivially related to $|\psi\rangle$; $\hat{Z}_L$ thus manipulates a second degree of freedom in the array. We show below that $\hat{X}_L$ and $\hat{Z}_L$ do not commute, so they manipulate two independent degrees of freedom just as for a physical qubit.

It may seem that we could choose other chains of single qubit operator products to define different $\hat{X}_L$ or $\hat{Z}_L$ operators. Consider for example the chain $\hat{X}'_L =$ $\hat{X}_1 \hat{X}_{10} \hat{X}_{11} \hat{X}_{12} \hat{X}_3 \hat{X}_4 \hat{X}_5$, also shown in Fig. 3. This chain of $\hat{X}$ operators satisfies the same conditions as $\hat{X}_L$, as the $\hat{X}$ data qubit operators are paired so as to bracket each measure-Z qubit. Hence $\hat{X}'_L$ commutes with all the stabilizers, and will generate a quiescent state $|\psi_{X'}\rangle = \hat{X}'_L |\psi\rangle$ with the same measurement outcomes as $|\psi\rangle$ and $|\psi_X\rangle$. However the state $|\psi_{X'}\rangle$ is actually linearly related to $|\psi_X\rangle$: First, note that we can manipulate the set of operators appearing in $\hat{X}'_L$, as

$$
\begin{aligned}
\hat{X}'_L &= \hat{X}_1 \hat{X}_{10} \hat{X}_{11} \hat{X}_{12} \hat{X}_3 \hat{X}_4 \hat{X}_5 \\
&= \left( \hat{X}_2 \hat{X}_{10} \hat{X}_{11} \hat{X}_{12} \right) \left( \hat{X}_1 \hat{X}_2 \hat{X}_3 \hat{X}_4 \hat{X}_5 \right) \\
&= \left( \hat{X}_2 \hat{X}_{10} \hat{X}_{11} \hat{X}_{12} \right) \hat{X}_L,
\end{aligned}
\tag{7}
$$

where we have used $\hat{X}_2^2 = \hat{I}$ in going from the first to the second line in Eq. (7). Now, the operator product in parentheses in the third line of Eq. (7) is just the stabilizer outlined in black in Fig. 3: In other words, we have simply multiplied $\hat{X}_L$ by an operator product that is stabilized to a $\pm 1$ eigenvalue by the surface code. If we operate on a quiescent state $|\psi\rangle$ with $\hat{X}'_L$, we can simply replace this operator product by its eigenvalue:

$$\hat{X}'_L|\psi\rangle = X_{2,10,11,12}\hat{X}_L|\psi\rangle = \pm\hat{X}_L|\psi\rangle = \pm|\psi_X\rangle, \quad (8)$$

where $X_{2,10,11,12} = \pm 1$ is the measurement outcome of the stabilizer $\hat{X}_2\hat{X}_{10}\hat{X}_{11}\hat{X}_{12}$. This holds for any operator $\hat{X}'_L$ that can be written as a stabilized operator product times $\hat{X}_L$, that is $\hat{X}'_L|\psi\rangle = \pm\hat{X}_L|\psi\rangle$, where the sign is determined by the measurement outcomes of the corresponding stabilizers. In fact, *any* $\hat{X}'_L$ chain that crosses the array in Fig. 3 can be written as $\hat{X}_L$ multiplied by a product of $\hat{X}$ stabilizers, as can easily be verified. Hence there is only one linearly independent $\hat{X}_L$ operator for this array.

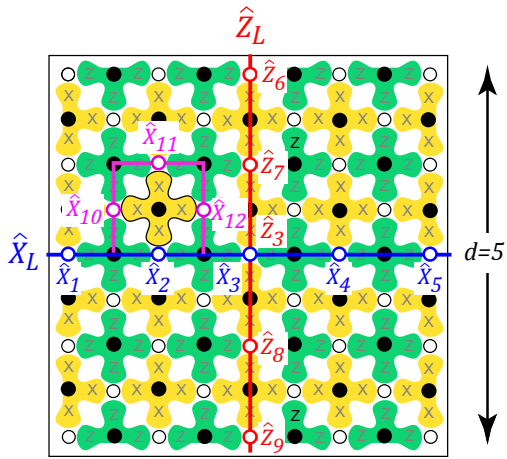

FIG. 3. (Color online) A square 2D array of data qubits, with X boundaries on the left and right, and Z boundaries on the top and bottom. The array has 41 data qubits, but only 40 $\hat{X}$ and $\hat{Z}$ stabilizers. A product chain $\hat{X}_L = \hat{X}_1 \hat{X}_2 \hat{X}_3 \hat{X}_4 \hat{X}_5$ of $\hat{X}$ operators connects the two X boundaries, commutes with all the array stabilizers and changes the array state from the quiescent state $|\psi\rangle$ to $|\psi_X\rangle = \hat{X}_L |\psi\rangle$ with the same measurement outcomes as $|\psi\rangle$. A second product chain $\hat{Z}_L = \hat{Z}_6 \hat{Z}_7 \hat{Z}_3 \hat{Z}_8 \hat{Z}_9$ connects the two Z boundaries and commutes with the array stabilizers; it changes the array state from $|\psi\rangle$ to $|\psi_Z\rangle = \hat{Z}_L |\psi\rangle$. The operator chains $\hat{X}_L$ and $\hat{Z}_L$ anti-commute. A modification of the $\hat{X}_L$ chain to the chain $\hat{X}'_L = \hat{X}_1 \hat{X}_{10} \hat{X}_{11} \hat{X}_{12} \hat{X}_3 \hat{X}_4 \hat{X}_5$ generates a quiescent state $|\psi_{X'}\rangle = X_{2,10,11,12} |\psi_X\rangle$, related to $|\psi_X\rangle$ by the result of the measurement $X_{2,10,11,12} = \pm 1$ of the encircled measure-X qubit (outlined in black).

This result may seem mysterious; after all, if we operate on the array with a loop of $\hat{X}$ operators corresponding to a stabilizer, e.g. the loop $\hat{X}_{\text{loop}} = \hat{X}_2\hat{X}_{10}\hat{X}_{11}\hat{X}_{12}$, we are bit-flipping four data qubits, so the state $|\psi'\rangle = \hat{X}_{\text{loop}}|\psi\rangle$ should be different from $|\psi\rangle$. However, the fact that $\hat{X}_{\text{loop}}$ is equal to a stabilizer means that $|\psi\rangle$ already includes superposition states of the un-bit-flipped and the bit-flipped states of the data qubits involved in $\hat{X}_{\text{loop}}$; this holds for any operator that is equal to the product of one or more stabilizers.[11]

The same argument applies to alternative $\hat{Z}'_L$ operators: Any such operator is equal to the original $\hat{Z}_L$ multiplied by a product of $\hat{Z}$ stabilizers, and thus any $\hat{Z}'_L|\psi\rangle$ differs from $\hat{Z}_L|\psi\rangle$ by $\pm 1$. Hence there is also only one linearly independent $\hat{Z}_L$ operator for this array.

We now show that the two operators $\hat{X}_L$ and $\hat{Z}_L$ satisfy the critical anti-commutation relation in Eq. (1). We have

$$\begin{aligned}
\hat{X}_L\hat{Z}_L &= \left(\hat{X}_1\hat{X}_2\hat{X}_3\hat{X}_4\hat{X}_5\right)\left(\hat{Z}_9\hat{Z}_{10}\hat{Z}_3\hat{Z}_{11}\hat{Z}_{12}\right) \\
&= \hat{X}_3\hat{Z}_3\left(\hat{X}_1\hat{X}_2\hat{X}_4\hat{X}_5\right)\left(\hat{Z}_9\hat{Z}_{10}\hat{Z}_{11}\hat{Z}_{12}\right) \\
&= -\hat{Z}_3\hat{X}_3\left(\hat{Z}_9\hat{Z}_{10}\hat{Z}_{11}\hat{Z}_{12}\right)\left(\hat{X}_1\hat{X}_2\hat{X}_4\hat{X}_5\right) \\
&= -\hat{Z}_L\hat{X}_L,
\end{aligned} \quad (9)$$

where all the single qubit operations are on different qubits, and thus commute, except for $\hat{X}_3$ and $\hat{Z}_3$, which anti-commute, $\hat{X}_3\hat{Z}_3 = -\hat{Z}_3\hat{X}_3$. This makes $\hat{X}_L$ and $\hat{Z}_L$ anti-commute. The logical operators $\hat{X}_L$ and $\hat{Z}_L$ therefore have exactly the same commutation relation as the physical qubit operators $\hat{X}$ and $\hat{Z}$. You can show that $\hat{X}_L^2 = \hat{Z}_L^2 = \hat{I}$. We can also construct the logical operator $\hat{Y}_L = \hat{Z}_L\hat{X}_L$, comprising two chains of $\hat{Z}$ and $\hat{X}$ operators that each cross the array. The set of logical operators $\hat{X}_L$, $\hat{Y}_L$ and $\hat{Z}_L$ satisfy all the commutation relations in Eq. (1), thus making this 2D array a logical qubit.

We will be using the logical operators $\hat{X}_L$ and $\hat{Z}_L$ to manipulate the degrees of freedom in the 2D array that are not constrained by the stabilizers. The state $|\psi\rangle$ describes the quantum state of all the data qubits; the surface code measurements ensure that $|\psi\rangle$ is an eigenstate of all the stabilizers, but in the case where we have more data qubits than stabilizers, $|\psi\rangle$ is not completely constrained. We can therefore write $|\psi\rangle$ as an outer product, $|\psi\rangle = |Q\rangle|q_L\rangle$, where $|Q\rangle$ is a vector in the $2^N$-dimensional Hilbert subspace on which the $N$ stabilizers operate, with $|Q\rangle$ constrained to a unique state in this Hilbert space, as determined by the $N$ stabilizer measurement outcomes. The remaining degrees of freedom in $|\psi\rangle$ are captured by $|q_L\rangle$. For the array in Fig. 3, with two unconstrained degrees of freedom, $|q_L\rangle$ is in a two-dimensional Hilbert space, i.e. represents a single qubit state. The stabilizers have no effect on $|q_L\rangle$, and the logical operators $\hat{X}_L$ and $\hat{Z}_L$ have no effect on $|Q\rangle$. The eigenstates of $\hat{Z}_L$ are $|q_L\rangle = |g_L\rangle$ and $|q_L\rangle = |e_L\rangle$, such that $\hat{Z}_L|g_L\rangle = +|g_L\rangle$ and $\hat{Z}_L|e_L\rangle = -|e_L\rangle$, while the corresponding eigenstates of $\hat{X}_L$ are $|q_L\rangle = |\pm_L\rangle = (|g_L\rangle \pm |e_L\rangle)/\sqrt{2}$, with eigenvalues $\pm 1$. Either pair, $|g_L\rangle$ and $|e_L\rangle$ or $|+_L\rangle$ and $|-_L\rangle$, can be used as a basis for the states $|q_L\rangle$. We can to some extent ignore the presence of $|Q\rangle$ when discussing manipulations of the logical state $|q_L\rangle$.

So far we have only shown how to define a single logical qubit in our array; later we will show how to make larger numbers of logical qubits within an array, which is done by creating more degrees of freedom within the array. Each logical qubit we create will increase the size of the logical Hilbert space of $|q_L\rangle$ by two and concomitantly reduce the size of the stabilizer Hilbert space of $|Q\rangle$ by two. With $n$ logical qubits, the Hilbert space for $|q_L\rangle$ has dimension $2^n$.

## VII. ERROR DETECTION

Now that we can construct surface code logical operators, we turn again to the consideration of errors in the surface code. The errors we consider here are on the *physical* qubits, either the data or the measure qubits. Errors occur due to single qubit errors (erroneous $\hat{X}$, $\hat{Y}$ or $\hat{Z}$ operations), measurement errors (reporting the incorrect outcome and projecting to the wrong state), initialization errors (setting a qubit to the wrong state), Hadamard errors (performing a Hadamard but in addition performing an erroneous $\hat{X}$, $\hat{Y}$ or $\hat{Z}$), and CNOT errors. Individual errors of these types are of course the most likely, but concatenated errors can also occur, for example two, three, or more adjacent data qubits suffering $\hat{X}$ errors in one surface code cycle, creating what are called *error chains*.

---

[11] In other words, if $|\psi\rangle$ includes qubits 2, 10, 11 and 12 in the state $|\ldots + - + + \ldots\rangle$, then $|\psi\rangle$ must also include the state where these qubits are in $|\ldots - + - - \ldots\rangle$, i.e. $|\psi\rangle = \pm|\ldots + - + + \ldots\rangle \pm |\ldots - + - - \ldots\rangle + \ldots$. Then operating with $\hat{X}_{\text{loop}}$ on $|\psi\rangle$ flips the first state to the second and vice versa without changing $|\psi\rangle$, $\hat{X}_{\text{loop}}|\psi\rangle = \mp|\ldots - + - - \ldots\rangle \mp |\ldots + - + + \ldots\rangle + \ldots = \mp|\psi\rangle$.

The surface code can deal with all these errors as long as the errors that occur during each surface code cycle can be identified (i.e. decoding which specific error(s) occurred on which particular qubit(s)).[12] Once identified, these errors can be tracked and the information used to correct any subsequent measurement outcomes using the classical control software. Edmonds' minimum-weight perfect-matching algorithm [43, 44] provides an automated method for doing this, and works perfectly for sufficiently sparse errors, but begins to fail as the error density increases, and as the length of the error chains increases. Numerical simulations using this type of algorithm can therefore provide estimates of the tolerance of the surface code to different types of errors. An example of a set of simulations of this kind is shown in Fig. 4.

The simulations used to generate Fig. 4 include the following types of errors, occurring during the surface code cycle shown in Fig. 1:

1. Attempting to perform a data qubit identity $\hat{I}$, but instead performing a single qubit operation $\hat{X}$, $\hat{Y}$, or $\hat{Z}$, each occurring with probability $p/3$.

2. Attempting to initialize a qubit to $|g\rangle$, but instead preparing $|e\rangle$ with probability $p$.

3. Attempting to perform a measure qubit Hadamard operation $\hat{H}$, but performing in addition one of the single qubit operations $\hat{X}$, $\hat{Y}$, or $\hat{Z}$, each with probability $p/3$.

4. Performing a measure qubit $\hat{Z}$ measurement, but reporting the wrong value and projecting to the wrong state with probability $p$.

5. Attempting to perform a measure qubit-data qubit CNOT, but instead performing one of the two-qubit operations $\hat{I}\otimes\hat{X}$, $\hat{I}\otimes\hat{Y}$, $\hat{I}\otimes\hat{Z}$, $\hat{X}\otimes\hat{I}$, $\hat{X}\otimes\hat{X}$, $\hat{X}\otimes\hat{Y}$, $\hat{X}\otimes\hat{Z}$, $\hat{Y}\otimes\hat{I}$, $\hat{Y}\otimes\hat{X}$, $\hat{Y}\otimes\hat{Y}$, $\hat{Y}\otimes\hat{Z}$, $\hat{Z}\otimes\hat{I}$, $\hat{Z}\otimes\hat{X}$, $\hat{Z}\otimes\hat{Y}$ or $\hat{Z}\otimes\hat{Z}$, each with probability $p/15$.[13]

The errors occur randomly during the simulation, with no correlation between errors. The probability $p$ is per step in the surface code cycle (there are eight steps in the cycle used for the simulation (Fig. 1), so the overall rate per cycle of the surface code is approximately $8p$, as discussed in more detail in Sect. VII B). Edmonds' matching algorithm maps changes detected in the stabilizer outcomes to physical qubit errors; the rate at which the algorithm makes mistakes, meaning it misidentifies the source of a particular error report, is displayed as $P_L$, the number of $\hat{X}_L$ errors appearing anywhere in the array, per surface code cycle.[14] This error rate is plotted as a function of the per-step error rate $p$.

The relation between $P_L$ and $p$ depends very strongly on a very important number, the array size $d$, which we will also call the *distance* of the array: $d$ is the minimum number of physical qubit bit-flips or phase-flips needed to define an $\hat{X}_L$ or $\hat{Z}_L$ operator. In Fig. 3, for example, we need a minimum of five physical operators to define a logical operator, so that array has a distance $d = 5$. At distance $d = 55$, the largest appearing in the simulations in Fig. 4, the equivalent array would have 55 data qubits along the horizontal and 55 data qubits along the vertical.

For small $p$, $P_L$ is small, and gets smaller as $d$ increases. For large $p$, $P_L$ is larger, and gets larger as $d$ increases. The cross-over between these two regimes occurs when $p$ crosses a threshold error rate $p_{th}$: For $p < p_{th}$, the logical error rate falls exponentially with $d$, while for $p > p_{th}$, $P_L$ increases with $d$. In Fig. 4, which applies to the stabilizer circuits in Fig. 1 simulated with the error sources listed above, the threshold rate is $p_{th} = 0.57\%$. We note this threshold is smaller than some of those mentioned earlier; this is in part because the threshold depends on the particular surface-code implementation, but mostly because we include more types of errors than were considered in earlier publications. As we discuss in more detail in Sect. VII B, the logical error rate $P_L$ responds in a strikingly different way to different types of errors.

For error rates $p < p_{th}$, the simulations show that the logical error rate scales with $p$ according to the power law $P_L \sim p^{d_e}$, where we define the error dimension for odd $d$ as

$$d_e = (d+1)/2. \tag{10}$$

For even $d$, we round down, so $d_e = d/2$. Using this, the error rate $P_L$ shown in Fig. 4 can be approximated more specifically by the empirical formula

$$P_L \cong 0.03 \ (p/p_{th})^{d_e}. \tag{11}$$

### A. Statistical model for the logical error rate

We can qualitatively understand the scaling in Eq. (11) by looking at the types of errors for which the matching algorithm fails. Consider Fig. 5a, in which two measure-Z qubits are reporting errors, marked by "E"s in the figure. Two examples of sets of data qubit errors that would

---

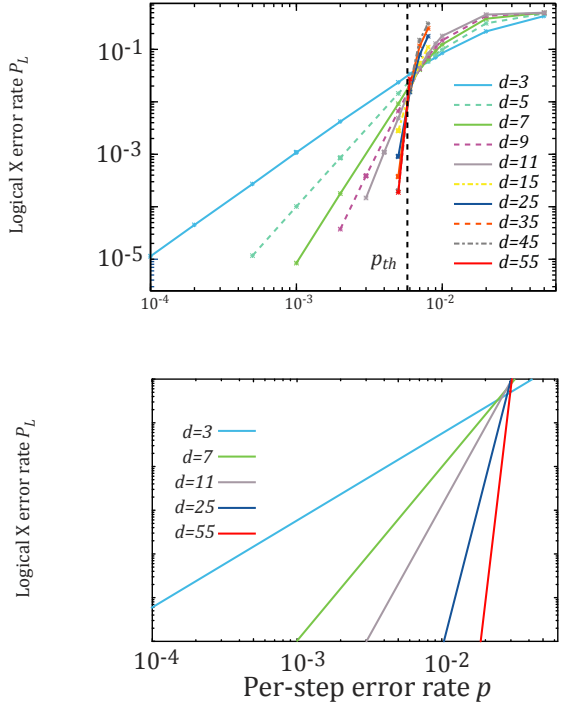

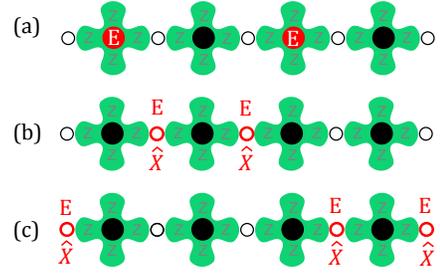

FIG. 4. (Color online) (a) Numerical simulations of surface code error rates, and how these error rates scale with the distance $d$ of the array. Per-step error rates $p$ less than the per-step threshold error rate of $p_{\mathrm{th}} = 0.57\%$ (dashed line) yield surface code logical error rates $P_L$ that vanish rapidly with increasing $d$. This threshold corresponds to roughly a 4% error rate for the entire surface code measurement cycle. (b) Estimated error rates, using statistical arguments given in main text, for array distances $d = 3, 7, 11, 25$ and 55. Note the estimate gives a logical error rate $P_L$ and a per-step threshold that are qualitatively similar to those from the more precise simulations.

FIG. 5. (Color online) (a) An example where two measure-Z qubits report errors in a single row of a 2D array, marked by "E"s. This error report could be generated by (b) two $\hat{X}$ errors appearing in the same surface code cycle on the 2nd and 3rd data qubit from the left, or (c) three $\hat{X}$ errors appearing in the other three data qubits in the row.

generate this error report are shown in Fig. 5b and c: In b, two of the five data qubits had $\hat{X}$ errors, while in c, the other three data qubits had $\hat{X}$ errors. These two events look identical in terms of measurement outcomes. A natural conclusion would be to assume that just two data qubit errors have occurred; this is of course more likely, as this will occur with probability proportional to $p^2$, while the triple error shown in Fig. 5c will occur at a rate scaling as $p^3$. More generally, for arrays with distance $d$, the most misidentifications will occur when $(d+1)/2$-fold qubit errors are misidentified as $(d-1)/2$-fold errors, where the rate these misidentifications occur scales as $p^{d_e}$. This explains the general scaling seen in Fig. 4 and in Eq. (11), and underlines the primary result that misidentification of qubit errors causes logical errors, and that large arrays are less error-prone than small arrays.

The magnitude of the logical error rate can be estimated using simple statistical arguments, considering only errors on the data qubits (and ignoring e.g. errors in the CNOTs between measure qubits and data qubits [22]). Chains of errors will give the same measurement outcomes if they are complementary in the way shown in Fig. 5; more formally, two chains of errors are complementary if their product is an array-crossing chain that commutes with all the stabilizers. Since the shortest error chains are the most likely, we need only consider minimum-length array-crossing chains, which have $d$ operators and can cross the array in any one of the $d$ data qubit rows. The most likely misidentifications occur between error chains with length $d_e - 1 = (d-1)/2$ that are complementary to error chains of length $d_e = (d+1)/2$. For a chain of errors in a given row, the number of possible $d_e$-fold errors is $d(d-1)\ldots d_e/d_e!$, where the denominator appears because error order does not matter. Given a per-step error rate $p$, the per-cycle individual error rate is $p_e \cong 8p$, as there are eight steps per cycle. The total $\hat{X}_L$ error rate $P_L^s$ from these statistical arguments is then

$$P_L^s = d\,\frac{d_e!}{(d_e-1)!d_e!}\,p_e^{d_e},\qquad(12)$$

where the factor of $d$ accounts for the $d$ independent rows in the array. A plot of this prediction is shown in Fig. 4b, and is seen to scale in a way similar to the simulations in Fig. 4a.

We can use these scaling relations to estimate the number of qubits needed to obtain a desired logical error rate. Using Eq. (11) for the error probability, we plot in Fig. 6 the total number $n_q$ of data and measurement qubits, $n_q = (2d-1)^2$, versus error rate normalized to the threshold $p/p_{\mathrm{th}} < 1$, for three values of the logical error rate $P_L$. We find that $n_q$ increases rapidly as $p$ approaches the threshold $p_{\mathrm{th}}$, so that a good target for the gate fidelity is above about 99.9% ($p \lesssim 10^{-3}$). In this case, a logical qubit will need to contain $10^3 - 10^4$ physical qubits in order to achieve logical error rates below $10^{-14} - 10^{-15}$,

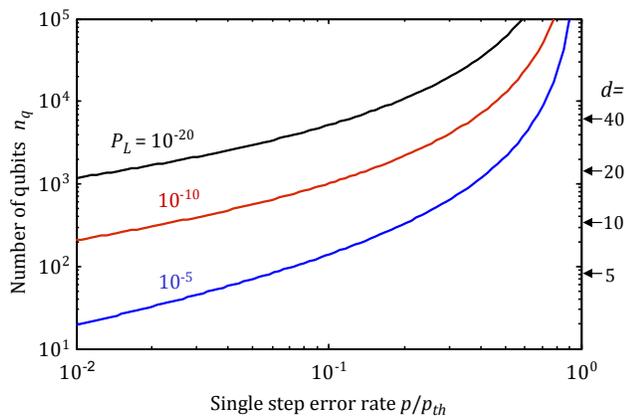

FIG. 6. (Color online) Estimated number of physical qubits $n_q$ per logical qubit, versus the single-step error rate $p$, the latter normalized to the threshold error rate $p_{th}$, plotted for different target logical error rates $P_L$. Notation on the right axis corresponds to the array distance $d$ for a single logical qubit.

sufficient to perform Shor's algorithm with a reasonable chance of success.[15]

Similar considerations apply when detecting and identifying errors in time as opposed to the spatially-correlated errors discussed above. If the measurement process gives errors at the rate $p_M$ per surface code cycle, the probability of double errors (the same error twice in a row) is $p_M^2$, and so on. A particular pattern of errors in time will suffer misidentifications following the same scaling as in Eq. (12), so we find that we need roughly the same distance $d$ in time (measured in complete surface code cycles) as we do in space to maintain the minimum logical error rate, assuming the measurement error rate $p_M$ is comparable to the physical qubit error rate $p$.

More complete calculations of this type can be found in Ref. [45].

### B. Logical error rate for different error classes

The logical error rates calculated from simulations such as in Fig. 4 depend on which types of errors are simulated, as well as the number of opportunities there are for each type of error. Errors that occur on the data qubits, primarily data qubit identity ("idle") operations that are replaced by erroneous $\hat{X}$, $\hat{Y}$ or $\hat{Z}$ operations, we term "class-0" errors. There are four opportunities per sur-face code cycle for such errors on each data qubit (there are a total of eight surface code cycle steps in Fig. 1, during four of which each data qubit is undergoing a CNOT, leaving four steps in which the data qubit is idle). Errors that occur on the measure qubits, namely initialization, measurement, and Hadamard operations, we term "class-1" errors; note that errors occurring during the two measure-Z identity $\hat{I}$ (idle) steps have no impact, as they are followed by a ground-state initialization of the measure-Z qubit, so these errors cannot propagate. Each measure-X qubit has four opportunities for class-1 errors per surface code cycle, and each measure-Z has two opportunities. There is thus an average of three opportunities per surface code cycle for a class-1 error. Finally, errors in the measure qubit-data qubit CNOT operations we term "class-2" errors. There are four steps per cycle during which class-2 errors can occur, corresponding to the four CNOTs for each measure qubit.

The logical error rate is least sensitive to class-1 errors, is more sensitive to class-0 errors, and is the most sensitive to class-2 errors. This is demonstrated by simulations in which we examine the dependence of the logical error rate $P_L$ on the error rate from each error class separately, with the error rate from the other error classes set to zero, shown in Fig. 7. It is readily apparent that the different error classes have very different thresholds, with the class-0 errors in Fig. 7a having a per-step threshold of 4.3% and a per-cycle threshold of 15.5% (the per-cycle value is not quite four times the per-step value, as multiple errors can cancel). The class-1 measure-Z qubit errors shown in Fig. 7b have a per-step threshold of 25%, translating to a per-cycle threshold of 50% as there are two opportunities for class-1 measure-Z errors per surface code cycle; the equivalent plot for measure-X qubits has a per-step threshold of 12.5%, and as there are four opportunities for class-1 errors per cycle in the measure-X qubits, the per-cycle threshold is again 50%. The class-2 errors shown in Fig. 7c have a per-step threshold of 1.25% (per-cycle threshold of $\sim 5\%$). Clearly the sensitivity to class-2 errors is greatest and to class-1 errors the smallest.

We can find empirical expressions for the scaling of the logical error rates due to each error class:

$$\text{Class-0}: P_{L0} \sim \left(\frac{p_0}{p_{th,0}}\right)^{d_e} \text{ with } p_{th,0} \cong 0.043,$$

$$\text{Class-1}: P_{L1} \sim \left(\frac{p_1}{p_{th,1}}\right)^{d_e} \text{ with } p_{th,1} \cong 0.12,$$

$$\text{Class-2}: P_{L2} \sim \left(\frac{p_2}{p_{th,2}}\right)^{d_e} \text{ with } p_{th,2} \cong 0.0125.$$

(13)

Here the rates $p_0$, $p_1$ and $p_2$ and their thresholds are per step in the surface code cycle, while the logical error rates are per surface code cycle. The logical error rates display the same dependence on the error dimension $d_e = (d + 1)/2$ (in terms of the array distance $d$) as the expression for the total logical error rate $P_L$ given in Eq. (11), but

---

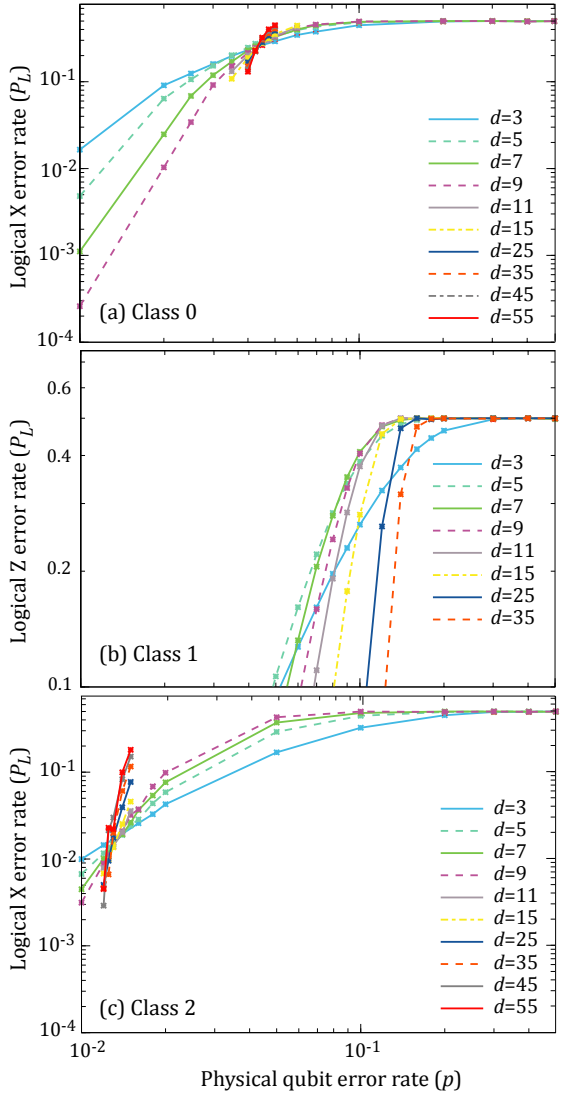

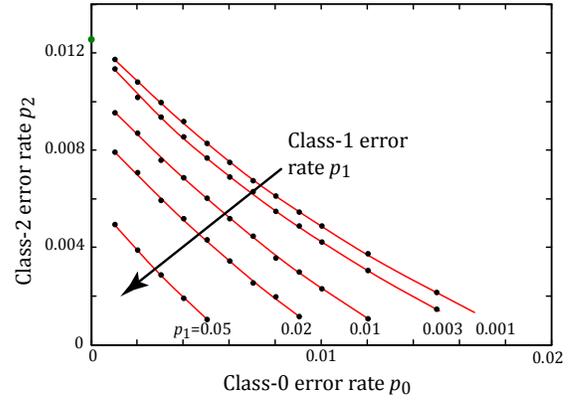

FIG. 8. (Color online) Contours of the three error class error rates $p_0$, $p_1$ and $p_2$ that combine to give the error threshold, corresponding to the total logical error rate $P_L = 0.02$. Data are from a distance $d = 9$ numerical simulation, and lines are guides to the eye. The green dot (upper vertical axis) is the class-2 error rate giving $P_L = 0.02$ with no class-0 or class-1 errors.

Returning to the general error rate discussion, we see that the surface code is able to handle a per-step physical qubit error rate up to the threshold $p_{th} = 0.57\%$, while still preserving the integrity of logical states in the array. This is achieved by identifying the sources of these errors and accounting for them. The error tolerance improves as the array distance $d$ increases; larger values of $d$ give lower logical error rates than do smaller $d$, as long as the error rate $p$ is smaller than the threshold $p_{th}$. We see that the array distance $d$ is an important parameter in the implementation. In fact, as we will see, the critical parameter is not the size of the entire 2D array, but instead the size of the logical qubits that are created within the array. We return to this topic in Section XII.

At the fundamental level, we have physical qubits which are being projectively measured on every surface code cycle, with errors being continuously detected and accounted for in software. Errors that are not erased through the surface code measurement cycle cause changes in the quiescent state $|\psi\rangle$, as indicated by changes in the stabilizer measurements. These errors do not affect the logical state $|q_L\rangle$, as they are restricted to the directly-stabilized part $|Q\rangle$ of $|\psi\rangle = |Q\rangle|q_L\rangle$. When however the matching algorithm makes a mistake, a logical error occurs, in which case typically both $|Q\rangle$ and $|q_L\rangle$ are affected; however, for sufficiently small $p$, these errors are very rare, with the directly-stabilized $|Q\rangle$ and the matching algorithm protecting the logical subspace of $|q_L\rangle$ from the vast majority of errors.

When we build on this platform, we can often ignore the details of this error-accounting apparatus, and instead discuss manipulating the physical qubits without worrying about errors. At this level of abstraction, we are with different thresholds for each class of error.

FIG. 7. (Color online) $\hat{X}_L$ error rate $P_L$ per surface code cycle as a function of the per-step physical error rate $p$, considering only (a) class-0 (data qubit) errors, (b) class-1 (measure qubit) errors and (c) class-2 (CNOT) errors (see text for full definition). For the class-1 errors in (b), results are for initialization and measurement errors in measure-Z qubits, with two opportunities per surface code cycle, while the equivalent for measure-X qubits includes Hadamard errors, with four opportunities per cycle.

The dependence of the overall logical error rate on the three classes of errors occurring concurrently is shown in Fig. 8, where we display the contours of error rates $(p_0, p_1, p_2)$ that give a total logical error rate $P_L = 0.02$, approximately the value of $P_L$ at threshold. This figure shows that class 0 and 2 errors have roughly equal impact on $P_L$, while $P_L$ is roughly five times less sensitive to class 1 errors.

thus dealing with "software-corrected" physical qubits. In the next few sections, we will be discussing the creation and operation of logical qubits; we will mostly ignore the error accounting for this discussion, to keep things simple, but we will return to this important issue when necessary.

# VIII. CREATING LOGICAL QUBITS

We have described one way to create the logical operators $\hat{X}_L$ and $\hat{Z}_L$, by building array-crossing chains of $\hat{X}$ and $\hat{Z}$ operators. These logical operators commute with all the stabilizers. However, as the chains of operators must cross the entire array, this becomes cumbersome for large arrays, and for arrays with only two X and two Z boundaries, only gives us a single logical qubit no matter how large the array. We can increase the number of logical qubits by creating more varied boundaries in the array; for example, using short alternating X and Z boundaries would allow the creation of multiple logical qubits, increasing the amount of information that can be stored. However, since you cannot deform in any non-trivial manner operators connecting sections of boundary of the same type in different locations as you cannot move sections of boundary of different types past one another, performing logical CNOTs between these boundary qubits is not possible, so this approach has limited utility. A much more powerful approach is to create holes (known as *defects* in the published literature) inside the boundaries of the array, which can be done by simply turning off one or more of the internal measure-X and measure-Z qubits ("turning off" means that the measure qubit no longer performs the surface code cycle of CNOTs followed by measurement).

In Fig. 9 we show how this can be done, by turning off a single measure-Z qubit and creating a hole, which we call a "Z-cut hole". Turning off the measure-Z qubit means that we no longer measure its stabilizer, which creates two additional degrees of freedom in the surface code array. We can manipulate these degrees of freedom in a similar fashion to the way we manipulate the array qubit, by defining logical operators $\hat{X}_L$ and $\hat{Z}_L$ that anticommute. We will call this logical qubit a "single Z-cut qubit".

We have positioned the Z-cut hole near an X boundary of the array. Turning off the $\hat{Z}$ stabilizer leaves a small hole surrounded by measure-X qubits, i.e. this creates an internal X boundary. We define the operator $\hat{X}_L = \hat{X}_1\hat{X}_2\hat{X}_3$, which connects the array's outer X boundary with the internal X boundary of the hole. As can be seen, this logical operator comprises a chain of operators that are paired across each $\hat{Z}$ stabilizer, so it commutes with all the $\hat{Z}$ stabilizers in the array, and trivially commutes with the $\hat{X}$ stabilizers as well. We also define $\hat{Z}_L = \hat{Z}_3\hat{Z}_4\hat{Z}_5\hat{Z}_6$, a set of data qubit $\hat{Z}$ operators that form a loop around the Z-cut hole; the $\hat{Z}$ operators are paired across each $\hat{X}$ stabilizer, commuting with each of these, and trivially with all the $\hat{Z}$ stabilizers as well. This operator loop is not constrained by a $\hat{Z}$ stabilizer as it would be without the Z-cut hole, so it can change the quiescent state $|\psi\rangle$ in a non-trivial way.

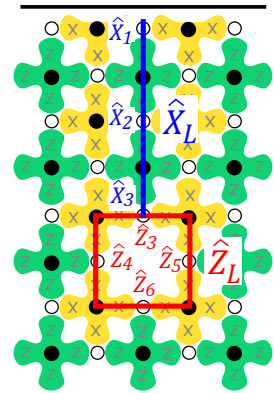

FIG. 9. (Color online) A single Z-cut qubit. The array boundary is indicated by a solid black line (top), while the other three sides of the array continue outwards indefinitely. Two logical operators, $\hat{X}_L = \hat{X}_1\hat{X}_2\hat{X}_3$ from the array's outer X boundary to the internal X boundary of the Z-cut hole, and $\hat{Z}_L = \hat{Z}_3\hat{Z}_4\hat{Z}_5\hat{Z}_6$ surrounding the Z-cut hole, are shown. Note that the operator chains for $\hat{X}_L$ and $\hat{Z}_L$ have one physical data qubit in common (data qubit 3).

The $\hat{X}_L$ chain and the $\hat{Z}_L$ loop share one data qubit, number 3 in Fig. 9. These two operators therefore anticommute:

$$\begin{aligned}
\hat{X}_L\hat{Z}_L &= (\hat{X}_1\hat{X}_2\hat{X}_3)(\hat{Z}_3\hat{Z}_4\hat{Z}_5\hat{Z}_6) \\
&= \hat{Z}_4\hat{Z}_5\hat{Z}_6(\hat{X}_3\hat{Z}_3)\hat{X}_1\hat{X}_2 \\
&= -\hat{Z}_4\hat{Z}_5\hat{Z}_6(\hat{Z}_3\hat{X}_3)\hat{X}_1\hat{X}_2 \\
&= -\hat{Z}_L\hat{X}_L.
\end{aligned} \tag{14}$$

You can show that $\hat{X}_L^2 = \hat{Z}_L^2 = \hat{I}$, and we can define $\hat{Y}_L = \hat{Z}_L\hat{X}_L$. Just as with the 2D array, we have created a set of anti-commuting operators associated with the Z-cut hole that satisfy all the requirements for a logical qubit, so the Z-cut hole is indeed a logical qubit. We note that if the outer perimeter of the array were a single Z boundary, there would be no way to build an appropriate $\hat{X}_L$ chain between the Z-cut hole and the array boundary; furthermore, the product of all the $\hat{Z}$ stabilizers in the array would fix the value of $\hat{Z}_L$, so creating a Z-cut hole in a Z-boundary array would not create any additional degrees of freedom: At least one X boundary is necessary to create a single Z-cut qubit. Note the distance $d$ of this qubit is $d = 3$, limited by the length of the $\hat{X}_L$ chain; it could be increased to $d = 4$ by moving the hole one stabilizer cell further away from the array boundary, but $d$ cannot be increased above 4 without creating a larger qubit hole (see Section XII). A single Z-cut qubit is called a *smooth defect* or a *dual defect* in the published literature.

An analogous single X-cut qubit can be formed by turning off a measure-X qubit in an array that has at least one Z boundary; a $\hat{Z}_L$ operator is then a chain of $\hat{Z}$ operators from the array Z boundary to the internal Z boundary created by turning off the measure-X qubit, and an $\hat{X}_L$ operator is a loop of $\hat{X}$ bit-flips surrounding the X-cut hole. You can show that these logical operators satisfy the requisite anti-commutation relations. A single X-cut qubit is called a *rough defect* or a *primal defect* in the published literature.

This concept can be made even more useful by making qubits that do not rely on operator chains that reach one of the array boundaries, simplifying their logical manipulation and greatly increasing the number of qubits that can be stored and manipulated simultaneously. An example of a "double Z-cut qubit," achieved by turning off *two* measure-Z qubits in the array, is shown in Fig. 10. Making these two Z-cut holes adds four additional degrees of freedom to the array. We can manipulate each Z-cut hole separately by defining $\hat{X}_{L1}$ and $\hat{Z}_{L1}$ for the upper Z-cut hole, and $\hat{X}_{L2}$ and $\hat{Z}_{L2}$ for the lower Z-cut hole, in a way completely analogous to the single Z-cut qubit. These four linearly independent operators manipulate all four degrees of freedom, with each logical operator pair $(\hat{X}_{Lj}, \hat{Z}_{Lj})$ commuting with the other pair but anti-commuting with each other.

We can manipulate the two qubit holes in a correlated fashion by replacing $\hat{X}_{L1}$ by the product $\hat{X}_{L1}\hat{X}_{L2}$; this operator product anti-commutes with both $\hat{Z}_{L1}$ and $\hat{Z}_{L2}$, and together with $\hat{X}_{L2}$ provides the same functionality as $\hat{X}_{L1}$ and $\hat{X}_{L2}$ (note that we can recover $\hat{X}_{L1}$ by multiplying the operator product by $\hat{X}_{L2}$, using $\hat{X}_{L2}^2 = \hat{I}_L$). Now, we can define a new $\hat{X}_L$ operator linking the two qubit holes, shown in Fig. 10; this operator is the product of three data qubit $\hat{X}$ operators, and links the internal $\hat{X}$ boundary of the upper qubit to the internal $\hat{X}$ boundary of the lower qubit. If you examine the figure, you will see that the product $\hat{X}_L\hat{X}_{L1}\hat{X}_{L2}$ is equal to the product of all the stabilizers enclosed by these operators (these stabilizers have black outlines in Fig. 10). Hence, if we multiply $\hat{X}_{L1}\hat{X}_{L2}$ by all these stabilizers, we generate $\hat{X}_L$, so in fact $\hat{X}_L$ is equivalent to $\hat{X}_{L1}\hat{X}_{L2}$ to within $\pm 1$ (the sign is equal to the product of all the enclosed stabilizers). Hence we can replace $\hat{X}_{L1}\hat{X}_{L2}$ by $\hat{X}_L$.

We now have the operator set $\{\hat{X}_L, \hat{X}_{L2}, \hat{Z}_{L1}, \hat{Z}_{L2}\}$, where the logical $\hat{X}$ operators commute with one another, as do the logical $\hat{Z}$ operators, but $\hat{X}_L$ anti-commutes with both $\hat{Z}_{L1}$ and $\hat{Z}_{L2}$, as these each have one data qubit in common; $\hat{X}_{L2}$ only anti-commutes with $\hat{Z}_{L2}$.

We are only interested in manipulating two of the four degrees of freedom in this double qubit, as using all four degrees of freedom necessitates using operators that connect to potentially distant boundaries. This is not consistent with a 2-D layout of logical qubits as all of those long-range operators will tangle around one another when attempting to perform gates. It is necessary to make each logical qubit local, which means its opera-tors must be local, to be able to perform logical gates in parallel. Consequently, we choose to use $\hat{X}_L$ along with either $\hat{Z}_{L1}$ or $\hat{Z}_{L2}$. If we choose to use $\hat{Z}_{L2} \equiv \hat{Z}_L$, and use this pair to define $\hat{Y}_L = \hat{Z}_L\hat{X}_L$, we have a complete set $\{\hat{X}_L, \hat{Z}_L, \hat{Y}_L\}$ of logical qubit operators for this double Z-cut qubit. By restricting ourselves to this set of logical operators, we are changing the two qubit holes in a correlated fashion, so that effectively we are manipulating a single two-level logical qubit; we will write the state of this composite logical qubit in the usual way, as e.g. $\alpha|g_L\rangle + \beta|e_L\rangle$. Note that in more advanced work [46], more general defect and logical operator configurations are permitted.

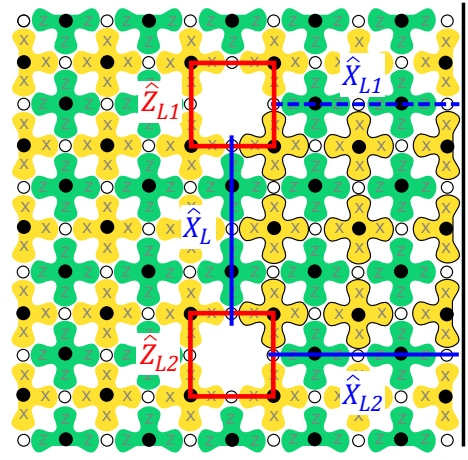

FIG. 10. (Color online) A double Z-cut qubit, formed by turning off two measure-Z qubits. For each Z-cut hole we can define the logical operators $\hat{X}_{L1}$ and $\hat{Z}_{L1}$, $\hat{X}_{L2}$ and $\hat{Z}_{L2}$, where $\hat{X}_{L1}$ and $\hat{X}_{L2}$ reach from the array X boundary on the right to the internal X boundary of each Z-cut hole, and $\hat{Z}_{L1}$ and $\hat{Z}_{L2}$ are loops surrounding each Z-cut hole. We can replace $\hat{X}_{L1}$ (dashed blue line) by the product $\hat{X}_{L1}\hat{X}_{L2}$, which performs an $\hat{X}_L$ bit flip of both qubit holes, and along with $\hat{X}_{L2}$ provides equivalent functionality to the two separate $\hat{X}$ operators. However, the product $\hat{X}_{L1}\hat{X}_{L2}$ can be multiplied by all the outlined $\hat{X}$ stabilizers, which at most give a sign change to the operator, to yield the equivalent operator $\hat{X}_L$ (solid blue line) that links the two qubit holes.

We can create a double X-cut qubit in a similar way to the double Z-cut qubit, by turning off two measure-X qubits. Both double Z-cut and double X-cut qubit types are shown in Fig. 11. The logical operators for the double X-cut qubit are defined in an analogous fashion to the double Z-cut qubit: The $\hat{Z}_L$ operator is a chain linking the two X-cut holes, and we choose either $\hat{X}_L = \hat{X}_{L1}$ or $\hat{X}_L = \hat{X}_{L2}$, the operator loops that surround one or the other of the X-cut holes. It is easy to show that these logical operators satisfy the necessary anti-commutation relations.

Double Z-cut (double X-cut) qubits are called *smooth* (*rough*) qubits or *dual* (*primal*) qubits in the literature. As we will almost exclusively discuss double-cut qubits

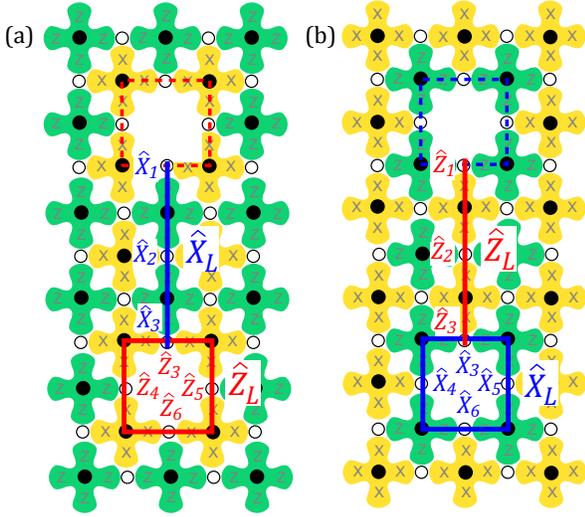

FIG. 11. (Color online) (a) A double Z-cut and (b) a double X-cut qubit, formed in a large array by turning off two measure-Z and two measure-X qubits, respectively; the array is assumed to extend outwards indefinitely. For the double Z-cut qubit the logical operators comprise an $\hat{X}_L = \hat{X}_1\hat{X}_2\hat{X}_3$ chain that links one Z-cut hole's internal X boundary to the other, and a $\hat{Z}_L = \hat{Z}_3\hat{Z}_4\hat{Z}_5\hat{Z}_6$ loop that encloses the lower Z-cut hole. For the X-cut qubit we have the $\hat{Z}_L = \hat{Z}_1\hat{Z}_2\hat{Z}_3$ chain that links the two internal X-cut holes' Z boundaries and the $\hat{X}_L = \hat{X}_3\hat{X}_4\hat{X}_5\hat{X}_6$ loop that encloses the lower X-cut hole. The $\hat{X}_L$ and $\hat{Z}_L$ operator chains share one data qubit, data qubit 3 for both examples, so the operators anti-commute. Note that the loop operators ($\hat{Z}_L$ for the Z-cut qubit and $\hat{X}_L$ for the X-cut qubit) can surround either of the two holes in the qubit, as discussed in the text.

from this point forward, we will simply call these Z-cut and X-cut qubits.

The two different types of logical qubits, Z-cut and X-cut, are needed to perform the topological braid transformation that provides the logical CNOT operation in the surface code: Only braids between mixed qubit types give the needed functionality. However, as we discuss below, a topological CNOT can be performed between two Z-cut qubits, by using an X-cut qubit as an intermediary; similarly, a CNOT between two X-cut qubits can be performed using a Z-cut as an intermediary. An arbitrary quantum computation can therefore be completed using mostly one flavor of logical qubit, with the other qubit type only making appearances in a supporting role.

The logical qubits we have introduced are "small" in that we only turn off a single measure qubit (removing a single stabilizer) in each of the two holes. For the Z-cut qubit shown in Fig. 11, the $\hat{Z}_L$ loop consists of four data qubit $\hat{Z}$ operators, and the $\hat{X}_L$ operator chain linking the two qubit holes have three data qubit $\hat{X}$ operators. Hence the distance for this qubit is only $d = 3$, meaning these logical qubits are relatively fault-intolerant. Moving the two holes further apart would increase the distance to $d = 4$, limited by the length of the $\hat{Z}_L$ loop. In Section

XII we describe how to create and initialize larger distance $d$ logical qubits, which as discussed in Sect. VII are significantly more fault-tolerant.

## IX. SOFTWARE-IMPLEMENTED $\hat{Z}_L$ AND $\hat{X}_L$

We now must inform the reader of a curious aspect of the surface code: The logical operators $\hat{X}_L$ and $\hat{Z}_L$ that we have spent so much time discussing are not actually implemented in the surface code hardware! These operations are handled entirely by the classical control software, as we shall describe in a moment. These logical gates could be implemented by performing physical bit-flip and phase-flip operations on the data qubits, as the surface code is fairly tolerant to errors in these types of single-qubit gates; however, a hardware-based solution will always have a higher error rate than one implemented in the control software. Instead, whenever a particular quantum algorithm calls for an $\hat{X}_L$ or $\hat{Z}_L$ operator, the operator is commuted through each subsequent logical operation in the algorithm until a second identical operation is called for, in which case the two cancel (since $\hat{X}_L^2 = \hat{Z}_L^2 = \hat{I}_L$), or until a measurement of the logical qubit is performed, in which case the operator is applied to the measurement outcome (hence an $\hat{X}_L$ would be applied by reversing the sign of a $\hat{Z}_L$ measurement, while it would have no effect on an $\hat{X}_L$ measurement, and similarly for a $\hat{Z}_L$ operator combined with an $\hat{X}_L$ or a $\hat{Z}_L$ measurement, respectively).

We demonstrate the concept with an example, accompanied by the warning that this example uses a number of results that we have not introduced; the reader may want to return to this example after reading the discussion of the Heisenberg representation of the CNOT in Sect. XIV C. We represent the two-qubit CNOT operation with the two-qubit operator $\hat{C}_L$, which transforms two-qubit operators such as $\hat{I}_L \otimes \hat{Z}_L$ according to the rules described there (so e.g. $\hat{C}_L(\hat{I}_L \otimes \hat{Z}_L) = (\hat{Z}_L \otimes \hat{Z}_L)\hat{C}_L$).

Consider the circuit fragment shown in Fig. 12a, in which a $\hat{Z}_L$ operator on logical qubit 1 and the identity $\hat{I}_L$ operator on qubit 2 is followed by a logical Hadamard on qubit 1, a logical CNOT in which qubit 1 is the control and qubit 2 the target, and a second logical Hadamard on qubit 1, with the circuit terminated by $\hat{X}_L$ measurements of both logical qubits.

Rather than actually applying the $\hat{Z}_L$ operator to qubit 1, the $\hat{Z}_L$ operator is instead commuted through the Hadamard, which transforms $\hat{Z}_L$ to $\hat{X}_L$ (using $\hat{H}_L\hat{Z}_L = \hat{X}_L\hat{H}_L$; see e.g. [2]). This is shown in Fig. 12b. The $\hat{X}_L$ on qubit 1 and $\hat{I}_L$ on qubit 2 are then commuted through the CNOT, using $\hat{C}_L(\hat{X}_L \otimes \hat{I}_L) = (\hat{X}_L \otimes \hat{X}_L)\hat{C}_L$, leaving $\hat{X}_L$ on qubit 1 and $\hat{X}_L$ on qubit 2 after the CNOT, as shown in Fig. 12c. Commuting the first qubit $\hat{X}_L$ through the second Hadamard, this becomes $\hat{Z}_L$ on qubit 1 with $\hat{X}_L$ on qubit 2 (Fig. 12d). We have now reached the terminal measurements at the end of the circuit. At this point the measurements are carried out, and the classical control software, which has calculated the commutations we have stepped through, corrects the measurements as needed using the pending operations: The $\hat{X}_L$ measurement of qubit 1, $M_X = \pm 1$, has its sign reversed by the pending $\hat{Z}_L$ on that qubit to $-M_X = \mp 1$, while the $\hat{X}_L$ measurement of qubit 2 is unaffected by its pending $\hat{X}_L$. The roles of the $\hat{Z}_L$ and $\hat{X}_L$ operators are ended by these logical measurements.

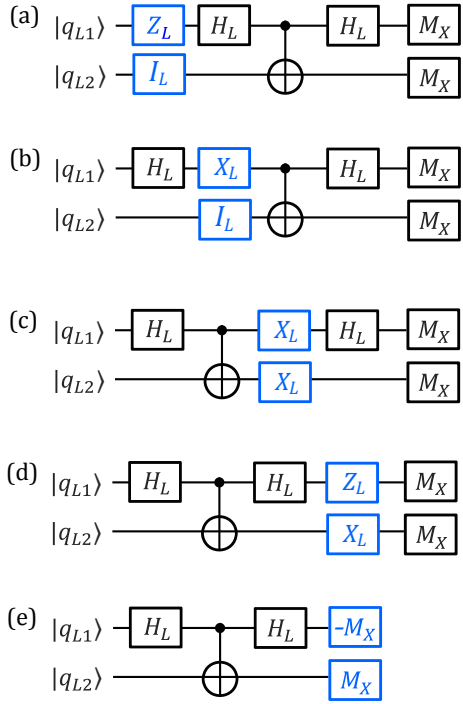

FIG. 12. (Color online) Example of how $\hat{Z}_L$ and $\hat{X}_L$ operators are handled in the control software. Here we commute a $\hat{Z}_L$ logical operator (colored blue (light)) on one qubit with an identity $\hat{I}_L$ (colored blue (light)) on a second qubit through a sequence of logical operations (colored black), here including two Hadamards and a CNOT, ending with a measurement of each qubit. (a) We start with $\hat{Z}_L$ operating on qubit 1 with the $\hat{I}_L$ identity on qubit 2. (b) The two operations are commuted one step to the right, with $\hat{Z}_L$ transformed by the Hadamard to an $\hat{X}_L$. (c) The two operations are commuted through the CNOT, with $\hat{X}_L$ on qubit 1 remaining unchanged while $\hat{I}_L$ on the second qubit transforms to $\hat{X}_L$. (d) $\hat{X}_L$ on qubit 1 is transformed by the second Hadamard to $\hat{Z}_L$, and finally (e) the $\hat{Z}_L$ on qubit 1 changes the sign of the $\hat{X}$ measurement from $M_X = \pm 1$ to $-M_X = \mp 1$, while the $\hat{X}_L$ on qubit 2 does nothing to that qubit's $\hat{X}$ measurement.

In general, single qubit $\hat{X}_L$ and $\hat{Z}_L$ operators can be commuted through the various one- and two-qubit operations in any topological quantum circuit, without actually performing these logical operations, until they either cancel with another identical single-qubit operator or are used to correct a measurement as described above.[16] We have shown how this works for a simple circuit involving Hadamards and a logical CNOT. We discuss how this is done for the $\hat{S}_L$ phase gate and the $\hat{T}_L$ gate in Sect. XVI.

## X. LOGICAL QUBIT INITIALIZATION AND MEASUREMENT

While we do not actually perform $\hat{X}_L$ or $\hat{Z}_L$ operations on the logical qubits, we do have to initialize and measure the qubits. In this section we described two methods to initialize logical qubits and two methods to measure the logical state.

### A. Initialization.

There are two ways to initialize a logical qubit state, which we will call "easy" and "difficult". The easy way is to initialize the qubit in an eigenstate of the qubit cut; for an X-cut qubit, easy initialization is in the $|+_L\rangle$ or $|-_L\rangle$ eigenstates of $\hat{X}_L$, while for a Z-cut qubit, easy initialization is in the $|g_L\rangle$ or $|e_L\rangle$ eigenstates. The process is illustrated for an X-cut qubit in Fig. 13, starting with a 2D array with no cuts, and stepping immediately to one with a double-X cut qubit, achieved by turning off two measure-X qubits to create two X-cut holes. We know the measurement outcomes of these two measure-X qubits just prior turning them off; if we define $\hat{X}_L$ as the loop surrounding the upper qubit hole, the initial state of the qubit corresponds to the measurement outcome of the upper measure-X qubit just before it is turned off, i.e. $|+_L\rangle$ if $X_{abcd} = +1$ and $|-_L\rangle$ if $X_{abcd} = -1$. If the initial state is not the desired one, we can apply a $\hat{Z}_L$ phase-flip to that logical qubit; as discussed above, this phase-flip would be applied "in software", meaning it would not actually be applied, but would instead be commuted through the quantum circuit until it is canceled by a second operator or the logical qubit is measured.

The alternative to initializing the X-cut logical qubit in the $\hat{X}_L$ basis is to perform a "difficult" initialization in the $\hat{Z}_L$ basis, i.e. in either $|g_L\rangle$ or $|e_L\rangle$. This can be done using a logical Hadamard after initializing in the $\hat{X}_L$ basis. Logical Hadamards are however somewhat complicated (see Section XV), so it is simpler to initialize in the $\hat{Z}_L$ basis directly. This is illustrated for an X-cut qubit in Fig. 14. Briefly, starting with a fully stabilized array, a strip of $\hat{X}$ stabilizers is turned off, where each end of the strip will serve as one of the qubit holes. In addition, the $\hat{Z}$ stabilizers that border the strip are switched from measuring four adjacent data qubits to measuring just three (by simply excluding one data qubit from the surface code CNOT cycle in Fig. 1). There are six $\hat{Z}$ stabilizers that are switched from four- to three-terminal operation in Fig. 14b.[17] The three isolated data qubits (numbered 1, 2 and 3 in Fig. 14) are measured once along $\hat{Z}$ to maintain error correction, and then set to the ground $|g\rangle$ state as in Fig. 14c.[18] Setting these three qubits to their ground state ensures that the quiescent state $|\psi\rangle$ will be in the logical ground state of the $\hat{Z}_L = \hat{Z}_1 \hat{Z}_2 \hat{Z}_3$ operator. Finally the two $\hat{X}$ stabilizers internal to the strip are turned back on, and the three-terminal $\hat{Z}$ stabilizers switched back to four-terminal measurements, completing the process.[19] Details are given in Appendix C.

---

[16] There is an important distinction between what are called "Clifford gates" and "non-Clifford gates"; the set of logical gates we are developing here, $\hat{X}_L$, $\hat{Z}_L$, $\hat{H}_L$, $\hat{S}_L$ and $\hat{S}_L^\dagger$, $\hat{T}_L$ and $\hat{T}_L^\dagger$, and the two-qubit logical CNOT, are all Clifford gates except $\hat{T}_L$ and $\hat{T}_L^\dagger$ (see Ref. [47]). We discuss how to commute $\hat{X}_L$ and $\hat{Z}_L$ through the $\hat{S}_L$ and $\hat{T}_L$ gates in Sect. XVI.

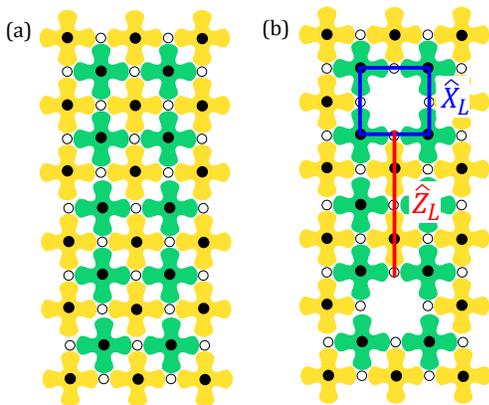

FIG. 13. (Color online) Easy initialization of an X-cut logical qubit. Using the logical qubit's upper hole, the stable measure-X outcome just prior to creating the qubit hole by turning off the measure-X qubit is equal to the initial logical eigenvalue of the qubit.

Note that the projective measurements of the two $\hat{X}$ stabilizers that were turned back on at the end of the initialization will leave the three data qubits in a $+1$ eigenstate of $\hat{Z}_L$, in other words a $|g_L\rangle$ eigenstate, even though the data qubits themselves will no longer be in their individual ground states $|g\rangle$. We can see this as follows: After the ground state re-set, the quiescent state $|\psi\rangle$ is transformed to $|\psi'\rangle = |ggg\rangle|\phi\rangle$, where $|ggg\rangle$ is the state of the three data qubits, and $|\phi\rangle$ the state of all the other data qubits in the array. The state $|\psi'\rangle$ is a $+1$ eigenstate of $\hat{Z}_L$, but is not an eigenstate of all the $\hat{X}$ stabilizers, as you can easily verify. However, we know that $\hat{Z}_L$ commutes with all the $\hat{X}$ stabilizers, so there are common eigenstates of both $\hat{Z}_L$ and these stabilizers. The state $|\psi'\rangle$ can be written as a superposition of $\hat{X}$ stabilizer eigenstates, where all these eigenstates are still $+1$ eigenstates of $\hat{Z}_L$, so the logical qubit remains in $|g_L\rangle$. The $\hat{X}$ stabilizer measurements will project $|\psi'\rangle$ onto one of these eigenstates, leaving us with a state $|\psi''\rangle$ that is still a $+1$ eigenstate of $\hat{Z}_L$ but is also an eigenstate of all the $\hat{X}$ (and $\hat{Z}$) stabilizers in the array.

---

[17] Note that even though these stabilizers only measure three data qubits, the geometry of the array ensures that two pairs of the three data qubits are also measured by $\hat{X}$ stabilizers, so the stabilizer measurements still all commute: If e.g. the $\hat{Z}$ stabilizer measures data qubits $a$, $b$ and $c$, one adjacent measure-X qubit will measure $a$ and $b$, and the other will measure $b$ and $c$.

[18] It might seem that the isolated data qubits need to be measured $d$ times along $\hat{Z}$ to maintain error detection in time, but the single measurements in this step are combined with the concurrent three-terminal measurements of the adjacent $\hat{Z}$ stabilizers, and compared to the prior four-terminal $\hat{Z}$ stabilizer measurements, to provide sufficient distance (continuity) in time.

[19] It turns out that easy initialization is never used in practice, as X-cut (Z-cut) logical qubits always need to be initialized in the $\hat{Z}_L$ ($\hat{X}_L$) basis; see for example the discussion of logical CNOTs between logical qubits of the same type in Sect. XIV D.

Initializing a Z-cut qubit is completely analogous to the X-cut qubit initialization. Initializing in a $\hat{Z}_L$ eigenstate is easy, because the $\hat{Z}$ stabilizers that are turned off to create the qubit holes ensure the logical qubit is in a $\hat{Z}_L$ eigenstate. Initializing in an $\hat{X}_L$ eigenstate is "difficult:" A strip of $\hat{Z}$ stabilizers is turned off, and the isolated data qubits then initialized to the $\hat{X}$ eigenstates $|+\rangle$ or $|-\rangle$, so that the state $|\psi\rangle$ is then in an eigenstate of $\hat{X}_L = \hat{X}_1 \hat{X}_2 \hat{X}_3$. The stabilizers are then turned back on, projecting the data qubits to an eigenstate of all the stabilizers, while maintaining it in an eigenstate of $\hat{X}_L$.

## B. Measurement

Measuring a logical qubit uses a procedure nearly the reverse of that used for initialization. As with initialization, measurement can be classified as "easy" or "difficult". For an easy measurement, the measure qubits in the qubit holes are simply turned on, with the stabilizer measurement projecting the logical qubit onto a stabilizer eigenstate, with the stabilizer eigenvalue equal to the logical qubit measurement outcome (determined in a fault-tolerant manner by completing $d$ surface code cycles). If this is done for the X-cut qubit shown in Fig. 15a, we turn on the two measure-X qubits in the holes, whose measurement projects the data qubits adjacent to them into an $\hat{X}_a \hat{X}_b \hat{X}_c \hat{X}_d$ product eigenstate. This constitutes an $\hat{X}_L$ measurement of the logical qubit, with the $\hat{X}_L$ eigenvalue equal to the stable value of $X_{abcd}$. If there is a pending $\hat{X}_L$ operator on that qubit, nothing happens, while if there is a pending $\hat{Z}_L$ operator, the sign of the measurement outcome is reversed. Note that this process destroys the logical qubit, which must be re-initialized to be used again.

For a difficult measurement, which would be used to e.g. measure the X-cut qubit in the $\hat{Z}_L$ basis, we use the measurement process illustrated in Fig. 15. Details are given in Appendix D.

A completely analogous process is used to perform an

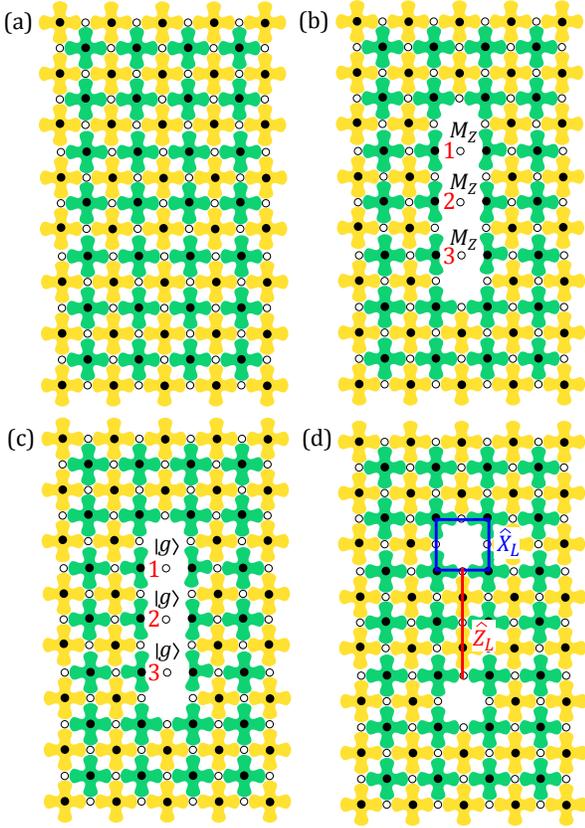

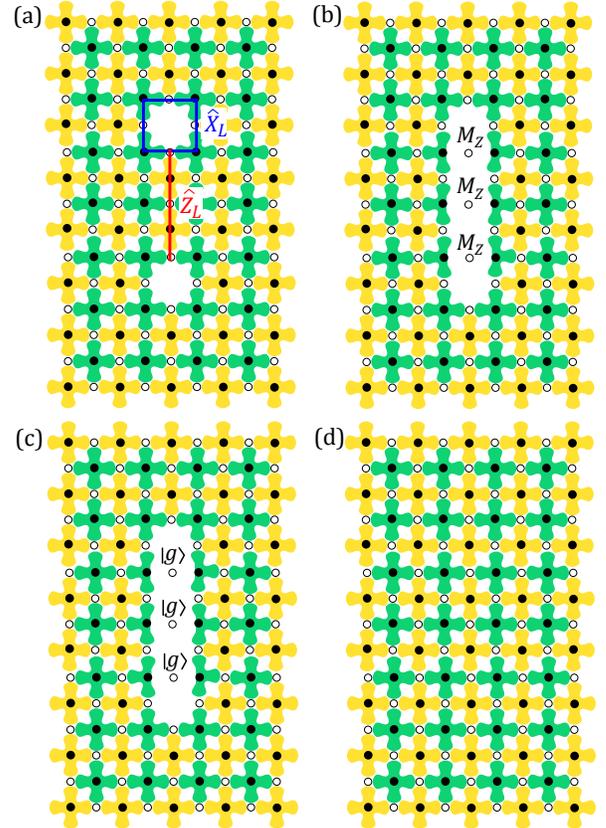

FIG. 14. (Color online) $\hat{Z}$ eigenstate ("difficult") initialization of an X-cut logical qubit. The process steps are as follows: (a) Starting with a fully stabilized array, we (b) turn off a column of four measure-X qubits, switch the adjacent measure-Z qubits from from 4- to 3-terminal measurements, and perform $\hat{Z}$ measurements of the data qubits numbered 1, 2 and 3, to maintain error tracking. (c) We reset (re-initialize) the data qubits 1, 2 and 3 to $|g\rangle$, and (d) turn two of the measure-X qubits back on, switch the adjacent measure-Z qubits back to 4-terminal measurements, leaving us with a logical qubit initialized in $|g_L\rangle$ due to step (c).

FIG. 15. (Color online) $Z$-axis ("difficult") measurement process for an X-cut logical qubit. We start (a) with an X-cut qubit, for which we display the logical operators $\hat{X}_L$ and $\hat{Z}_L$, and begin the measurement (b) by turning off the measure-X qubits in the two qubit cuts while also switching the neighboring measure-Z qubits from four- to three-terminal measurements. We measure the un-stabilized data qubits along $\hat{Z}$. The product of the measurement outcomes is the measurement of $\hat{Z}_L$. A pending $\hat{Z}_L$ operator would have no effect, while a pending $\hat{X}_L$ would be used to reverse the sign of this measurement outcome. We then (c) reset the qubits to their ground states $|g\rangle$ and (d) destroy the logical qubit by turning on all the stabilizers.

"easy" measurement of $\hat{Z}_L$ or a "difficult" measurement of $\hat{X}_L$ for a Z-cut qubit.

## XI. ERRORS DURING STABILIZER MANIPULATIONS

Our initialization and measurement procedures, especially the "difficult" ones, involve turning stabilizers on and off, switching stabilizers from four- to three-terminal measurements and back, measuring individual data qubits, and re-setting (re-initializing) data qubits to eigenstates of $\hat{X}$ or $\hat{Z}$. These processes are not used just for initialization and measurement, but also form the heart of the topologically-implemented CNOT operation as well as the logical Hadamard. These manipulations would seem to allow errors to occur in some undetectable fashion, as we are manipulating the very structure that stabilizes the quiescent states. However, as long as this kind of manipulation is done in the appropriate way, it turns out the quiescent state is still protected.

Consider for example the $\hat{Z}$ eigenstate "difficult" initialization of an X-cut qubit that appears in Fig. 14. Passing from panel a to panel b, we turn off a number of $\hat{X}$ stabilizers and convert some of the $\hat{Z}$ stabilizers from four- to three-terminal measurements, isolating data qubits 1, 2 and 3 from the array. Passing from panel b to c, we perform data qubit ground state re-sets, and passing from panel c to d we turn the stabilizers back on. It would seem that errors could occur on the isolated data qubits 1, 2 or 3, or on the data qubits bordering the long cut in panels b and c.

Consider first the isolated data qubits: $\hat{Z}$ errors on these data qubits will have no effect, as these data qubits are measured along $\hat{Z}$ in panel b of Fig. 15, a measurement that is not affected by $\hat{Z}$ errors, and are then re-set to $|g\rangle$ in panel c, a process that is also immune to $\hat{Z}$ errors. These data qubits are then re-joined with the array and fully stabilized in panel d, with $\hat{Z}$ errors taken care of in the usual way. An $\hat{X}$ error on one of these data qubits will however have an effect: If this occurs between panel a and b, the $\hat{X}$ error will reverse the measurement $M_Z$ of the affected data qubit, but by computing the product of this $\hat{Z}$ measurement with the two three-terminal $\hat{Z}$ measurements adjacent to that data qubit, and comparing these products with the corresponding four-terminal measurements that occurred prior to panel b, this error can be detected and localized, and we can correct the data qubit $\hat{Z}$ measurement outcome. An $\hat{X}$ error between panel b and c will be erased by the ground-state re-set of the data qubit, and an $\hat{X}$ error between panels c and d will be detected by the usual stabilizer cycle, given that we know the data qubit state prior to the stabilizers being turned on.

Now consider the data qubits bordering the cut: $\hat{X}$ errors will be detected in the usual way, as each of these data qubits is stabilized by two $\hat{Z}$ stabilizers, which is sufficient to detect and locate these kinds of errors. A $\hat{Z}$ error on one of these qubits will generate a change of sign in the one $\hat{X}$ stabilizer that monitors that data qubit; this could then be interpreted either as a measurement error on that stabilizer (as it is an isolated error), or as a data qubit error. A measurement error would however disappear as the surface code cycle continues, so by comparing results in time, the error can be identified as a data qubit error, and any measurements then corrected.

In summary, maintaining the surface code error detection is done by careful combinations of measurements before and after a stabilizer manipulation, sometimes involving measurements of individual isolated qubits that are combined with three-terminal stabilizers, sometimes by slightly modifying the error locating algorithms when errors occur on only partly-stabilized data qubits (such as those on the edge of a cut). This can however be done with the same fault tolerance as in the rest of the surface code array, and this means that these manipulations, as long as they respect the distances needed for fault tolerance, both in space and in time, do not affect the stabilization required for proper operation.

## XII. LARGER LOGICAL QUBITS

We can create logical qubits by turning off e.g. two $\hat{Z}$ stabilizers some distance apart, creating two small holes with X boundaries. As shown in Fig. 11a, $\hat{Z}_L$ is a chain of $\hat{Z}$ operators surrounding one of the holes, involving only four physical qubit $\hat{Z}$ operators, equal to the number of data qubits on the boundary of the hole. Similarly the $\hat{X}_L$ operator is a chain of three $\hat{X}$ operators linking the two Z-cut holes. We could put the holes further apart, increasing the distance for $\hat{X}_L$, but this would of course not increase the length of the $\hat{Z}_L$ loop. An error chain with the same length as $\hat{Z}_L$ or $\hat{X}_L$ can emulate these logical operators, corrupting the logical qubit in an undetectable manner; this therefore sets the distance $d$ that determines the relation between the physical qubit error rate $p$ and the resulting logical error rate $P_L$ emerging from the matching algorithm, as discussed in Section VII. For the logical qubits in Fig. 11 we have $d = 3$, and if we separate the holes further, $d = 4$ is the limit, determined by number of data qubits bordering a qubit hole.

The error-handling ability is significantly improved if we increase the size and spacing of the two holes, as this will increase the number of physical qubits involved in $\hat{Z}_L$ and $\hat{X}_L$. Here we describe how to create a five-stabilizer Z-cut hole, increasing the number of physical qubit operators in the $\hat{Z}_L$ loop from four to eight, significantly improving the error tolerance. The two larger holes that now make up the logical qubit can also be moved further apart, lengthening the operator chain for $\hat{X}_L$, and improving its stability as well; if the $\hat{X}_L$ chain is increased to eight physical qubit operators, then we have a distance $d = 8$ logical qubit. This distance must be preserved in all logical qubit operations, and operations that involve turning stabilizers on or off must also be spaced in time by at least $d = 8$ surface code cycles, in order to preserve the temporal spacing needed for error matching of e.g. measurement errors. Even larger qubits can of course be created; turning off sixteen $\hat{Z}$ stabilizers and nine $\hat{X}$ stabilizers in a square pattern would create a $d = 16$ Z-cut logical qubit (the identity of the qubit is determined by whether it is a loop of $\hat{Z}$ operators that encloses the qubit holes or a loop of $\hat{X}_L$ operators, the former corresponding to a Z-cut qubit and the latter an X-cut).

The process to make a larger Z-cut qubit is outlined in Fig. 16; a completely analogous process is used to make larger X-cut qubits. Details are given in the figure caption and in Appendix E.

The process shown in Fig. 16 initializes the qubit in an eigenstate of $\hat{Z}_L$, with an eigenvalue equal to the product of the stable measurement outcomes of the four $\hat{Z}$ stabilizers in the cut. Initializing the qubit instead in an $\hat{X}_L$ eigenstate can be done using a procedure analogous to that described in Section X.

Measuring a larger qubit is also similar to the procedure for measuring small qubits. If we want to measure the logical qubit in Fig. 16 along $\hat{Z}_L$, an "easy" measurement, we turn on the four $\hat{Z}$ and the one $\hat{X}$ stabilizer in the hole; the $\hat{Z}_L$ eigenvalue is equal to the stable product of the four $\hat{Z}$ stabilizers, $Z_{s1}Z_{s2}Z_{s3}Z_{s4} = \pm1$ (where e.g. $Z_{s1} = Z_{1,abcd}$ is the measurement outcome of the stabilizer $\hat{Z}_{s1}$). If we want instead to measure the logical qubit along $\hat{X}_L$, a "difficult" measurement, we follow a procedure analogous to that for the difficult measurement of a small qubit, shown in Fig. 15: We open the strip of $\hat{Z}$ stabilizers through which the $\hat{X}_L$ chain passes, turn the adjacent $\hat{X}$ stabilizers from four- to three-terminal measurements, and measure each of the isolated data qubits along $\hat{X}$.[20] The resulting $\hat{X}$ eigenvalues of each data qubit are multiplied together, and their product is the value of $X_L$. Finally, we reset these data qubits to their ground states $|g\rangle$, and turn all the stabilizers back on, destroying the logical qubit.

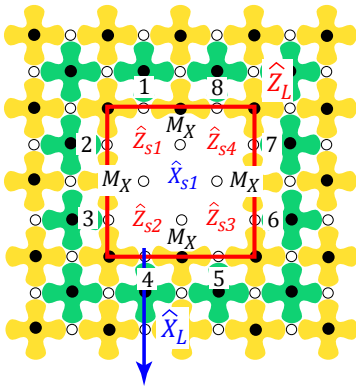

FIG. 16. (Color online) Z-cut qubit with a five-stabilizer hole, showing only the upper qubit multi-cell hole. The logical operator $\hat{Z}_L$ is the chain $\hat{Z}_L = \hat{Z}_1 \hat{Z}_2 \hat{Z}_3 \hat{Z}_4 \hat{Z}_5 \hat{Z}_6 \hat{Z}_7 \hat{Z}_8$, shown in red (gray). The four $\hat{Z}$ stabilizers $\hat{Z}_{s1}$, $\hat{Z}_{s2}$, $\hat{Z}_{s3}$ and $\hat{Z}_{s4}$ are turned off along with the $\hat{X}$ stabilizer $\hat{X}_{s1}$, and the four internal data qubits measured along X for error tracking. The initial value of $\hat{Z}_L$ is equal to the product of the four $\hat{Z}$ stabilizer measurements just prior to turning the stabilizers off (note this stabilizer product commutes with each of the individual data qubit measurements $M_X$).

## XIII. MOVING QUBITS

We have shown how to create and initialize logical qubits in the surface code array. We will discuss later how the single logical qubit gates $\hat{S}_L$, $\hat{T}_L$ and the Hadamard $\bar{H}_L$, all required to perform quantum algorithms, can be performed on these logical qubits. We turn now to the very interesting method used to entangle logical qubits. This can be done in a very elegant way by moving the logical qubit holes around on the 2D array, providing a central functionality of the surface code. By moving one logical qubit hole between the two holes of a second logical qubit, thus "braiding" the logical qubits together, we can perform a logical CNOT. We first however need to learn how to move a logical qubit hole.

Moving a logical qubit is best described in two stages, first outlining what we physically do to the surface code array, and second describing how we transform the logical operators $\hat{X}_L$ and $\hat{Z}_L$, in a way that preserves their identity while matching what is happening to the data qubits in the array. The transformations of the logical operators are described in the context of the Heisenberg representation, so we start by briefly reviewing the Heisenberg representation of quantum mechanics (see e.g. [48] for a more complete introduction, and [47] in relation to quantum computing).

In quantum mechanics, many processes can be described by unitary transformations $\hat{U}$, with $\hat{U}\hat{U}^\dagger = \hat{U}^\dagger\hat{U} = \hat{I}$. These include for example changes of basis, changes of representation, and the temporal evolution of the system. In the Schrödinger representation, these unitary transformations are applied to the wavefunction, so that $|\psi\rangle \to \hat{U}|\psi\rangle$, with the operators kept static. Hence for example wavefunctions are time-dependent, arising from the application of the evolution operator $\hat{U}(t)$, while operators are not. Inner products are invariant under a unitary transformation, since

$$\begin{aligned} \langle\phi|\psi\rangle &\to \langle\phi|\hat{U}^\dagger\hat{U}|\psi\rangle \\ &= \langle\phi|\psi\rangle. \end{aligned} \quad (15)$$

However for the matrix elements of an operator $\hat{A}$ we have

$$\begin{aligned} \langle\phi|\hat{A}|\psi\rangle &\to \left(\langle\phi|\hat{U}^\dagger\right)\hat{A}\left(\hat{U}|\psi\rangle\right) \\ &= \langle\phi|\left(\hat{U}^\dagger\hat{A}\hat{U}\right)|\psi\rangle. \end{aligned} \quad (16)$$

As in general $\hat{A} \neq \hat{U}^\dagger\hat{A}\hat{U}$, the matrix elements can change under a unitary transformation. The second line in Eq. (16) provides the basis for the Heisenberg representation: We can equivalently assume the wavefunction $|\psi\rangle$ does not change under the unitary transformation $\hat{U}$, and instead modify the operator $\hat{A} \to \hat{A}' = \hat{U}^\dagger\hat{A}\hat{U}$. Hence for example in the Heisenberg picture, operators rather than wavefunctions are modified under a change of basis, and operators are time-dependent. The evolution of any measurable quantity will be the same in the Heisenberg representation as in the Schrödinger representation, so there is no detectable difference between the two pictures.

In the surface code, moving and braiding qubits are transformations that affect both the physical array of qubits as well as the logical operators. The physical transformations are not unitary, as these involve projective measurements of the data qubits, but the transformations of the logical operators are unitary, and are best described using the Heisenberg representation. We will assume for this discussion that no errors occur on the physical qubits or during measurements; we discuss error handling after we have worked through the error-free description.

---

[20] In our initial description, we only required the ability to measure qubits along $\hat{Z}$; an $\hat{X}$ measurement in addition would be useful here, but the equivalent can be performed by first performing a Hadamard operation on the qubit, followed by a $\hat{Z}$ measurement.

## A. One-cell logical qubit move

In Fig. 17 we illustrate how to move a logical Z-cut qubit hole downwards by one cell in the 2D array. We remind the reader that a logical Z-cut qubit is created by turning off a pair of $\hat{Z}$ stabilizers, creating two Z-cut holes. The logical qubit has the logical operators $\hat{X}_L = \hat{X}_1\hat{X}_2\hat{X}_3$ and $\hat{Z}_L = \hat{Z}_3\hat{Z}_4\hat{Z}_5\hat{Z}_6$, where we define $\hat{Z}_L$ as the loop of $\hat{Z}$ operators surrounding the lower Z-cut hole. A detailed step-by-step description of the move process is given in Appendix F.

We first describe the physical operations involved in the move, which take two complete surface code cycles. To begin, we wait for the current surface code cycle to finish over the entire array. Before starting the next cycle, we instruct the control software not to measure the $\hat{Z}$ stabilizer just below the qubit hole that we are moving, that is, the control circuitry will turn off that measure-Z qubit in the next surface code cycle. This means that data qubit 6 will not be measured by a $\hat{Z}$ stabilizer (see Fig. 17). We also convert the two measure-X qubits adjacent to data qubit 6 from four-terminal to three-terminal measurements, that is, these measure-X qubits will not include data qubit 6 in their CNOTs in the next surface code cycle. We then perform the next surface code cycle, during which we measure data qubit 6 along $\hat{X}$ and obtain its eigenvalue $X_6$; this measurement is used for tracking errors, as discussed below, and for monitoring the sign of the redefined $\hat{X}_L$ operator, also discussed below. After completing this cycle, we turn on the measure-Z qubit in the original lower Z-cut hole, so it begins to perform a normal CNOT cycle. We also convert the two measure-X qubits that monitor data qubit 6 back to four-terminal measurements, so they again include data qubit 6 in their CNOT cycles. We then perform the next surface code cycle, which completes the physical operations for the move. To establish all stabilizer values in time, we then wait an additional $d - 1$ surface code cycles. Altogether this one-cell move takes $1 + d$ surface code cycles.

While performing the physical operations, we also need to manipulate the definitions of the two logical operators $\hat{Z}_L$ and $\hat{X}_L$ to preserve their functionality. Before we perform any physical operations, we redefine $\hat{Z}_L$ so that it encloses two stabilizer cells, its original cell plus the one below it, by multiplying $\hat{Z}_L$ by the four $\hat{Z}$ operators, $\hat{Z}_6\hat{Z}_7\hat{Z}_8\hat{Z}_9$, that make up the $\hat{Z}$ stabilizer below the qubit hole. We term the new extended operator $\hat{Z}_L^e$, which is given by

$$\begin{aligned}\hat{Z}_L^e &= (\hat{Z}_3\hat{Z}_4\hat{Z}_5\hat{Z}_6) \times (\hat{Z}_6\hat{Z}_7\hat{Z}_8\hat{Z}_9) \\ &= \hat{Z}_3\hat{Z}_4\hat{Z}_5\hat{Z}_7\hat{Z}_8\hat{Z}_9,\end{aligned} \tag{17}$$

as shown in Fig. 17a. We complete the first surface code cycle of the move, turning off the lower $\hat{Z}$ stabilizer and measuring data qubit 6. We redefine the $\hat{X}_L$ operator by multiplying it by $\hat{X}_6$, giving

$$\hat{X}_L' = \hat{X}_1\hat{X}_2\hat{X}_3\hat{X}_6; \tag{18}$$

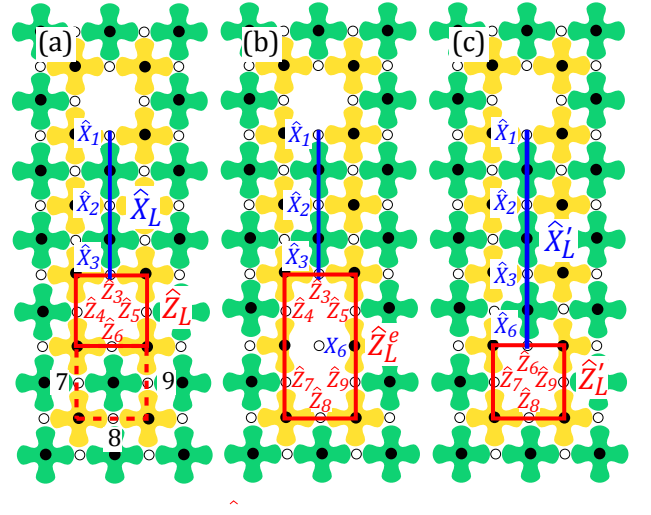

FIG. 17. (Color online) Process for moving a Z-cut logical qubit hole vertically down by one cell in the 2D array. (a) Logical qubit with $\hat{X}_L = \hat{X}_1\hat{X}_2\hat{X}_3$ and $\hat{Z}_L = \hat{Z}_3\hat{Z}_4\hat{Z}_5\hat{Z}_6$. (b) We extend the $\hat{Z}_L$ operator by multiplying it by the $\hat{Z}$ stabilizer just below the lower qubit hole: $\hat{Z}_L^e = (\hat{Z}_6\hat{Z}_7\hat{Z}_8\hat{Z}_9)\hat{Z}_L = \hat{Z}_3\hat{Z}_4\hat{Z}_5\hat{Z}_7\hat{Z}_8\hat{Z}_9$. We next turn off this $\hat{Z}$ stabilizer, and also turn the surrounding four-terminal $\hat{X}$ stabilizers into three-terminal stabilizers, leaving data qubit 6 without any stabilizer measurements. We perform an $\hat{X}$ measurement of data qubit 6 (using either an idle measure-Z qubit, or directly measuring the data qubit, as shown). (c) We define the extended operator $\hat{X}_L'$ by multiplying $\hat{X}_L$ by $\hat{X}_6$: $\hat{X}_L' = \hat{X}_6\hat{X}_L = \hat{X}_1\hat{X}_2\hat{X}_3\hat{X}_6$. We now turn on the $\hat{Z}$ stabilizer just above qubit 6; we wait $d$ surface code cycles in order to properly establish the value of this stabilizer. We define $\hat{Z}_L'$ as the product of this $\hat{Z}$ stabilizer and $\hat{Z}_L^e$: $\hat{Z}_L' = \hat{Z}_6\hat{Z}_7\hat{Z}_8\hat{Z}_9$.

see Fig. 17b.

We next complete the second surface code cycle of the move, turning the original qubit hole stabilizer back on. We redefine $\hat{Z}_L^e$ by multiplying it by the four $\hat{Z}$ operators that make up the stabilizer that was just turned on, $\hat{Z}_3\hat{Z}_4\hat{Z}_5\hat{Z}_6$, giving

$$\begin{aligned}\hat{Z}_L' &= (\hat{Z}_3\hat{Z}_4\hat{Z}_5\hat{Z}_6) \times \hat{Z}_L^e \\ &= (\hat{Z}_3\hat{Z}_4\hat{Z}_5\hat{Z}_6) \times (\hat{Z}_3\hat{Z}_4\hat{Z}_5\hat{Z}_7\hat{Z}_8\hat{Z}_9) \\ &= \hat{Z}_6\hat{Z}_7\hat{Z}_8\hat{Z}_9.\end{aligned} \tag{19}$$

The two new logical operators $\hat{Z}_L'$ and $\hat{X}_L'$ are now as drawn in Fig. 17c. Note the new logical operators still share a single data qubit and commute with all the stabilizers, so these continue to define a logical qubit. The process of moving the qubit hole involves operations similar to those used for the "easy" and "difficult" initializations, so error processes are as discussed in Sect. XI.

## B. Byproduct operators

One problem that is created in the move transformation is that the extension of $\hat{X}_L$ by multiplication by $\hat{X}_6$ can yield an extended $\hat{X}'_L$ that differs in sign from $\hat{X}_L$; this will occur if the measurement outcome $X_6 = -1$. Similarly the final $\hat{Z}'_L$ can differ in sign from $\hat{Z}_L$, depending on the product of the two $\hat{Z}$ stabilizer measurement outcomes involved in the move. Given the initial state $|\psi\rangle$ prior to the transformation, the transformation coupled with these sign changes results in

$$|\psi\rangle \rightarrow \hat{X}'^{p_X}_L \hat{Z}'^{p_X}_L |\psi'\rangle, \qquad (20)$$

where $|\psi'\rangle$ is the desired state, which would result if there were no sign changes in the logical operators. Hence we must operate on the resulting state with the operator product $\hat{Z}'^{p_Z}_L \hat{X}'^{p_Z}_L$ to regain the desired state $|\psi'\rangle$.

The additional bit- and phase-flip operators appearing in Eq. (20) are called "byproduct operators". The power $p_X$ appearing in Eq. (20) is determined by whether a sign change occurred in $\hat{X}_L$ during the move, with $p_X = 0(1)$ if a sign change did not (did) occur; the power $p_Z$ is determined similarly for sign changes in $\hat{Z}_L$. The byproduct operator $\hat{Z}'_L$ corrects the sign of the transformed $\hat{X}'_L$ through their anti-commutation relation, and the byproduct operator $\hat{X}'_L$ similarly corrects $\hat{Z}'_L$.

We do not actually apply the byproduct operators as explicit gates; instead, we simply track these additional $\hat{X}_L$ and $\hat{Z}_L$ operators in the control software, as discussed in Sect. IX. These corrective operators are thus only applied when the logical qubit is measured, reversing the sign of a $\hat{Z}_L$ measurement if that qubit has an $\hat{X}_L$ byproduct operator, and reversing the sign of an $\hat{X}_L$ measurement if that qubit has a $\hat{Z}_L$ byproduct operator. Two corrective $\hat{X}_L$s or two corrective $\hat{Z}_L$s cancel, as $\hat{X}^2_L = \hat{Z}^2_L = \hat{I}_L$.

We can also look at what happens to our description of the 2D array wavefunction $|\psi\rangle$ in terms of the stabilizer and logical qubit subspaces. Prior to the move, we have $|\psi\rangle = |Q\rangle|q_L\rangle$; after the move, we have $\hat{X}'^{p_Z}_L \hat{Z}'^{p_Z}_L |\psi'\rangle = \hat{X}'^{p_Z}_L \hat{Z}'^{p_Z}_L |Q'\rangle|q'_L\rangle$, where $|Q'\rangle|q'_L\rangle$ is the desired state. The byproduct operators only affect the logical state, so we can write the equivalent to Eq. (20) as

$$|q_L\rangle \rightarrow \hat{X}'^{p_X}_L \hat{Z}'^{p_X}_L |q'_L\rangle, \qquad (21)$$

where $|q'_L\rangle$ is the desired logical state.

## C. Multi-cell logical qubit move

We can easily generalize the one-cell move to a multi-cell move. A multi-cell move can be done in the same number of surface code clock cycles as a one-cell move, and we can translate the logical qubit hole over an unlimited number of cells. Multi-cell moves are performed by extending the one-cell move to a contiguous strip of cells. A multi-cell move is shown in Fig. 18 for a Z-cut qubit hole, using a somewhat more abstract representation than we used in Fig. 17; we have mostly dropped identification of the individual physical qubits, but the analogy with Fig. 17 should be clear. Details for the multi-cell move are given in the figure caption and in Appendix G.

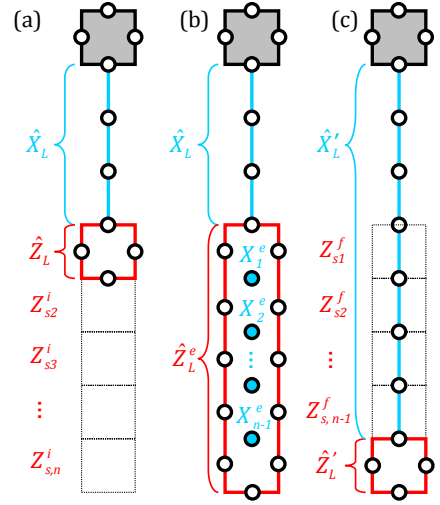

FIG. 18. (Color online) Process for moving a Z-cut logical qubit hole through multiple cells, using a simplified representation where the $\hat{X}_L$ chain is a blue (light) line linking the logical qubit holes and passing through the data qubits (open circles), and the $\hat{Z}_L$ loop is in red (gray). The upper (gray) and lower (white) Z-cut holes enclose idle $\hat{Z}$ stabilizers; the outlines of these holes have measure-X qubits on the vertices with data qubits on the edges. (a) Initial state, showing the logical operators $\hat{X}_L$ and $\hat{Z}_L$, with the initial (pre-move) software-corrected $\hat{Z}$ stabilizer measurement outcomes $Z^i_{s2}$, $Z^i_{s3} \ldots Z^i_{s,n}$. The $\hat{Z}$ stabilizers involved in the next step are shown with thin black outlines; the $j$th stabilizer $\hat{Z}_{s,j}$ is the stabilizer cell marked by that stabilizer's initial measurement outcome $Z^i_{s,j}$. (b) Extension of $\hat{Z}_L$ to $\hat{Z}^e_L$, with $\hat{X}$ measurements of the isolated data qubits yielding the measurement eigenvalues $X^e_1$, $X^e_2, \ldots X^e_{n-1}$ (data qubits filled in blue (light)). (c) Final state, with $\hat{Z}'_L$ the shifted $\hat{Z}_L$ logical operator and $\hat{X}'_L$ the extended chain for $\hat{X}_L$, with the $\hat{Z}$ stabilizers (thin black outlines) turned back on; after waiting $d$ surface code cycles, we establish the software-corrected final (post-move) measurement outcomes $Z^f_{s1}$, $Z^f_{s2}, \ldots Z^f_{s,n-1}$.

Briefly, $\hat{Z}_L$ is extended by multiplying it by all the stabilizers through which the logical qubit is to be moved, giving the extended $\hat{Z}^e_L$:

$$\hat{Z}^e_L = (\hat{Z}_{s2}\hat{Z}_{s3} \ldots \hat{Z}_{s,n})\hat{Z}_L. \qquad (22)$$

Here each stabilizer operator $\hat{Z}_{sj}$ represents the product of the four $\hat{Z}$ operators on the four data qubits surrounding the $j$th stabilizer cell. These stabilizers are then turned off, and the four-terminal $\hat{X}$ stabilizers adjacent to the strip switched to three-terminal measurements. The un-stabilized data qubits in the strip are all measured along $\hat{X}$, and $\hat{X}_L$ is extended by multiplying it by all the $\hat{X}$ operators on these data qubits,

$$\hat{X}'_L = (\hat{X}_1 \ldots \hat{X}_{n-1})\hat{X}_L. \qquad (23)$$

We complete the move for $\hat{Z}_L$ by then multiplying $\hat{Z}^e_L$ by all the stabilizers except the one that is in the new shifted qubit hole,

$$
\begin{aligned}
\hat{Z}'_L &= (\hat{Z}_{s1}\hat{Z}_{s2}\ldots\hat{Z}_{s,n-1})\hat{Z}^e_L \\
&= (\hat{Z}_{s1}\ldots\hat{Z}_{s,n-1})(\hat{Z}_{s2}\ldots\hat{Z}_{s,n})\hat{Z}_L \\
&= \hat{Z}_{s1}\hat{Z}_{s,n}\hat{Z}_L \\
&= \hat{Z}_{s,n}
\end{aligned}
\qquad (24)
$$

where we use the fact that $\hat{Z}^2_{sj} = \hat{I}$ and that $\hat{Z}_L = \hat{Z}_{s1}$, i.e. the original $\hat{Z}_L$ is just the set of four $\hat{Z}$ data qubit operators that define $\hat{Z}_{s1}$. After the move, we wait an additional $d-1$ surface code cycles to establish all stabilizer values in time.

Byproduct operators may be appear due to any sign changes that occur to the logical operators $\hat{X}'_L$ and $\hat{Z}'_L$ during this move; see Appendix G for details.

A completely analogous process is used to move X-cut logical qubit holes, exchanging the roles of the $\hat{X}$ and $\hat{Z}$ stabilizers and measurements.

We note that multi-cell moves can be done very quickly, as a very long cut can be made in just one step of the surface code cycle, and the qubit holes moved in $1+d$ surface code cycles, the same as for a one-cell move. This therefore enables long-distance interaction and communication between logical qubits, a very powerful capability.

### D. Errors during move transformations

In Sect. XI we discussed the handling of errors when the surface code stabilizers are being manipulated, where we focussed on errors occurring when an X-cut logical qubit is initialized in a $\hat{Z}_L$ eigenstate. A very similar discussion applies to errors that occur during the move transformations, with some differences in the details. The data qubits that are left isolated during the move can suffer from both $\hat{Z}$ and $\hat{X}$ errors; $\hat{X}$ errors have no impact, as they are erased by the single data qubit $\hat{X}$ measurements during the move. $\hat{Z}$ errors will be detected by computing the product of each data qubit $\hat{X}$ measurement with its corresponding three-terminal $\hat{X}$ stabilizer measurement, and comparing this product with that stabilizer's prior four-terminal $\hat{X}$ measurement (this is how $\hat{X}$ errors were handled in Sect. XI). $\hat{Z}$ errors on the data qubits bordering the cut are detected and localized by the two $\hat{X}$ stabilizers that monitor each of these data qubits. An $\hat{X}$ error on one of these data qubits changes the sign of the solitary $\hat{Z}$ stabilizer that monitors that data qubit, but this error will persist in time, and can thus be distinguished from a stabilizer measurement error by waiting $d$ surface code cycles.

### XIV. THE BRAIDING TRANSFORMATION AND THE LOGICAL CNOT

A braiding transformation is a particularly important type of move. A braid comprises a pair of multi-cell moves involving one of the two holes in a double-cut logical qubit. The shifted hole travels over a path that forms a closed loop in the 2D array, starting and ending in the same location, with the first move taking the hole partway around the loop, and the second move completing the loop. As with the move transformations, braids are described in terms of their effect on the logical qubit operators.

Most importantly, as we shall see, the braid transformation can entangle two logical qubits, in a way that makes the braid transformation equivalent to a logical CNOT.

### A. Braid transformation of one logical qubit

We begin by describing the braid transformation of a single logical qubit, using as an example the Z-cut logical qubit. In Fig. 19 we show the transformation of the $\hat{X}_L$ operator and in Fig. 20 that of the $\hat{Z}_L$ operator. The braid shifts the qubit hole along a path that, in this example, encloses an area containing only fully stabilized cells; we will discuss below what happens when the braid path encloses another logical qubit hole. The braid's first move shifts the qubit hole eight cells along the path (for a general braid transformation, this can be any number of cells between one and the array distance $d$ less than the total number of cells in the braid path).[21] This move is completed before the second move is started, meaning that all the stabilizers that were switched off during the move have been turned back on (except for the last cell). After waiting $d$ surface code cycles, to catch measurement errors during the first move, the second multi-cell move is then performed, shifting the qubit hole the remaining four cells in the path, and returning the hole to where it started. The braid cannot be completed in a single multi-cell move, as this would isolate the part of the surface code array enclosed by the braid.

As can be seen in Figs. 19 and 20, the braid transforms the two logical operators $\hat{X}_L$ and $\hat{Z}_L$ in somewhat different ways. The $\hat{X}_L$ operator for this qubit is a chain of three data qubit $\hat{X}$ operators linking the two holes (Fig. 19a). As shown in Fig. 19b-e, this operator is extended in each of the moves, such that after the second move, the $\hat{X}_L''$ operator in Fig. 19e includes the original chain of three data qubit operators in addition to a closed loop of $\hat{X}$ operators, $\hat{X}_{\mathrm{loop}} = \hat{X}_1 \ldots \hat{X}_{12}$. This closed loop encloses only fully stabilized cells, with stabilizers $\hat{X}_{s1}$, $\hat{X}_{s2} \ldots \hat{X}_{s9}$, so $\hat{X}_{\mathrm{loop}} = \hat{X}_{s1} \ldots \hat{X}_{s9}$. The quiescent state $|\psi\rangle$ is necessarily an eigenstate of $\hat{X}_{\mathrm{loop}}$, with eigenvalue given by the measurement outcomes of the enclosed stabilizers, $X_{\mathrm{loop}} = X_{s1} \ldots X_{s9}$. Another way to understand this is that the closed loop of operators $\hat{X}_{\mathrm{loop}}$ can be deformed through all the enclosed $\hat{X}$ stabilizers, leaving only a product of measurement outcomes.

---

[21] Moving the hole so that after this move it ends up less than $d$ cells from the start point would effectively reduce the array distance, and thus reduce the fault tolerance during the braid transformation.

The $\hat{Z}_L$ operator is the loop of physical qubit $\hat{Z}$ operators that encloses the lower qubit hole. As shown in Fig. 20, the braid transformation alternately extends and then collapses the operator loop that surrounds the Z-cut hole, once for each of the two moves; other than sign changes from $\hat{Z}$ stabilizers that the loop passes over, the final $\hat{Z}_L$ loop is the same as the initial loop. The difference between the transformation of $\hat{X}_L$ and $\hat{Z}_L$ is the key to how the braid acts as a CNOT.

Details of the braid transformation involving just one logical qubit are given in the captions for Figs. 19 and 20, with a discussion of the sign changes and byproduct operators given in Appendix H. Note that in Figs. 19 and 20, while we separate the actions that affect the $\hat{X}_L$ operator from those that affect $\hat{Z}_L$, all $\hat{X}$ measurements of the isolated data qubits and measurements of the $\hat{Z}$ stabilizers are performed as part of the braid, independent of what logical operators are involved in the transformation, just as they are for the one-cell and multi-cell moves.

### B. Braiding two qubits

We now turn to the less trivial situation where the braid path encloses another logical qubit hole. The easiest way to understand this is to examine how operators on the two logical qubits are transformed by the braid. The two-qubit operators will be outer products of the single-qubit operators $\hat{X}_L$, $\hat{Z}_L$ and $\hat{I}_L$. The $\hat{Y}_L$ operator is the product of $\hat{Z}_L$ and $\hat{X}_L$, and its transformation can be understood in terms of the $\hat{X}_L$ and $\hat{Z}_L$ transformations.

We quickly review two-qubit operators, using as an example $\hat{X}_L \otimes \hat{Z}_L$. We write the two-qubit logical state as $|a_L b_L\rangle$, where $|a_L\rangle$ represents the state of the first logical qubit and $|b_L\rangle$ that of the second qubit. The notation $\hat{X}_L \otimes \hat{Z}_L$ means that $\hat{X}_L$ acts on the first qubit state $|a_L\rangle$, and $\hat{Z}_L$ acts on the second qubit state $|b_L\rangle$. Using the eigenstates $|g_L\rangle$ and $|e_L\rangle$ as basis states, the outer product $\hat{X}_L \otimes \hat{Z}_L$ can be represented in the standard two-qubit basis $|g_L g_L\rangle$, $|g_L e_L\rangle$, $|e_L g_L\rangle$, $|e_L e_L\rangle$ by the $4 \times 4$ matrix

$$
\begin{aligned}
\hat{X}_L \otimes \hat{Z}_L &= \begin{pmatrix} 0 & 1 \\ 1 & 0 \end{pmatrix} \otimes \begin{pmatrix} 1 & 0 \\ 0 & -1 \end{pmatrix} \\
&= \begin{pmatrix} 0 & 0 & 1 & 0 \\ 0 & 0 & 0 & -1 \\ 1 & 0 & 0 & 0 \\ 0 & -1 & 0 & 0 \end{pmatrix}.
\end{aligned} \tag{25}
$$

We will see that the braid transforms each two-qubit operator into some other two-qubit operator, with the outcome depending on the particular pair of operators, and on the order in which their product is formed (so that e.g. $\hat{X}_L \otimes \hat{Z}_L$ does not transform the same way as $\hat{Z}_L \otimes \hat{X}_L$). The outcome of a braid also depends on what types of qubits we are braiding together; we focus here on braiding a Z-cut qubit through an X-cut qubit, as shown in Fig. 21. As we will see, only a braid between a Z-cut and an X-cut qubit yields an operator transformation that is equivalent to a logical CNOT. It is not possible to obtain the desired CNOT transformations when braiding two Z-cut qubits together, or two X-cut qubits; we discuss these situations in Sect. XIV D.

In Fig. 21, the Z-cut qubit is in the upper part of the figure, with the braid moving the lower Z-cut hole of this qubit around a closed loop. The X-cut qubit is in the lower part of the figure, with the braid taking the Z-cut hole between the two X-cut holes, enclosing the upper X-cut hole in the braid loop. As we will work out later, proving that a braid is equivalent to a CNOT only involves showing that four of the sixteen possible two-qubit operator combinations transform correctly; these are $\hat{X}_L \otimes \hat{I}_L$, $\hat{I}_L \otimes \hat{X}_L$, $\hat{Z}_L \otimes \hat{I}_L$ and $\hat{I}_L \otimes \hat{Z}_L$. The transformations for all the other two-qubit combinations of $\hat{X}_L$, $\hat{Z}_L$ and $\hat{I}_L$ can be constructed from these four. Here we give a brief outline of these transformations, with details given in Appendix I.

$\hat{X}_{L1} \otimes \hat{I}_{L2} \to \hat{X}_{L1} \otimes \hat{X}_{L2}$: Figure 21 shows the transformation of $\hat{X}_{L1}$ on the first, Z-cut qubit, with no operation ($\hat{I}_{L2}$) on the second qubit. As with the empty loop braid, the extension of $\hat{X}_{L1}$ operator chain in the two moves creates an operator $\hat{X}_{L1}''$ that comprises a closed loop of $\hat{X}$ operators in addition to the original operator chain $\hat{X}_{L1}$. The loop encloses the X-cut qubit's upper hole. For the empty loop braid, we could move the loop through all the enclosed cells, as these were all stabilized, so the loop operator resolved to a simple product of measurement outcomes. Here, we cannot do this, as the X-cut hole is inside the loop and is not stabilized; instead, we can transform the loop through each stabilized cell, until it wraps tightly around the X-cut hole. This loop of $\hat{X}$ operators is then equivalent to $\hat{X}_{L2}$, a logical $\hat{X}$ operation on the second qubit. The remaining chain of operators in $\hat{X}_{L1}''$ is the same as the original $\hat{X}_{L1}$ chain prior to the braid. We see therefore that the braid takes $\hat{X}_{L1}$ on the first qubit and $\hat{I}_{L2}$ on the second qubit to $\hat{X}_{L1}$ on the first qubit and $\hat{X}_{L2}$ on the second qubit, i.e.

$$
\hat{X}_{L1} \otimes \hat{I}_{L2} \to \hat{X}_{L1} \otimes \hat{X}_{L2}, \tag{26}
$$

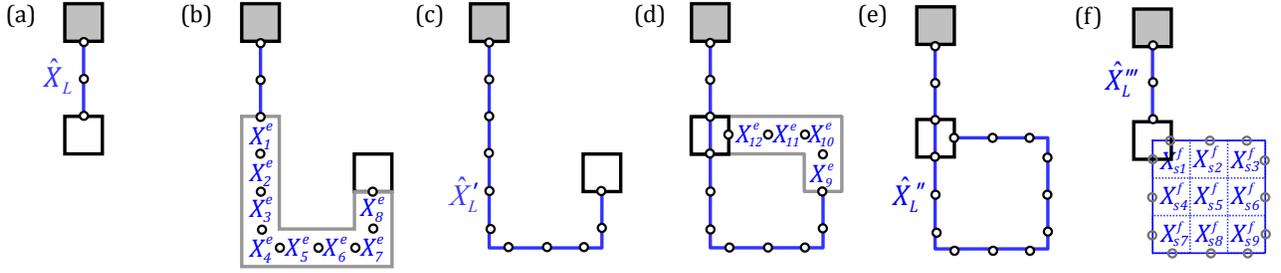

FIG. 19. (Color online) Braid transformation of a single Z-cut qubit in a fully-stabilized array, showing steps affecting the $\hat{X}_L$ operator (delineated in blue (dark)). (a) Double Z-cut qubit with the logical operator $\hat{X}_L$, showing the three data qubits (open circles) that define the logical operator. (b) Extended opening (heavy grey) for first move in the braid. Data qubits isolated by the extension are measured in $\hat{X}$ with outcomes $X_1^e, \ldots, X_8^e$. (c) The $\hat{X}_L$ operator (blue, dark) is extended to $\hat{X}_L' = \hat{X}_1 \ldots \hat{X}_8 \hat{X}_L$. (d) Second move in the braid, which again involves measuring isolated data qubits in $\hat{X}$. (e) The $\hat{X}_L'$ operator is extended to $\hat{X}_L''$ and now includes the original chain linking the two qubit holes, and in addition a closed loop of operators. (f) The closed loop is completely stabilized by the nine $\hat{X}$ stabilizers $\hat{X}_{s1} \ldots \hat{X}_{s9}$ (blue (dark) squares). We can thus reduce the operator chain $\hat{X}_L''$ to a chain $\hat{X}_L'''$ that is identical to the original $\hat{X}_L$, other than possible sign changes. The sign changes are captured as in the multi-cell move by defining the power $p_X$ through $(-1)^{p_X} \equiv X_{s1}^f X_{s2}^f \ldots X_{s9}^f = \pm 1$, given by the product of all the $\hat{X}$ stabilizer measurement outcomes enclosed by the $\hat{X}_L''$ loop. The power $p_X = 0(1)$ determines whether a $\hat{Z}_L$ byproduct operator does not (does) appear, multiplying the final wavefunction.

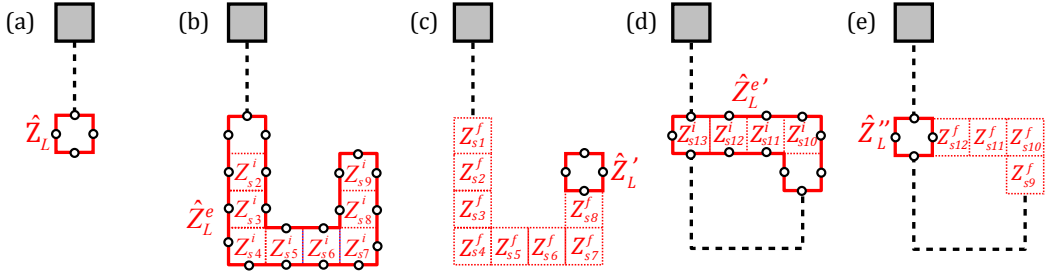

FIG. 20. (Color online) Braid transformation of a $\hat{Z}_L$ operator on a single Z-cut qubit in a fully-stabilized array. (a) Double Z-cut qubit with $\hat{Z}_L$ in red (gray); data qubits are open circles; dashed line linking the two qubit holes does not represent an operator. (b) Extended opening (solid red (gray) loop) for first move in the braid. The pre-move $\hat{Z}$ stabilizer outcomes $Z_{s2}^i, \ldots, Z_{s9}^i$ are all equal to $\pm 1$. The $\hat{Z}_L^e$ operator (solid red (gray) loop) is extended from $\hat{Z}_L$, $\hat{Z}_L^e \equiv \hat{Z}_{s2} \ldots \hat{Z}_{s9} \hat{Z}_L$. (c) The extended opening is closed up by turning on the stabilizers $\hat{Z}_{s1}, \ldots, \hat{Z}_{s8}$, and waiting $d$ surface code cycles to obtain the error-corrected outcomes $Z_{s1}^e, \ldots, Z_{s8}^e$. The new $\hat{Z}_L'$ operator is the red (gray) loop formed by four data qubit $\hat{Z}$ operators, equal to $\hat{Z}_L' = (\hat{Z}_{s1} \ldots \hat{Z}_{s8})(\hat{Z}_{s2} \ldots \hat{Z}_{s9}) \hat{Z}_L$. (d) Second move in the braid, which involves extending $\hat{Z}_L'$ to $\hat{Z}_L^{e'}$, using the associated $\hat{Z}$ stabilizer outcomes. (e) The final $\hat{Z}_L''$ is a closed loop of four data qubit operators identical to the original $\hat{Z}_L'$. $\hat{Z}_L''$ may differ in sign with respect to $\hat{Z}_L$, captured by defining $(-1)^{p_Z} \equiv (Z_{s9}^f \ldots Z_{s12}^f)(Z_{s10}^i \ldots Z_{s13}^i)(Z_{s1}^f \ldots Z_{s8}^f)(Z_{s2}^i \ldots Z_{s9}^i) = \pm 1$. The power $p_Z$ determines whether an $\hat{X}_L$ byproduct operator appears, multiplying the final wavefunction.

with details provided in Appendix I.

$\hat{I}_{L1} \otimes \hat{X}_{L2} \rightarrow \hat{I}_{L1} \otimes \hat{X}_{L2}$: Figure 22 shows the braid transformation of $\hat{I}_{L1}$ on the first Z-cut qubit with $\hat{X}_{L2}$ on the second X-cut qubit. Braiding the Z-cut qubit hole through the X-cut qubit does not generate any chains of operators from either qubit that wrap around or otherwise interact with the other qubit, so the braid transformation in this case does nothing (other than sign changes). Hence we find

$$\hat{I}_{L1} \otimes \hat{X}_{L2} \rightarrow \hat{I}_{L1} \otimes \hat{X}_{L2} \qquad (27)$$

(details and sign changes in Appendix I).

$\hat{I}_{L1} \otimes \hat{Z}_{L2} \rightarrow \hat{Z}_{L1} \otimes \hat{Z}_{L2}$: Figure 23 shows the transformation of $\hat{I}_{L1}$ on the first Z-cut qubit and $\hat{Z}_{L2}$ on the second X-cut qubit. In Fig. 23b, $\hat{Z}_{L2}$ is extended as shown in the figure, by multiplying it by the $\hat{Z}$ stabilizers outlined by the dashed boxes. The first qubit's hole is then moved through the path left open by the extension of $\hat{Z}_{L2}$ (Fig. 23c). $\hat{Z}_{L2}$ is then multiplied by all the stabilizers shown in the dashed boxes (Fig. 23d), leaving behind a loop of $\hat{Z}$ operators surrounding the first qubit's hole, a loop that is exactly a $\hat{Z}_{L1}$ operation on the first Z-cut qubit, as well as the original $\hat{Z}_{L2}$ on qubit 2, unchanged from prior to the braid (other than possible sign

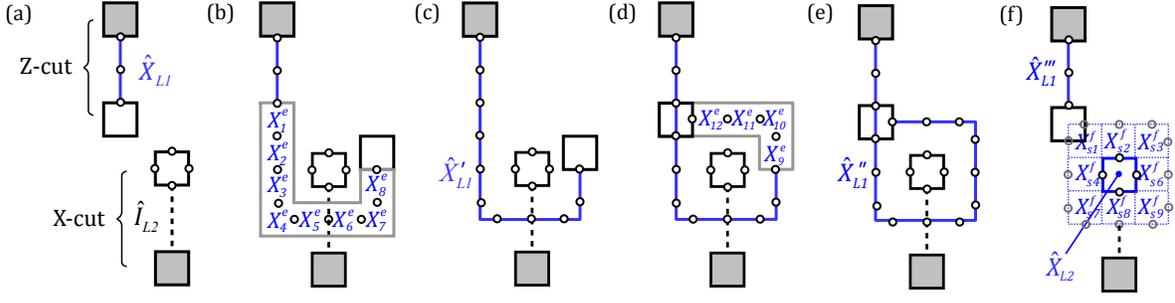

FIG. 21. (Color online) A braid on a Z-cut qubit, where the lower Z-cut hole of the qubit is braided around the upper X-cut hole of an X-cut qubit. The braid's effect on the $\hat{X}_{L1}$ operator on Z-cut qubit 1 is displayed. (a) Z-cut qubit (above) and X-cut qubit (below) with corresponding logical operators; data qubits are shown as open circles. (b) Extension for first move in braid, where data qubits are measured along $\hat{X}$, with measurement outcomes $X_1^e$, $X_2^e$, ... $X_8^e$. (c) $\hat{X}_{L1}$ operator is extended in length to $\hat{X}_{L1}'$. (d) Extension for second move in braid, where data qubits are measured along $\hat{X}$ with measurement outcomes $X_9^e$, ... $X_{12}^e$. (e) $\hat{X}_{L1}''$ operator is extended in length, comprising the original chain plus a closed loop of data qubit operators, the loop enclosing the upper hole of the X-cut qubit. (f) The loop part of the $\hat{X}_{L1}''$ operator is moved through the enclosed stabilized cells, leaving a loop of $\hat{X}$ data qubit operators that is equal to $\hat{X}_{L2}$ on the second qubit, with $\hat{X}_{L1}$ unchanged from before the braid (other than possible sign changes).

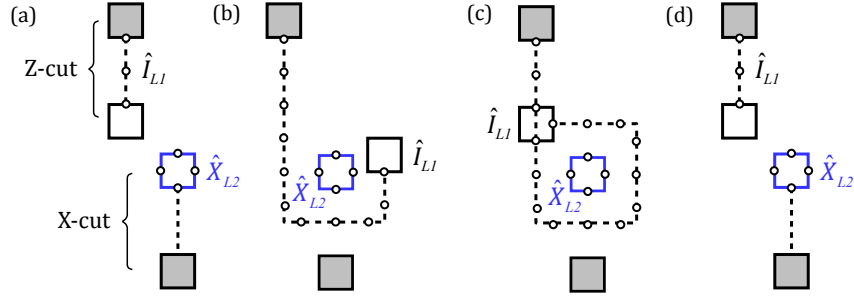

FIG. 22. (Color online) (a)-(d) Illustration of the braid transformation of the $\hat{I}_{L1} \otimes \hat{X}_{L2}$ operator product. After the move, there is no net change of either operator.

changes). The braid therefore performs the transformation

$$\hat{I}_{L1} \otimes \hat{Z}_{L2} \to \hat{Z}_{L1} \otimes \hat{Z}_{L2}. \tag{28}$$

$\hat{Z}_{L1} \otimes \hat{I}_{L2} \to \hat{Z}_{L1} \otimes \hat{I}_{L2}$: Finally we consider the braid transformation of $\hat{Z}_{L1}$ on the first qubit with $\hat{I}_{L2}$ on the second qubit. Braiding the first qubit hole through the second qubit drags the loop of $\hat{Z}$ operators along as it did for the empty loop braid, but as the loop preserves its closed form during the two moves, it does not generate a chain or loop of operators that can act on the second qubit, so the braid transformation does nothing (other than sign changes). Hence we find

$$\hat{Z}_{L1} \otimes \hat{I}_{L2} \to \hat{Z}_{L1} \otimes \hat{I}_{L2}. \tag{29}$$

In general, a braid transformation leaves logical operators that are built from closed loops of data qubit operators unchanged, and there is no braid-induced interaction with the other qubit hole. By contrast, logical operators that are built from open chains of data qubit operators linking the two qubit holes end up leaving a loop of operators surrounding the other qubit hole. The different braid outcomes arise because the first logical qubit is a Z-cut qubit, for which the $\hat{X}_L$ operator is an open chain that interacts with the second qubit, while the second logical qubit is an X-cut qubit, for which the $\hat{Z}_L$ operator is an open chain that interacts with the first qubit.

The braid is made of two move transformations that induce sign changes in the first qubit's logical operators, which appear as qubit 1 byproduct logical operators $\hat{X}_{L1}$ and $\hat{Z}_{L1}$ acting on the surface code wavefunction, as was discussed for the one-cell and multi-cell moves. The braid transformation also generates sign changes in the second qubit's logical operators, even though that qubit is not displaced during the move. These sign changes generate qubit 2 byproduct logical operators $\hat{X}_{L2}$ and $\hat{Z}_{L2}$ acting on the wavefunction. Details regarding the byproduct operators are given in Appendix I.

We now turn to a discussion of the CNOT gate, and make clear why the transformations we have detailed ac-tually identify the braid as a CNOT between two logical qubits.

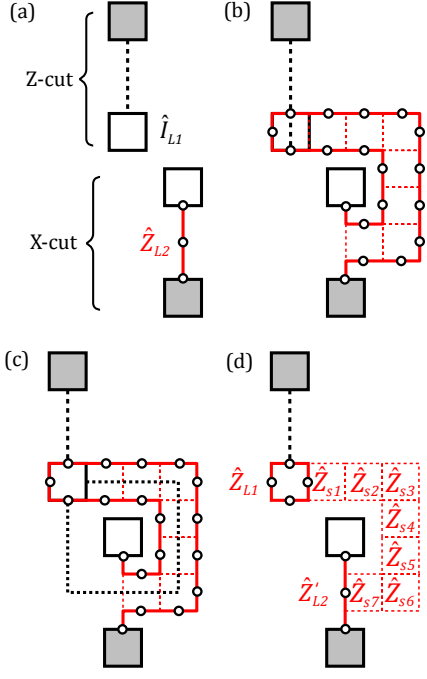

FIG. 23. (Color online) Braid transformation of $\hat{I}_{L1} \otimes \check{Z}_{L2}$. (a) Prior to the move, displaying $\hat{I}_{L1}$ on qubit 1 and $\check{Z}_{L2}$ on qubit 2. Data qubits are shown as open circles where relevant. (b) The $\check{Z}_{L2}$ operator is extended by multiplying it by a set of stabilizers (dotted boxes), resulting in the heavy red (gray) line that is equivalent to the original $\check{Z}_{L2}$, with the exception of possible sign changes. (c) Qubit 1's hole is moved through the path opened up by extending $\check{Z}_{L2}$. (d) $\check{Z}_{L2}$ is moved back to its original form, leaving behind a loop of $\hat{Z}$ physical qubit operators that comprise a $\check{Z}_{L1}$ operator on qubit 1.

## C. The CNOT gate

The two-qubit CNOT gate is a fundamental gate for quantum computation. One of the two qubits in the CNOT serves as the control, and the other as the target. In the standard two-qubit basis $|gg\rangle$, $|ge\rangle$, $|eg\rangle$, $|ee\rangle$, the CNOT is represented by the $4 \times 4$ real matrix

$$\hat{C} = \begin{pmatrix} 1 & 0 & 0 & 0 \\ 0 & 1 & 0 & 0 \\ 0 & 0 & 0 & 1 \\ 0 & 0 & 1 & 0 \end{pmatrix}. \tag{30}$$

If the control qubit is in $|g\rangle$, the target qubit state is unchanged, while if the control qubit is in $|e\rangle$, the target qubit undergoes an $\hat{X}$ bit-flip, exchanging $|g\rangle$ and $|e\rangle$. Note that $\hat{C}$ is Hermitian, $\hat{C}^\dagger = \hat{C}$, and unitary, $\hat{C}\hat{C}^\dagger = \hat{C}^\dagger\hat{C} = \hat{C}\hat{C} = \hat{I}$.

One way to test an experimental CNOT gate is to perform the CNOT on each of the four two-qubit basis states, and then do projective measurements of the result onto each of the four basis states. These sixteen experiments can be compared to the matrix in Eq. (30) to verify that the CNOT has been implemented correctly. This is essentially a Schrödinger picture test of the CNOT.

An equivalent method is to use the Heisenberg picture, by examining the transformation of the various two-qubit operators due to the action of the CNOT. This would seem to involve showing that the CNOT performs the correct transformation for all sixteen outer products of pairs of the four single qubit operators, $\hat{I}$, $\hat{X}$, $\hat{Y}$ and $\hat{Z}$. It turns out you only need to show that the CNOT transforms four of these outer products correctly:

$$\hat{C}^\dagger \left( \hat{I} \otimes \hat{X} \right) \hat{C} = \hat{I} \otimes \hat{X}, \tag{31}$$

$$\hat{C}^\dagger \left( \hat{X} \otimes \hat{I} \right) \hat{C} = \hat{X} \otimes \hat{X}, \tag{32}$$

$$\hat{C}^\dagger \left( \hat{I} \otimes \hat{Z} \right) \hat{C} = \hat{Z} \otimes \hat{Z}, \text{ and} \tag{33}$$

$$\hat{C}^\dagger \left( \hat{Z} \otimes \hat{I} \right) \hat{C} = \hat{Z} \otimes \hat{I}. \tag{34}$$

The other twelve relations are either trivial ($\hat{I} \otimes \hat{I}$ transforms to itself), or can be written in terms of these four transformations. Hence $\hat{C}^\dagger(\hat{X} \otimes \hat{X})\hat{C} = \hat{X} \otimes \hat{I}$ is the same as Eq. (32), as can be seen by multiplying both sides of Eq. (32) by $\hat{C}^\dagger$ on the left and by $\hat{C}$ on the right, and using the fact that $\hat{C}$ is unitary. Combinations such as $\hat{X} \otimes \hat{Z}$ transform according to

$$\begin{aligned} \hat{C}^\dagger \left( \hat{X} \otimes \hat{Z} \right) \hat{C} &= \hat{C}^\dagger \left( \hat{X} \otimes \hat{I} \right) \left( \hat{I} \otimes \hat{Z} \right) \hat{C} \\ &= \hat{C}^\dagger \left( \hat{X} \otimes \hat{I} \right) \hat{C}\hat{C}^\dagger \left( \hat{I} \otimes \hat{Z} \right) \hat{C} \\ &= \left( \hat{X} \otimes \hat{X} \right) \left( \hat{Z} \otimes \hat{Z} \right) \\ &= \hat{X}\hat{Z} \otimes \hat{X}\hat{Z} \\ &= \hat{Y} \otimes \hat{Y}. \end{aligned} \tag{35}$$

Other combinations involving $\hat{Y}$ can be worked out using the identity $\hat{Y} = \hat{Z}\hat{X}$; hence

$$\begin{aligned} \hat{C}^\dagger \left( \hat{Y} \otimes \hat{I} \right) \hat{C} &= \hat{C}^\dagger \left( \hat{Z}\hat{X} \otimes \hat{I} \right) \hat{C} \\ &= \hat{C}^\dagger \left( \hat{Z} \otimes \hat{I} \right) \left( \hat{X} \otimes \hat{I} \right) \hat{C} \\ &= \hat{C}^\dagger \left( \hat{Z} \otimes \hat{I} \right) \hat{C}\hat{C}^\dagger \left( \hat{X} \otimes \hat{I} \right) \hat{C} \\ &= \left( \hat{Z} \otimes \hat{I} \right) \left( \hat{X} \otimes \hat{X} \right) \\ &= \hat{Z}\hat{X} \otimes \hat{I}\hat{X} \\ &= \hat{Y} \otimes \hat{X}. \end{aligned} \tag{36}$$

We can therefore validate a CNOT implementation by verifying that it satisfies the four relations given by Eqs. (31-34). However, these are precisely the four transformations that we worked out for the braid, so indeed a braid is a CNOT.

Note that the full braid transformation, including all the operations on the physical data qubits, involves a number of projective measurements and is therefore not unitary. However, if we consider the effect of the braid on the product $|Q\rangle|q_L\rangle$, while the transformation of the stabilized state $|Q\rangle$ is not unitary, the transformation of the logical state $|q_L\rangle$ is indeed a unitary one.

### D. CNOT between two Z-cut qubits

We have only shown how to perform CNOTs using a braid with a Z-cut qubit as the control and an X-cut qubit as the target. We can extend this to a CNOT between two Z-cut qubits using the circuit shown in Fig. 24a-c. The circuit performs a CNOT of the logical Z-cut "target-in" qubit using the logical Z-cut control qubit, with the Z-cut "target-out" qubit carrying the result. An ancillary X-cut qubit is the target for the three logical CNOTs in the circuit, so all the CNOTs are braid transformations between a Z-cut and an X-cut qubit. The circuit includes two logical measurements, with outcomes $M_Z$ and $M_X$. If the target-in qubit is measured to be in $|+_L\rangle$ ($|-_L\rangle$), with $M_X = +1(-1)$, then the target-out qubit does not (does) have a $\hat{Z}_L$ applied to it prior to the CNOT. If the X-cut qubit is measured to be in $|g_L\rangle$ ($|e_L\rangle$), with $M_Z = +1(-1)$, then the target does not (does) have an $\hat{X}_L$ applied to it after the CNOT.

You can verify that this circuit works properly by using operator transformations, or alternatively by testing it with the four input basis states $|gg\rangle$, $|ge\rangle$, $|eg\rangle$ and $|ee\rangle$ for the control and target-in Z-cut qubits (dropping the logical subscript $L$ for now). For example, consider $|eg\rangle$, where the control is in $|e\rangle$ and the target-in is in $|g\rangle$. We write the four-qubit state $|abcd\rangle$ where $a$ is the top, Z-cut control qubit, $b$ the X-cut qubit, $c$ the target-out and $d$ the target-in qubit. The state is initially $|eg+g\rangle$. This state is unchanged after the first CNOT in the circuit, while the second CNOT yields $|ee+g\rangle$. The third CNOT generates the entangled state $|eegg\rangle + |egeg\rangle$. We now measure the second (X-cut) qubit in $\hat{Z}$, which can yield two outcomes:

1. If the $\hat{Z}$ measurement of the second (X-cut) qubit yields $M_Z = +1$ (meaning the second qubit was measured in $|g\rangle$), the state is projected to $|egeg\rangle$ with the target-out qubit in $|e\rangle$; with $M_Z = +1$ we do not apply $\hat{X}_L$ to this qubit. The outcome of the target-in qubit measurement has no effect (a $\hat{Z}_L$ applied to the target doesn't affect it), so we see that the target-out ends up in $|e\rangle$, and we have the circuit performing the transformation $|e_L g_L\rangle \rightarrow |e_L e_L\rangle$ as desired.

2. If instead the $\hat{Z}$ measurement of the second qubit yields $M_Z = -1$ (meaning projection of the second qubit to $|e\rangle$), after the measurement the state is $|eegg\rangle$ with the target-out in $|g\rangle$. As we have $M_Z = +1$, we need to apply an $\hat{X}_L$ to the target-out qubit, which transforms it to $|e\rangle$. Hence the circuit performs the same transformation, $|eg\rangle \rightarrow |ee\rangle$.

We encourage the reader to test the other possible input states, to show that $|gg\rangle \rightarrow |gg\rangle$, $|ge\rangle \rightarrow |ge\rangle$, and $|ee\rangle \rightarrow |eg\rangle$.

There is an analogous circuit for performing a CNOT between two X-cut qubits, using a Z-cut qubit as an intermediate ancillary to perform the braid. This is shown in Fig. 24d-f, with a state transfer between the control-in and control-out qubits.

### E. Single-control, multi-target CNOTs

It is frequently necessary to implement single-control, multi-target CNOTs; these appear for example in the distillation circuits used to purify imperfect states, as we shall see when we discuss the $\hat{S}$ and $\hat{T}$ gates. It turns out that these kinds of CNOTs are actually no more complicated than a single-control, single-target CNOT, and in fact can be implemented in the same number of surface code cycles as a CNOT between two Z-cut or between two X-cut qubits. We display such a circuit in Fig. 25, implementing a single-control, triple-target logical CNOT between X-cut qubits. This circuit uses a single Z-cut qubit as an intermediary for the braid transformation, and an X-cut control-out ancillary to which the control-in state is transferred during the circuit operation. The Z-cut intermediary is braided through the control and the three target qubit holes, a process that can be performed in just $2d$ surface code cycles, followed by measurements that are used to interpret the result. This can of course be extended to any number of target qubits, without any additional surface code cycles needed to complete the operation.

## XV. THE HADAMARD TRANSFORMATION

The Hadamard transformation is a single-qubit gate that in the standard qubit $|g\rangle$, $|e\rangle$ basis is represented by the $2 \times 2$ matrix

$$\hat{H} = \frac{1}{\sqrt{2}} \begin{pmatrix} 1 & 1 \\ 1 & -1 \end{pmatrix}. \tag{37}$$

In the Schrödinger picture, this unitary Hermitian operator[22] takes the $\hat{Z}$ eigenstates $|g\rangle$ and $|e\rangle$ and transforms

---

[22] There is an unfortunate collision between the notation for the Hadamard and the Hamiltonian operators, which both use $\hat{H}$; however, we have managed to avoid invoking the Hamiltonian at any point in this article, so this should not cause confusion.

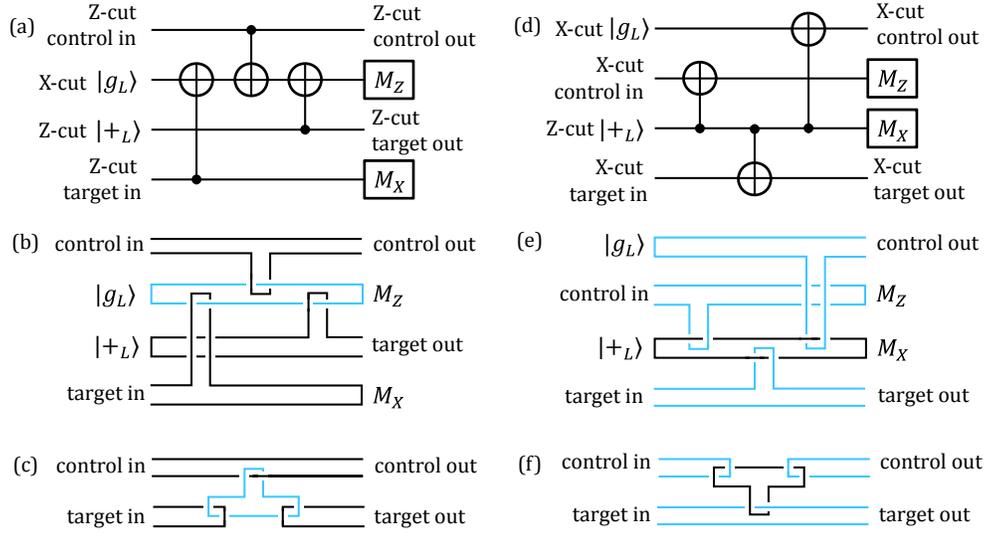

FIG. 24. (Color online) (a) Logical CNOT between two Z-cut qubits. The control, target in, and the second ancilla (in $|+_L\rangle$) are all Z-cut qubits, while the first ancilla is an X-cut qubit. The logical CNOTs are all between Z- and X-cut qubits, generated by braiding transformations. The measure outcomes $M_Z$ and $M_X$ signal how the output state should be interpreted, as described in the main text. (b) Simplified representation using braids, with black lines for Z-cut qubits and blue (light) lines for the X-cut qubit. There are a pair of lines for each logical qubit, one line for each qubit hole. The two lines join when a logical qubit is created or measured. (c) Even more condensed representation. (d) Logical CNOT between two X-cut qubits. The control, target in, and the first ancilla are all X-cut qubits, while the second ancilla is a Z-cut qubit. The logical CNOTs are all between X- and Z-cut qubits, generated by braiding transformations with the Z-cut as the control. The measure outcomes $M_Z$ and $M_X$ signal how the output state should be interpreted. (e) Simplified representation, with black lines for the Z-cut qubits and blue (light) for the X-cut qubit. (f) Even more simplified representation.

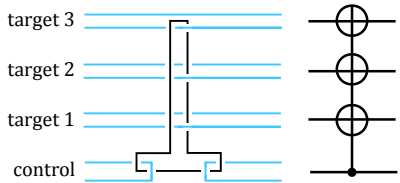

FIG. 25. (Color online) Schematic of circuit used to implement a single-qubit control, triple target logical CNOT (circuit equivalent on right), using the same notation as was used in Fig. 24. The blue (light) lines are the logical X-cut qubits used for the control and targets, and the black line is an ancillary logical Z-cut qubit that serves as an intermediary between the X-cut qubits. The sequence involves a total of six logical qubits, analogous to Fig. 24.

them to the $\hat{X}$ eigenstates $|+\rangle$ and $|-\rangle$, respectively. It similarly takes the $|+\rangle$ and $|-\rangle$ eigenstates to $|g\rangle$ and $|e\rangle$. In terms of the Bloch sphere, the Hadamard is represented by a 180° rotation about an axis in the $x-z$ plane, at 45° to the $z$ axis. In the surface code, such arbitrary rotations cannot be performed, so an equivalent logical rotation is not possible.

In the Heisenberg representation, the Hadamard takes $\hat{X}$ to $\hat{Z}$ and vice versa, so it satisfies

$$\hat{H}^\dagger \hat{X} \hat{H} = \hat{Z},$$
$$\hat{H}^\dagger \hat{Z} \hat{H} = \hat{X}. \tag{38}$$

A logical Hadamard is implemented in the surface code as shown in Figs. 26, 27 and 28, with details of the process given in Appendix J and in Ref. [49]. We start in Fig. 26 with an array of $d = 7$ logical qubits. The first step in the process is to isolate the logical qubit to be transformed from the larger 2D array, by turning off a ring of stabilizers, isolating the logical qubit in a separate patch of the 2D array, as shown in Fig. 27a. The $\hat{Z}_L$ operator loop is transformed to a patch-crossing chain by multiplying it by a number of stabilizers in the patch. By widening the ring that isolates the logical qubit into a "moat", so that the moat engulfs the two qubit holes, the logical qubit is transformed to a simple "patch" qubit as shown in Fig. 27b-d, similar to the $d = 5$ array qubit we discussed in Section VI.

The key to the logical Hadamard is now implemented, by performing *physical* Hadamards on all the data qubits in the patch; this exchanges the $\hat{X}_L$ operator for $\hat{Z}_L$ and vice versa, as well as swapping the identities of the X and Z stabilizers (Fig. 27e). This however results in a misalignment of the stabilizers in the patch with those in the larger 2D array, so we then perform two swaps, from data qubit to measure qubit, then from measure qubit to data qubit, shifting the patch by one stabilizer cell and realigning the stabilizers (Fig. 27f). The two Z-cut holes are then recreated (Fig. 28g), positioned so that the Hadamard-transformed $\hat{X}_L$ chain ends on the internal boundary of each hole, and the $\hat{Z}_L$ chain is multiplied by a set of stabilizers that returns it to a loop around one of the qubit holes. The qubit holes are then moved to realign them with their original positions (Fig. 28h-j), a move that is split into two steps to preserve the array distance $d$. In the final step, the qubits are rejoined with the main array (as in Fig. 26).

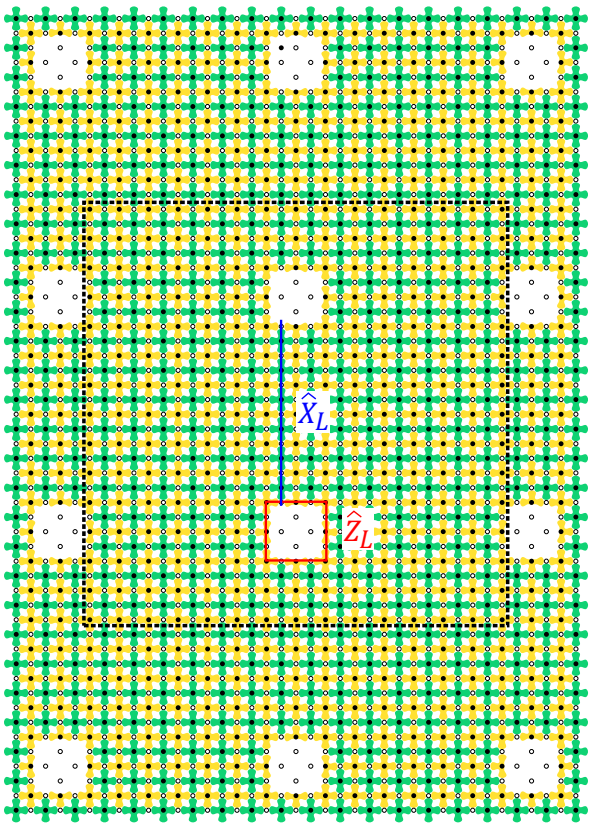

FIG. 26. (Color online) Initial 2D array for the logical Hadamard example. The spacing and size of the Z-cut qubit holes corresponds to a distance $d = 7$. The two holes in the center of the array form the logical qubit that is the target for the logical Hadamard, and for which we display the $\hat{X}_L$ and $\hat{Z}_L$ logical operators. The dashed box outlines the limits for what is shown in Figs. 27 and 28.

The surface code is used to correct and stabilize any errors that occur in this process, including potential errors from the physical Hadamards. This will maintain error correction as long as the distance $d$ of the "patch qubit" that undergoes the Hadamard is sufficiently large, and errors in the physical Hadamards sufficiently rare. The distance $d$ determines the number of times some of the stabilizer measurements need to be repeated in the Hadamard process, as noted in the figure captions.

These repeat numbers are therefore specific to this distance $d = 7$ geometry [49], and larger distance codes will require more repetitions.

In this process, we performed physical Hadamards on the data qubits, and then did two swaps to realign the stabilizers. A completely equivalent process is to first perform a data qubit-measure qubit swap, then perform a physical Hadamard on all the measure qubits, which now store the data qubit states. The process is completed by swapping the Hadamard-transformed measure qubit states to the appropriate data qubits, after which the measure qubits regain their original $\hat{X}$ or $\hat{Z}$ stabilizer roles.

## XVI. SINGLE QUBIT $\hat{S}_L$ AND $\hat{T}_L$ OPERATORS

To complete our set of logical gates, we need surface-code implementations of the $\hat{S}_L$ and $\hat{T}_L$ operators and their adjoints. These operators are represented in the standard $|g_L\rangle$, $|e_L\rangle$ basis by the $2 \times 2$ matrices

$$\hat{S}_L = \begin{pmatrix} 1 & 0 \\ 0 & i \end{pmatrix}, \tag{39}$$

and

$$\hat{T}_L = \begin{pmatrix} 1 & 0 \\ 0 & e^{i\pi/4} \end{pmatrix}. \tag{40}$$

The $\hat{S}_L$ gate is also sometimes called the $P$ or *phase* gate, and $\hat{T}_L$ is also called the $\pi/8$ gate.[23]

We note the important identities

$$\hat{T}_L \hat{T}_L = \hat{S}_L \tag{41}$$

and

$$\hat{S}_L \hat{S}_L = \hat{Z}_L. \tag{42}$$

We also have the identities

$$\hat{T}_L = \hat{S}_L \hat{T}_L^\dagger, \tag{43}$$

and correspondingly

$$\hat{T}_L^\dagger = \hat{S}_L^\dagger \hat{T}_L = \hat{Z}_L \hat{S}_L \hat{T}_L. \tag{44}$$

As we will see, the circuit used to implement $\hat{T}_L$ is probabilistic, and half the time generates $\hat{T}_L^\dagger$. From Eq. (43), when this occurs $\hat{T}_L^\dagger$ can be converted to $\hat{T}_L$ by application of an $\hat{S}_L$ gate. $\hat{T}_L$ can be similarly converted to $\hat{T}_L^\dagger$ by application of $\hat{S}_L$, with the result having a byproduct $\hat{Z}_L$ operator; in other words, $\hat{S}_L\hat{T}_L|\psi_L\rangle = \hat{Z}_L\hat{T}_L^\dagger|\psi_L\rangle$, and the $\hat{Z}_L$ byproduct operator is handled in software.

_______

[23] The $\pi/8$ name originates from defining $\hat{T}_L$ by the matrix in Eq. (40) multiplied by an overall (and unimportant) phase factor of $e^{-i\pi/8}$.

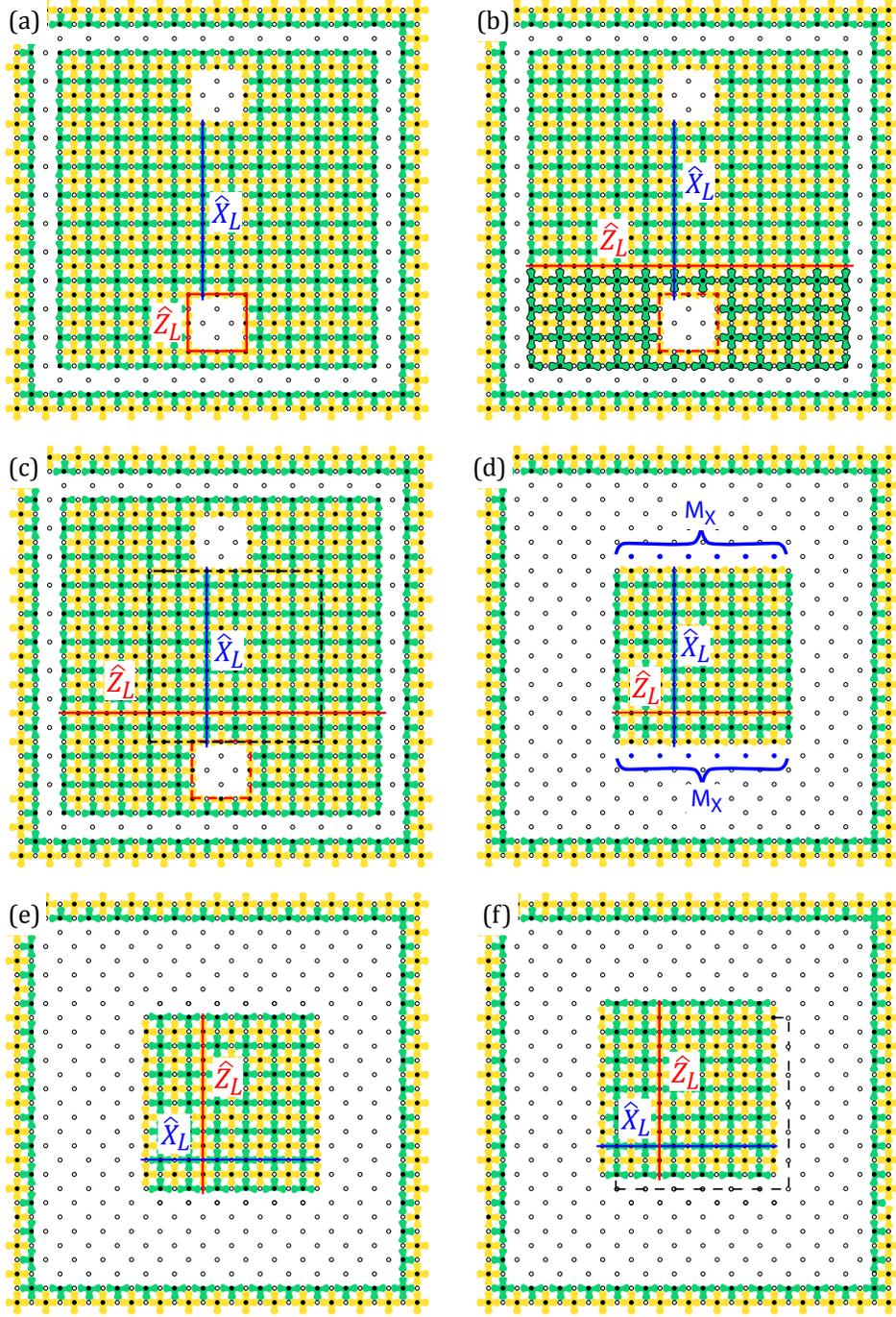

FIG. 27. (Color online) (a) In the first surface code cycle for the Hadamard, a ring of $\hat{X}$ stabilizers surrounding the target logical qubit is turned off, and the four-terminal $\hat{Z}$ stabilizers on the ring's borders are reduced to three- and two-terminal measurements. The data qubits isolated within the ring itself are measured in the $\hat{Z}$ basis to maintain error tracking with the reduced-size $\hat{Z}$ stabilizers. This stabilizer measurement pattern is applied three times (three is specific to the distance-7 qubits used here; see Ref. [49]). (b) The $\hat{Z}_L$ operator (dashed red line) is multiplied by all the black outline $\hat{Z}$ stabilizers, transforming $\hat{Z}_L$ to a chain of operators (solid red line) going from left to right. (c) In the next surface code cycle, all $\hat{X}$ and $\hat{Z}$ stabilizers in 2D patch outside the dashed box are turned off, widening the ring into a "moat" as in (d), eliminating the two qubit holes. In the same cycle, all isolated data qubits are measured in the $\hat{Z}$ basis, except those colored blue (light), which are measured in the $\hat{X}$ basis to preserve error tracking with the adjacent three-terminal $\hat{X}$ stabilizers. Note that the other un-stabilized data qubits do not need to be measured, as changes in their quantum states will be accounted for once their respective stabilizers are turned back on. (e) Before the next surface code cycle starts, a physical Hadamard is performed on all data qubits, swapping $\hat{X}_L$ and $\hat{Z}_L$ and the stabilizer identities. (f) Following the Hadamard, a pair of swap operations is performed between each patch data qubit and its neighboring measure qubits, first between each data qubit and the measure qubit above it, then between each measure qubit and the data qubit to its left. This shifts the patch by one stabilizer cell in the array, aligning the measure qubits in the patch with those in the larger array. The dashed box shows the location of the patch prior to these swaps. This sequence continues in Fig. 28.

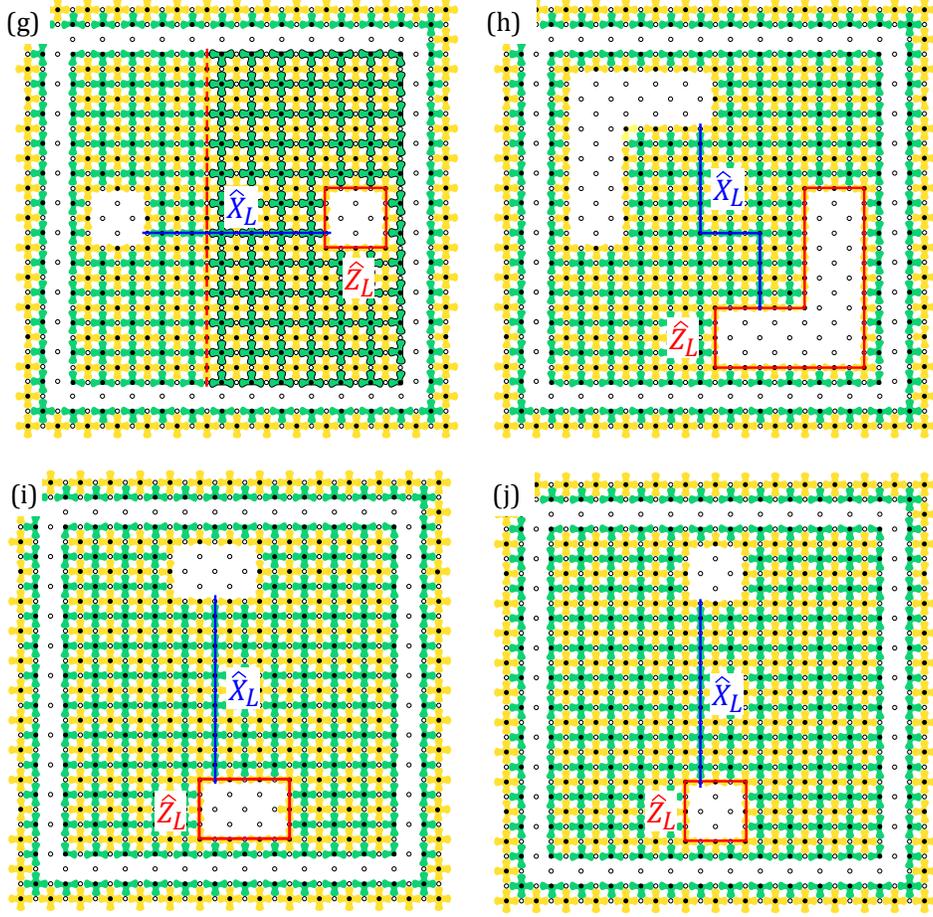

FIG. 28. (Color online) (continued from Fig. 27) (g) In the next surface code cycle, most of the $\hat{X}$ and $\hat{Z}$ stabilizers are turned back on, leaving two Z-cut enlarged holes and a ring, one stabilizer cell wide, isolating the 2D patch from the array, as in (a). The Z-cut holes are positioned so that the $\hat{X}_L$ chain ends on the internal boundary of each hole. The $\hat{Z}_L$ chain (dashed red line) is multiplied by all the black outline $\hat{Z}$ stabilizers, leaving a $\hat{Z}_L$ loop (solid red line) that encloses the right Z-cut hole. (h) In the subsequent surface code cycle, and the first step of returning the Z-cut holes to original locations, the Z-cut holes are expanded and $\hat{Z}_L$ and $\hat{X}_L$ modified accordingly. This move is split in two steps to preserve the distance $d$ during this process; the process pauses here for $d$ surface code cycles to establish all values in time. (i) In the second step of returning the Z-cut holes to their original locations, the Z-cuts are expanded to encompass the original positions. The stabilizer measurements are performed twice. (j) In the final step, the cuts are reduced to their original size, as in Fig. 26. This stabilizer measurement pattern is applied three times (see Ref. [49]). Following this, the isolated patch is reconnected by turning on the appropriate stabilizers.

A high-fidelity logical implementation of $\hat{S}_L$ and $\hat{T}_L$ involves special ancilla states. Implementing $\hat{S}_L$ relies on the $|Y_L\rangle$ ancilla state

$$|Y_L\rangle = \frac{1}{\sqrt{2}}\left(|g_L\rangle + i|e_L\rangle\right), \qquad (45)$$

while implementing $\hat{T}_L$ relies on the $|A_L\rangle$ ancilla state

$$|A_L\rangle = \frac{1}{\sqrt{2}}\left(|g_L\rangle + e^{i\pi/4}|e_L\rangle\right). \qquad (46)$$

The $|Y_L\rangle$ and $|A_L\rangle$ ancilla states are created in a special "short qubit," which can be put in an arbitrary but concomitantly imperfect state, a process known as "state injection." Once the state has been injected, the "short qubit" is increased to the standard distance $d$ to make it less error-prone, and the imperfect logical state of the standard-distance qubit is then purified by a high-fidelity process known as "distillation"; we discuss these steps below. The $\hat{S}_L$ and $\hat{T}_L$ gates are then implemented with circuits using logical CNOTs and Hadamards involving these ancilla states, as shown in Fig. 29 and Fig. 30, respectively.

The $\hat{S}_L$ gate implementation shown in Fig. 29 involves two logical CNOTs and two logical Hadamard opera-tions. An input state $|\psi_L\rangle$ in one logical qubit is deterministically transformed into $\hat{S}_L|\psi_L\rangle$ by interacting with the ancilla qubit in the $|Y_L\rangle$ state. This can be most easily seen by testing the circuit with the input state $|\psi_L\rangle = \alpha|g\rangle + \beta|e\rangle$, where we drop the $L$ subscript for simplicity. We consider the states $|g\rangle$ and $|e\rangle$ of $|\psi_L\rangle$ separately. For $|g\rangle$, the CNOT does nothing, so the $|Y_L\rangle$ ancilla state transforms according to $\hat{H}\hat{H}|Y_L\rangle = |Y_L\rangle$. For $|e\rangle$, the CNOT performs an $\hat{X}_L$ on the ancilla state, which then transforms according to $\hat{H}\hat{X}\hat{H}\hat{X}|Y_L\rangle = \hat{Z}\hat{X}|Y_L\rangle = -|e\rangle + i|g\rangle$, where we have dropped normalization. Writing the circuit state as $|ab\rangle$ where $a$ represents the input state and $b$ the ancilla, the circuit produces the output

$$\alpha|g\rangle(|g\rangle + i|e\rangle) + \beta|e\rangle(-|e\rangle + i|g\rangle) \tag{47}$$
$$= \alpha|gg\rangle + i\alpha|ge\rangle + i\beta|eg\rangle - \beta|ee\rangle \tag{48}$$
$$= (\alpha|g\rangle + i\beta|e\rangle)(|g\rangle + i|e\rangle) \tag{49}$$
$$= (\hat{S}_L|\psi_L\rangle)|Y_L\rangle, \tag{50}$$

as desired.

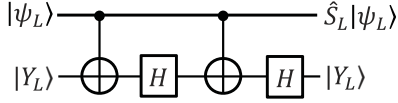

FIG. 29. Logic circuit that implements the $\hat{S}_L$ gate. The circuit uses the ancilla state $|Y_L\rangle = (|g_L\rangle + i|e_L\rangle)/\sqrt{2}$, on which two controlled logical CNOTs and two logical Hadamards are performed, resulting in the input state $|\psi_L\rangle$ being transformed to $\hat{S}_L|\psi_L\rangle$. Note that as $\hat{S}_L^\dagger = \hat{Z}_L\hat{S}_L$, the same circuit transforms $|\psi_L\rangle$ to $\hat{Z}_L\hat{S}_L^\dagger|\psi_L\rangle$ where $\hat{Z}_L$ is a byproduct operator.

To instead apply $\hat{S}_L^\dagger$, we use the identity $\hat{S}_L^\dagger = \hat{Z}_L\hat{S}_L$, which means we use the circuit shown in Fig. 29 and have a byproduct $\hat{Z}_L$ appear on the output, in other words $|\psi_L\rangle$ transforms to $\hat{Z}_L(\hat{S}_L^\dagger|\psi_L\rangle)$.

The $\hat{T}_L$ gate is implemented with the non-deterministic circuit shown in Fig. 30. Given the input ancilla state $|\theta_L\rangle = |g_L\rangle + e^{i\theta}|e_L\rangle$, the output $|\phi_L\rangle$ of this circuit is

$$|\phi_L\rangle = \hat{X}_L^{p_Z}\hat{R}_Z\left((-1)^{p_Z}\theta\right)|\psi_L\rangle. \tag{51}$$

The first operator is a byproduct operator, whose power $p_Z$ is equal to $0(1)$ if the $\hat{Z}_L$ measurement $M_Z$ of the logical qubit state is $+1(-1)$. The second operator is a rotation by the angle $\theta$,

$$\hat{R}_Z(\theta) = \begin{bmatrix} 1 & 0 \\ 0 & e^{i\theta} \end{bmatrix}. \tag{52}$$

For the $\hat{T}_L$ gate, the rotation angle is $\theta = +\pi/4(-\pi/4)$ depending on the sign $+1(-1)$ of the $\hat{Z}_L$ measurement.

We can test this circuit with the input state $|\psi\rangle = \alpha|g\rangle + \beta|e\rangle$ (dropping the $L$ subscript) and the ancilla state with $\theta = \pi/4$ (this is the state $|A_L\rangle$). Hence the initial state is

$$(|g\rangle + e^{i\pi/4}|e\rangle)|\psi\rangle$$
$$= \alpha|gg\rangle + \beta|ge\rangle + e^{i\pi/4}\alpha|eg\rangle + e^{i\pi/4}\beta|ee\rangle. \tag{53}$$

The CNOT transforms this to

$$\alpha|gg\rangle + \beta|ge\rangle + e^{i\pi/4}\alpha|ee\rangle + e^{i\pi/4}\beta|eg\rangle =$$
$$= (\alpha|g\rangle + e^{i\pi/4}\beta|e\rangle)|g\rangle + (\beta|g\rangle + e^{i\pi/4}\alpha|e\rangle)|e\rangle. \tag{54}$$

If we measure the second qubit in $|g\rangle$, i.e. we have the outcome $M_Z = 1$, the first qubit state is $\alpha|g\rangle + e^{i\pi/4}\beta|e\rangle$, which is precisely $\hat{T}_L|\psi_L\rangle$, so we have succeeded. If instead we measure the second qubit in $|e\rangle$, i.e. the measurement outcome is $M_Z = -1$, the first qubit state is

$$\beta|g\rangle + e^{i\pi/4}\alpha|e\rangle = \hat{X}\left(e^{i\pi/4}\alpha|g\rangle + \beta|e\rangle\right)$$
$$= e^{i\pi/4}\hat{X}\left(\alpha|g\rangle + \beta e^{-i\pi/4}|e\rangle\right), \tag{55}$$

which is $\hat{X}_L\hat{T}_L^\dagger|\psi_L\rangle$ multiplied by an overall phase, which can be neglected.

Hence approximately half of the times we run the circuit we will succeed in generating $\hat{T}_L|\psi_L\rangle$, signaled by the measurement $M_Z = +1$. However, the other times we run the circuit the measurement $M_Z = -1$ will signal the output state

$$|\phi_L\rangle = \hat{X}_L\hat{R}_Z(-\pi/4)|\psi_L\rangle = \hat{X}_L\hat{T}_L^\dagger|\psi_L\rangle. \tag{56}$$

In this case, to achieve the desired result, we must fix the output using $\hat{S}_L\hat{T}_L^\dagger = \hat{T}_L$. There is a slight complication if there are byproduct $\hat{X}_L$ or $\hat{Z}_L$ operators, which we discuss in the next section; here we assume there are no byproduct operators. The correction to the output is achieved by applying $\hat{S}_L$ to $|\phi_L\rangle$:

$$\hat{S}_L|\phi_L\rangle = \hat{S}_L(\hat{X}_L\hat{T}_L^\dagger|\psi_L\rangle)$$
$$= \hat{X}_L\hat{S}_L^\dagger\hat{T}_L^\dagger|\psi_L\rangle$$
$$= \hat{X}_L(\hat{Z}_L\hat{S}_L)\hat{T}_L^\dagger|\psi_L\rangle$$
$$= (\hat{X}_L\hat{Z}_L)\hat{T}_L|\psi_L\rangle, \tag{57}$$

where we use the identity $\hat{S}_L\hat{X}_L = i\hat{X}_L\hat{S}_L^\dagger$ and drop the unimportant phase factor of $i$. Hence passing the output state $|\phi_L\rangle$ through the $\hat{S}_L$ circuit gives the result $(\hat{X}_L\hat{Z}_L)\hat{T}_L|\psi_L\rangle$, where $\hat{X}_L\hat{Z}_L$ are byproduct operators that are handled by the control software.

If we need to perform $\hat{T}_L^\dagger$, we use the circuit in Fig. 30, and about half the time we will get the measurement outcome $M_Z = -1$, which indicates the circuit produced $\hat{X}_L\hat{T}_L^\dagger$, which is the desired output (with a byproduct operator $\hat{X}_L$). The other half of the time we get the measurement outcome $M_Z = +1$, and we can correct the circuit output $\hat{T}_L|\psi_L\rangle$ by applying $\hat{S}_L$, as $\hat{S}_L\hat{T}_L = \hat{Z}_L\hat{T}_L^\dagger$; this is done using the $\hat{S}_L$ circuit, with the output including the byproduct operator $\hat{Z}_L$ that is handled by the control software.

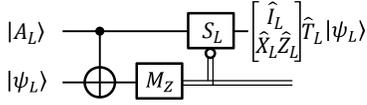

FIG. 30. Logic circuit for the $\hat{T}_L$ gate. The circuit uses the ancilla state $|A_L\rangle = (|g_L\rangle + e^{i\pi/4}|e_L\rangle)/\sqrt{2}$, which is used to control a CNOT on the target state $|\psi_L\rangle$, transforming the ancilla to the output state $|\phi_L\rangle$. The CNOT target state is measured along $\hat{Z}_L$, resulting in a projective, probabilistic outcome. Following this, the CNOT control qubit is processed by a conditional $\hat{S}_L$ gate: If the $\hat{Z}_L$ measurement $M_Z$ yields a $+1$ outcome, the output state is the desired one, $|\phi_L\rangle = \hat{T}_L|\psi_L\rangle$, and the $\hat{S}_L$ gate is not applied. If however the measurement yields a $-1$ outcome, the output is $|\phi_L\rangle = \hat{X}_L\hat{T}_L^\dagger$ and the $\hat{S}_L$ gate is applied, resulting in the output $\hat{X}_L\hat{Z}_L\hat{T}_L|\psi_L\rangle$. Note the double lines represent the classical measurement data with a probabilistic outcome, with these data controlling the $\hat{S}_L$ gate. The bracketed operators in the output correspond to $M_Z = +1$ ($\hat{I}_L$) and $M_Z = -1$ ($\hat{X}_L\hat{Z}_L$).

## A. Commuting $\hat{X}_L$ and $\hat{Z}_L$ through $\hat{S}_L$ and $\hat{T}_L$

We turn briefly to the issue of commuting the $\hat{X}_L$ and $\hat{Z}_L$ operators through the $\hat{S}_L$ and $\hat{T}_L$ gates, which is necessary if the classical control hardware is to handle all $\hat{X}_L$ and $\hat{Z}_L$ operations, as discussed in Sect. IX. The commutation relations are most easily worked out using the circuits that implement $\hat{S}_L$ and $\hat{T}_L$, Figs. 29 and 30.

First we show that $\hat{Z}_L$ and $\hat{S}_L$ commute, which can be seen by applying the $\hat{S}_L$ circuit to the input state $\hat{Z}_L|\psi_L\rangle$. We have an implicit $\hat{I}_L$ operation on the CNOT target state $|Y_L\rangle$, and we can commute the operator pair $\hat{Z}_L \otimes \hat{I}_L$ through the first CNOT, as $\hat{Z}_L \otimes \hat{I}_L \to \hat{Z}_L \otimes \hat{I}_L$. The Hadamard and the identity operator commute, and $\hat{Z}_L \otimes \hat{I}_L$ can then be commuted through the second CNOT. The second Hadamard likewise commutes with $\hat{I}_L$, so the operator pair $\hat{Z}_L \otimes \hat{I}_L$ can be commuted through the entire circuit. Hence $\hat{Z}_L$ and $\hat{S}_L$ commute. This result should not be surprising, as the $\hat{S}_L$ gate represents a form of logical rotation about the $\hat{Z}$ axis, and therefore should commute with $\hat{Z}_L$.

When passing the $\hat{X}_L$ operator through the $\hat{S}_L$ circuit, we have for the first CNOT transformation $\hat{X}_L \otimes \hat{I}_L \to \hat{X}_L \otimes \hat{X}_L$, and commuting through the Hadamard, which acts on the second operator, we have $\hat{X}_L \otimes \hat{X}_L \to \hat{X}_L \otimes \hat{Z}_L$. The second CNOT performs the transformation $\hat{X}_L \otimes \hat{Z}_L \to \hat{Y}_L \otimes \hat{Y}_L$, and the second Hadamard does not affect the second $\hat{Y}_L$ (other than an unimportant sign change). Hence $\hat{X}_L \otimes \hat{I}_L$ is transformed to $\hat{Y}_L \otimes \hat{Y}_L$, and the control software must note the byproduct operator transformation $\hat{X}_L \to \hat{Y}_L = \hat{Z}_L\hat{X}_L$ on the output state, with the two byproduct operators handled in software. An additional $\hat{Y}_L$ also appears on the ancilla $|Y_L\rangle$, but as $\hat{Y}_L|Y_L\rangle = i|Y_L\rangle$, this does nothing other than the unimportant phase factor $i$, which can be ignored.

In summary then, the $\hat{S}_L$ gate commutes with $\hat{Z}_L$ and $\hat{X}_L$ as follows:

$$\begin{cases} \hat{S}_L\hat{Z}_L \to \hat{Z}_L\hat{S}_L, \\ \hat{S}_L\hat{X}_L \to \hat{Z}_L\hat{X}_L\hat{S}_L. \end{cases} \tag{58}$$

To show that $\hat{Z}_L$ and $\hat{T}_L$ commute, we use similar reasoning using the circuit in Fig. 30. As $\hat{T}_L$ also represents a form of rotation about $\hat{Z}$, this makes sense. There is a slight wrinkle when we get the measurement outcome $M_Z = -1$, which means that a $\hat{S}_L$ gate will be applied, in this case to the state $|\phi_L\rangle = \hat{X}_L\hat{T}_L^\dagger(\hat{Z}_L|\psi_L\rangle)$. Then we have

$$\begin{aligned} \hat{S}_L|\phi_L\rangle &= \hat{S}_L\hat{X}_L\hat{T}_L^\dagger(\hat{Z}_L|\psi_L\rangle) \\ &= \hat{X}_L\hat{S}_L^\dagger\hat{Z}_L\hat{T}_L^\dagger|\psi_L\rangle \\ &= \hat{X}_L\hat{Z}_L\hat{S}_L^\dagger\hat{T}_L^\dagger|\psi_L\rangle \\ &= \hat{X}_L\hat{Z}_L\hat{Z}_L\hat{S}_L\hat{T}_L^\dagger|\psi_L\rangle \\ &= \hat{X}_L\hat{T}_L|\psi_L\rangle, \end{aligned} \tag{59}$$

which is the desired result, with a byproduct operator $\hat{X}_L$ that is handled by the control software.

The commutation of $\hat{X}_L$ through the $\hat{T}_L$ gate is similar. We take the input state $\hat{X}_L|\psi_L\rangle$, and commute $\hat{X}_L$ through the CNOT with an implicit $\hat{I}_L$ operating on the CNOT control qubit. The CNOT takes $\hat{I}_L \otimes \hat{X}_L \to \hat{I}_L \otimes \hat{X}_L$. Hence we have an additional $\hat{X}_L$ that reverses the sign of the terminal $M_Z$ measurement. If after applying this $\hat{X}_L$, the measurement outcome is $M_Z = -1$, we have the correct output, $\hat{T}_L|\psi_L\rangle$.

However, if after applying $\hat{X}_L$, the measurement gives $M_Z = +1$, the circuit will apply $\hat{S}_L$ to $|\phi_L\rangle = \hat{X}_L\hat{T}_L^\dagger|\psi_L\rangle$. Using $\hat{S}_L\hat{T}_L^\dagger = \hat{T}_L$, and dropping global phase factors, we have

$$\begin{aligned} \hat{S}_L|\phi_L\rangle &= \hat{S}_L(\hat{X}_L\hat{T}_L^\dagger|\psi_L\rangle) \\ &= \hat{X}_L\hat{S}_L^\dagger\hat{T}_L^\dagger|\psi_L\rangle \\ &= \hat{X}_L\hat{Z}_L\hat{S}_L\hat{T}_L^\dagger|\psi_L\rangle \\ &= \hat{X}_L\hat{Z}_L\hat{T}_L|\psi_L\rangle, \end{aligned} \tag{60}$$

which is as desired, with the appearance of two byproduct operators $\hat{X}_L\hat{Z}_L$.

We summarize the $\hat{T}_L$ logic circuit outcomes, as well as the commutation of $\hat{T}_L$ with $\hat{Z}_L$ and $\hat{X}_L$ in Table IV.

| Input | Output | Measurement |
|-------|--------|-------------|
| $|\psi_L\rangle$ | $\hat{T}_L|\psi_L\rangle$ | $M_Z = +1$ |
|  | $\hat{X}_L\hat{Z}_L\hat{T}_L|\psi_L\rangle$ | $M_Z = -1$ |
| $\hat{Z}_L|\psi_L\rangle$ | $\hat{Z}_L\hat{T}_L|\psi_L\rangle$ | $M_Z = +1$ |
|  | $\hat{X}_L\hat{T}_L|\psi_L\rangle$ | $M_Z = -1$ |
| $\hat{X}_L|\psi_L\rangle$ | $\hat{X}_L\hat{Z}_L\hat{T}_L|\psi_L\rangle$ | $M_Z = +1$ |
|  | $\hat{T}_L|\psi_L\rangle$ | $M_Z = -1$ |

TABLE IV. $\hat{T}_L$ logic circuit outcomes and commutation relations for $\hat{T}_L$ with $\hat{Z}_L$ and $\hat{X}_L$. The measurement $M_Z$ indicates which state is output from the circuit.

.

## B. Short qubits and state distillation

The ancilla $|Y_L\rangle$ and $|A_L\rangle$ states used in the $\hat{S}_L$ and $\hat{T}_L$ gates are $\hat{X}_L$ rotations of $|g_L\rangle$ by $90°$ and $45°$, respectively. These are not easy to perform in the surface code. Consider a distance $d = 3$ Z-cut qubit, with $\hat{X}_L = \hat{X}_1\hat{X}_2\hat{X}_3$ the operator chain connecting the two Z-cut holes. There is no obvious way to perform an arbitrary rotation of this logical qubit. For example, how would we perform a differential rotation of the state $|\psi_L\rangle$ using $\epsilon\hat{X}_L$, with the desired final state $(\hat{I}_L + \epsilon\hat{X}_L)|\psi_L\rangle$? You cannot perform this rotation by operating with $\hat{I} + \epsilon\hat{X}_j$ on each of the $j = 1, 2, 3$ data qubits, as this will give the result $(\hat{I} + \epsilon\hat{X}_1 + \epsilon\hat{X}_2 + \epsilon\hat{X}_3))|\psi_L\rangle$, to first order in $\epsilon$, which is not what we want. To get around this problem, we need to create a qubit in which the $\hat{X}_L$ chain is just one qubit in length; such a "short qubit" would of course be error-prone, but if we minimize the number of surface code cycles in which the qubit remains small, these errors will not accumulate significantly.

We can create and initialize a short X-cut qubit using the procedure shown in Fig. 31, which is known as "state injection"; details are in Appendix K.

The state injected in the short qubit is necessarily imprecise; even the best possible classically-controlled state preparation will have insufficient precision to achieve error rates below $10^{-14}$, needed for our example Shor's algorithm problem. The two target states we are interested in, $|Y_L\rangle$ and $|A_L\rangle$, can however be made significantly more precise through distillation. This is a probabilistic process in which an imperfect input state can be purified by multiple executions of a particular logic circuit, with the output state from each execution rapidly approaching the perfect desired result. Circuits for distilling the $|Y_L\rangle$ and $|A_L\rangle$ ancilla states are shown in Figs. 32 and 33, which implement the 7-qubit Steane [50] and the 15-qubit Reed-Muller encodings [17], respectively. These circuits may look very complicated, but the reader should note that these circuits only involve single-control, multi-target logical CNOTs, and as discussed in Sect. XIV E, these CNOTs take the same number of surface code cycles as a single-target logical CNOT. Hence, to complete the $|Y_L\rangle$ distillation, only six logical CNOT cycles are needed, and for the $|A_L\rangle$ distillation, at most eight such cycles are needed. Hence the temporal overhead to complete a distillation is actually relatively small; the surface code can implement these circuits very efficiently.

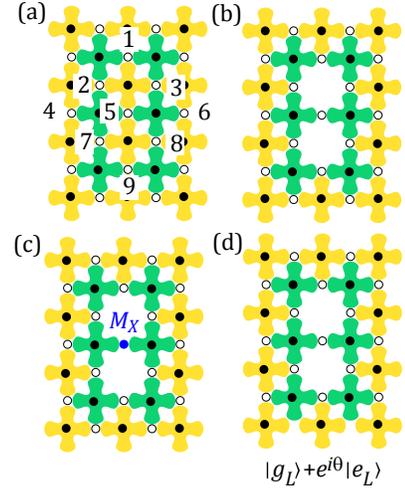

FIG. 31. (Color online) Sequence to generate a short X-cut qubit. (a) Surface code array, with the array extending indefinitely outside the region shown here. (b) Two $\hat{X}$ stabilizers are turned off, creating a short X-cut qubit, with a single data qubit separating the two holes. (c) In the next surface code cycle, the central data qubit (qubit 5) is measured along $\hat{X}$ simultaneously with the measurement of the four-terminal $\hat{Z}$ stabilizers adjacent to it; the remainder of the array has the standard surface code cycle for all stabilizers. This creates a logical qubit initialized in either $|+_L\rangle$ ($X_5 = +1$) or $|-_L\rangle$ ($X_5 = -1$). Prior to starting the next surface code cycle, the central data qubit is rotated using $\hat{R}_Z(\theta)$ to the desired final state $|g_L\rangle + e^{i\theta}|e_L\rangle$. (d) The surface code stabilization is restarted. After this the short qubit is enlarged to protect against errors.

In the circuit shown in Fig. 32, a logical Bell pair is created, and one logical qubit from the pair is entangled with six other logical qubits using the Steane code. Each of the seven encoded qubits is then rotated with an approximate $\hat{S}_L$ gate, each of these gates using an approximate $|Y_L\rangle$ states as an ancilla. In the first stage of distillation, each of these ancilla states is directly produced by state injection in a short qubit, while in later distillation stages the ancillae will be purified versions of these states. The distillation circuit generates one output state $|\psi_L\rangle$ in the other logical qubit of the Bell pair, and performs seven $\hat{X}_L$ measurements of the other qubits. The measurement outcomes are used to interpret the output state $|\psi_L\rangle$, by evaluating the three Steane code $\hat{X}$ stabilizers $\hat{X}_{S1} = \hat{X}_{L3}\hat{X}_{L4}\hat{X}_{L5}\hat{X}_{L6}$, $\hat{X}_{S2} = \hat{X}_{L2}\hat{X}_{L5}\hat{X}_{L6}\hat{X}_{L7}$ and $\hat{X}_{S3} = \hat{X}_{L1}\hat{X}_{L4}\hat{X}_{L6}\hat{X}_{L7}$ (numbering as in Fig. 32). These products are evaluated by taking the appropriate products of the individual $\hat{X}_{Lj}$ measurements.[24] If each of the stabilizer measurement outcomes $\{X_{S1}, X_{S2}, X_{S3}\}$ is equal to $\{+1, +1, +1\}$, then the output state $|\psi_L\rangle$ is a purified version of $|Y_L\rangle$ and will be kept (otherwise the state is discarded). If the product of all the individual logical measurements is $X_{L1}X_{L2}\ldots X_{L7} = -1$, nothing additional is needed, but if this product is $+1$, then the output will include a $\hat{Z}_L$ byproduct operator.

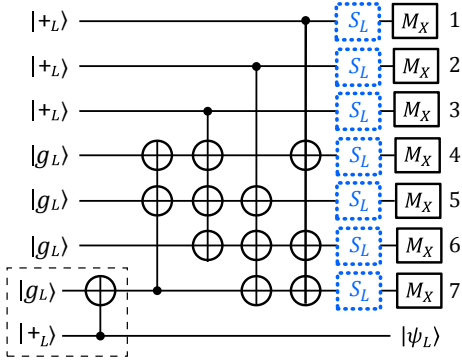

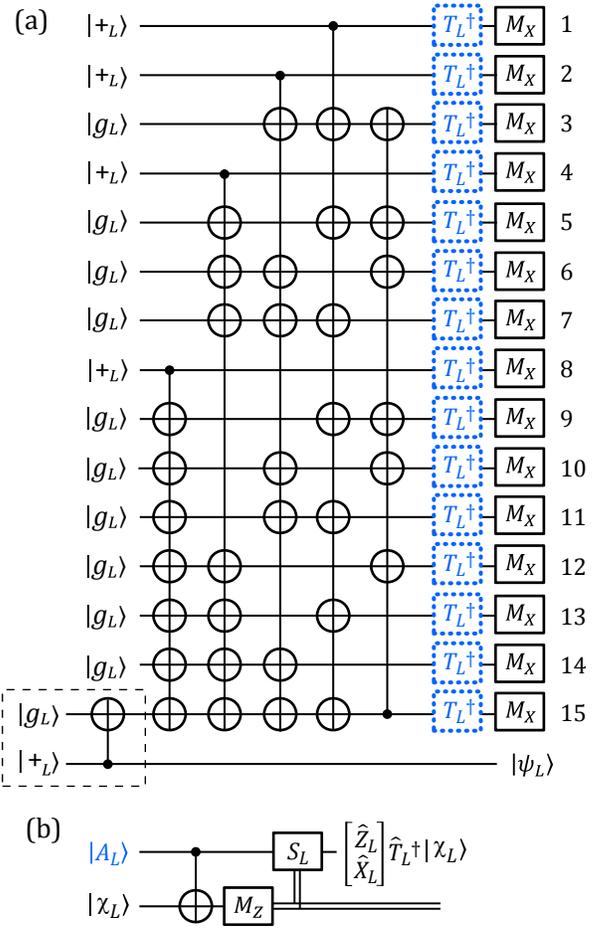

FIG. 32. (Color online) Distillation circuit for the $|Y_L\rangle$ ancilla state. A logical Bell pair is created (dashed box, bottom left), and one qubit from the pair is encoded with 6 ancilla logical qubits using the Steane code [17, 51]. The seven encoded qubits are then each rotated with an $\hat{S}_L$ gate (dotted blue); the ancilla states used by the $\hat{S}_L$ gates are precisely the $|Y_L\rangle$ states that are being purified, produced in the first round of distillation by state injection in a short qubit, or are from a prior round of distillation. The output from each $\hat{S}_L$ gate is then measured along $\hat{X}_L$, and the results indicate whether the output state $|\psi_L\rangle$ in the other qubit of the Bell pair should be discarded, or is a purified version of $|Y_L\rangle$, possibly involving a $\hat{Z}_L$ phase-flip (in software) as discussed in the main text.

If the ancilla $|Y_L\rangle$ states used in the $\hat{S}_L$ gates in Fig. 32 are perfect, and the circuit is operated flawlessly, the output state $|\psi_L\rangle$ will always be a perfect $|Y_L\rangle$. The ancilla $|Y_L\rangle$ states can however suffer from errors; a $\hat{Y}_L$ error does nothing, as $\hat{Y}_L|Y_L\rangle = i|Y_L\rangle$, while $\hat{Z}_L|Y_L\rangle = |Y_L^\star\rangle$ and $\hat{X}_L|Y_L\rangle = -i|Y_L^\star\rangle$. If there is a probability $p$ of having an $\hat{X}_L$ or $\hat{Z}_L$ error, and the circuit is operated flawlessly, then the output state will have a probability $7p^3 \ll p$ of having an error. The output will be successfully distilled with a probability $1 - 7p$.

Clearly the distillation converges rapidly to a nearly perfect output state. If one cycle of distillation does not result in a sufficiently accurate output, more cycles can be added. To run the circuit twice, one needs to prepare at least $7^2 = 49$ approximate $|Y_L\rangle$ input states, with the outputs $|\psi_L\rangle$ of each distillation circuit providing the seven input states for the second distillation circuit.[25] If the original states have a probability $p$ of $\hat{X}$ or $\hat{Z}$ errors, the first set of output states will have an error rate $7p^3$, and the output from the second distillation an error rate $7(7p^3)^3 = 7^4p^9$, with exponential improvement. To run the circuit three times, $7^3 = 343$ input states are required, with an output error rate $7^{13}p^{27}$, and so on. Note that if the error rate $p$ is of order 1%, running the distillation circuit twice generates states with an error rate of about $10^{-15}$, below the target of $10^{-14}$ set by the size of the Shor's algorithm problem we consider here.

FIG. 33. (Color online) Distillation circuit for the $|A_L\rangle$ ancilla state [19, 51]. (a) A logical Bell pair is created (dashed box, bottom left), and one qubit from the pair encoded with 14 ancilla logical qubits using the Reed-Muller code [17, 51]. The fifteen encoded logical qubits are then each rotated with a $\hat{T}_L^\dagger$ gate (dotted blue). The ancillae for the $\hat{T}_L^\dagger$ gates are the $|A_L\rangle$ states that are being purified, prepared either by state injection in a short qubit or are produced in a previous distillation round. Following the $\hat{T}_L^\dagger$ gates, fifteen $\hat{X}_L$ measurements $M_X$ are made, with the measurement pattern indicating whether to discard the output state $|\psi_L\rangle$ of the other qubit of the Bell pair, or indicating that $|\psi_L\rangle$ is a purified $|A_L\rangle$ state, possibly with an additional $\hat{Z}_L$ byproduct operator. (b) Diagram for the $\hat{T}_L^\dagger$ gate, which is similar to the $\hat{T}_L$ gate in Fig. 30, with a CNOT using the imperfect $|A_L\rangle$ (blue, light) as the control on the input state $|\chi_L\rangle$. When the measurement $M_Z = -1$, the output is $\hat{X}_L\hat{T}_L^\dagger|\chi_L\rangle$; the $\hat{X}_L$ has no effect when the $M_X$ measurement is made in panel (a). When the measurement $M_Z = +1$, the output is $\hat{T}_L|\chi_L\rangle$, and must be corrected (up to byproduct operators) using the $\hat{S}_L$ circuit in Fig. 29, since $\hat{S}_L\hat{T}_L|\chi_L\rangle = \hat{Z}_L\hat{T}_L^\dagger|\chi_L\rangle$. The byproduct operator $\hat{Z}_L$ will reverse the sign of the measurement $M_X$ that occurs after this gate in panel (a).

---

[24] Note that the Steane code stabilizers $\hat{X}_{Sj}$ are formed from products of *logical* operators, and are not to be confused with the surface code stabilizers!

A similar discussion applies to the $|A_L\rangle$ distillation circuit shown in Fig. 33. In this circuit, a logical Bell pair is created, and one qubit from the pair encoded with fourteen other logical qubits using the Reed-Muller code. The fifteen qubits are then rotated with $\hat{T}_L^\dagger$ gates, where each $\hat{T}_L^\dagger$ gate uses an approximation of the $|A_L\rangle$ state as an ancilla, either from a short qubit injection in the first round of distillation, or a previously purified $|A_L\rangle$ in later rounds.

The $\hat{T}_L^\dagger$ circuit is shown in Fig. 33b. The $\hat{T}_L^\dagger$ gate is probabilistic, where given the input $|\chi_L\rangle$, about half the time it generates $\hat{T}_L|\chi_L\rangle$, signalled by the measurement $M_Z = +1$, and $\hat{S}_L$ is then applied using the circuit in Fig. 29, giving $\hat{S}_L\hat{T}_L|\chi_L\rangle = \hat{Z}_L\hat{T}_L^\dagger|\chi_L\rangle$, with a $\hat{Z}_L$ byproduct operator appearing in the output state. The rest of the time the circuit generates $\hat{X}_L\hat{T}_L^\dagger|\chi_L\rangle$, signalled by $M_Z = -1$, with a byproduct $\hat{X}_L$.

The outputs of the $\hat{T}_L^\dagger$ gates are all measured along $\hat{X}_L$, where the sign of the measurement result $M_X$ is reversed due to the byproduct $\hat{Z}_L$ if the $\hat{S}_L$ circuit had to be used. The pattern of measurement outcomes is used to evaluate the four $\hat{X}$ stabilizers for the Reed-Muller code, $\hat{X}_{R1} = \hat{X}_{L4}\hat{X}_{L5}\ldots\hat{X}_{L11}$, $\hat{X}_{R2} = \hat{X}_{L1}\hat{X}_{L2}\ldots\hat{X}_{L7}\hat{X}_{L15}$, $\hat{X}_{R3} = \hat{X}_{L2}\ldots\hat{X}_{L5}\hat{X}_{L10}\ldots\hat{X}_{L13}$ and $\hat{X}_{R4} = \hat{X}_{L1}\hat{X}_{L2}\hat{X}_{L5}\hat{X}_{L6}\hat{X}_{L9}\hat{X}_{L10}\hat{X}_{L13}\hat{X}_{L14}$ (using the qubit numbering in Fig. 33).[26] If each of the measurement outcomes $\{X_{R1}, X_{R2}, X_{R3}, X_{R4}\}$ is $\{+1, +1, +1, +1\}$, the state $|\psi_L\rangle$ of the other qubit of the Bell pair is a purified $|A_L\rangle$ state. If the product of all the measurements $X_{L1}X_{L2}\ldots X_{L15} = +1$, no postprocessing is needed, while if this product is $-1$, then a $\hat{Z}_L$ byproduct operator will appear.

If the probability of an error containing a $\hat{Z}_L$ component in the approximate $|A_L\rangle$ states is $p$, the output state $|\psi_L\rangle$ will have an error rate $35p^3$. The distillation will succeed with a probability $1 - 15p$, so if $p$ is of order 1%, the distillation will have to be re-run about one-sixth of the attempts. Additional distillation cycles, built in the same way as for the $|Y_L\rangle$ distillation, will improve the output exponentially.

## XVII. PHYSICAL IMPLEMENTATIONS

We have now covered all the basic aspects of the surface code approach to quantum computing. We have described all the gates that are required to implement e.g. Shor's algorithm or Grover's search algorithm. The discussion has been mostly theoretical, while the motivation for developing the surface code is of course to find a realistic and practical *physical* implementation for a quantum computer. There are a number of physical systems in which this scheme could in principle be implemented, ranging from cold atoms [52, 53] and ions [54–56], to semiconductor-based approaches [57], to superconducting integrated circuits [58–63]. Each of these systems has certain advantages and certain disadvantages. For any system to be a candidate for a surface code implementation, it must of course meet the requirements for single-qubit and two-qubit gate and measurement fidelities, which is not true for any system to date, although a number of systems are close to these requirements. The surface code clearly also requires a very large number of physical qubits (of order $10^8$ is probably the smallest number needed for a practical factoring computer), so a separate requirement is the ability to assemble and integrate a large number of nominally identical qubits. Furthermore, the operation and error detection of the surface code assumes classical logic support, with the classical logic operating significantly faster than the qubits, in order that state preparation, qubit interactions, and error tracking can be maintained with a high level of fidelity.

Given classical processor clock rates of order 3 GHz, it is difficult to imagine performing the surface code classical processing if the rounds of error detection are applied much faster than $10^6$ to $10^7$ Hz. This implies physical quantum gates of duration 10-100 ns, and hence physical qubit coherence times of at least 1-10 $\mu$s, if the minimum physical gate fidelities of 99% are to be achieved. Gates operating at these speeds, and physical qubits with these coherence times, have already been achieved experimentally in superconducting systems [64, 65]. We believe that a surface-code cycle time of 200 nanoseconds is thus a reasonable target. Slower surface code operation implies higher qubit overhead or longer factoring times (a surface code cycle time of 2 ms would take 30 years to factor a 600 digit number, as per the scaling in Table I, if no additional qubits were devoted to the task).

The intimate intermingling of classical and quantum logic requires high-speed interconnects to each qubit, which makes very dense qubit geometries harder to implement. It will be challenging to integrate qubits with spacings of order of or less than 1 $\mu$m with classical electronics, while spacings of tens to hundreds of $\mu$m should be more straightforward. Note that a $10^4 \times 10^4$ array of physical qubits with inter-qubit spacing of 100 $\mu$m implies a physical size for the 2D array of about $1 \times 1$ m$^2$, which is perhaps manageable, so "large" qubits are not unrealistic.

---

[25] Note that this may take more than 49 attempts, as a fraction of the circuits, of order $p$, will fail to successfully distill their input states.

[26] Again, the Reed-Muller stabilizers here are defined in terms of logical qubits, and should not be confused with the surface code stabilizers.

These considerations appear to point to superconducting circuits as one of the best candidates for implementing the surface code: All of the physical and operating parameters for superconducting circuits fall into the ranges discussed here. There are clearly significant challenges in achieving sufficient gate and measurement fidelities. This is compounded by the need for tight integration of superconducting quantum circuitry with classical logic, all operating in tandem. However, we believe that with a continued, and significant, investment in superconducting quantum circuits, and in developing the necessary accompanying high-speed and closely-integrated classical logic, a quantum computer could be built.

# XVIII. ACKNOWLEDGEMENTS

We thank John Preskill for providing us with a brief history of surface code theory development, and James Wenner for assistance with numerical simulations. We also thank Frank K. Wilhelm, Michael R. Geller, Daniel Sank and James Wenner for their critical readings of a draft of this article. M.M. acknowledges support from an Elings Postdoctoral Fellowship. Support for this work was provided by IARPA under ARO award W911NF-08-1-0336 and under ARO award W911NF-09-1-0375. A.G.F. acknowledges support from the Australian Research Council Centre of Excellence for Quantum Computation and Communication Technology (project number CE110001027), with support from the US National Security Agency and the US Army Research Office under contract number W911NF-08-1-0527, and by the Intelligence Advanced Research Projects Activity (IARPA) via Department of Interior National Business Center contract number D11PC20166. The U.S. Government is authorized to reproduce and distribute reprints for Governmental purposes notwithstanding any copyright annotation thereon. Disclaimer: The views and conclusions contained herein are those of the authors and should not be interpreted as necessarily representing the official policies or endorsements, either expressed or implied, of IARPA, DoI/NBC, or the U.S. Government.

# Appendix A: Notation

The notation in this article is mostly standard. We use the following conventions:

1. Ground state for $\hat{Z}$ quantization axis: $|g\rangle = \begin{pmatrix} 1 \\ 0 \end{pmatrix}$.

2. Excited state $|e\rangle = \begin{pmatrix} 0 \\ 1 \end{pmatrix}$.

3. In the same $\hat{Z}$ axis basis, the operator $\hat{Z} = \hat{\sigma}_z = \begin{pmatrix} 1 & 0 \\ 0 & -1 \end{pmatrix}$, with eigenvalues $+1$, $-1$ for $|g\rangle$, $|e\rangle$. The measurement $M_Z$ returns these eigenvalues and projects to these respective eigenstates. Note the $\hat{Z}$ eigenvalue for $|e\rangle$ is smaller than that for $|g\rangle$; the Hamiltonian is positively proportional to $-\hat{Z}$.

4. Operator $\hat{X} = \hat{\sigma}_x = \begin{pmatrix} 0 & 1 \\ 1 & 0 \end{pmatrix}$ with eigenvalues $+1$ and $-1$ for eigenstates $|+\rangle = \frac{1}{\sqrt{2}} \begin{pmatrix} 1 \\ 1 \end{pmatrix} = \frac{1}{\sqrt{2}}(|g\rangle + |e\rangle)$ and $|-\rangle = \frac{1}{\sqrt{2}} \begin{pmatrix} 1 \\ -1 \end{pmatrix} = \frac{1}{\sqrt{2}}(|g\rangle - |e\rangle)$, respectively. The measurement $M_X$ returns these eigenvalues and projects to these respective eigenstates.

5. Operator $\hat{Y}$ is *real*, unlike the Pauli $\sigma_y$, with $\hat{Y} = -i\hat{\sigma}_y = \hat{Z}\hat{X} = \begin{pmatrix} 0 & 1 \\ -1 & 0 \end{pmatrix}$.

6. We have the commutation relations $[\hat{X}, \hat{Y}] = -2\hat{Z}$, $[\hat{Y}, \hat{Z}] = -2\hat{X}$, and $[\hat{Z}, \hat{X}] = +2\hat{Y}$ (not having an $i$ in the definition for $\hat{Y}$ interrupts the cyclic permutation of the Pauli operators).

7. Hadamard operator $\hat{H} = \frac{1}{\sqrt{2}} \left( \hat{X} + \hat{Z} \right) = \frac{1}{\sqrt{2}} \begin{pmatrix} 1 & 1 \\ 1 & -1 \end{pmatrix}$. Note that $\hat{H}|g\rangle = |+\rangle$, $\hat{H}|e\rangle = |-\rangle$, $\hat{H}|+\rangle = |g\rangle$, $\hat{H}|-\rangle = |e\rangle$, $\hat{H}^2 = \hat{I}$.

8. The $\hat{S}$ gate is another $\hat{Z}$-axis rotation, represented in the $\hat{Z}$ basis by the matrix $\hat{S} = \begin{pmatrix} 1 & 0 \\ 0 & i \end{pmatrix}$.

9. The $\hat{T}$ gate is a $\hat{Z}$-axis rotation, represented in the $\hat{Z}$ basis by the matrix $\hat{T} = \begin{pmatrix} 1 & 0 \\ 0 & e^{i\pi/4} \end{pmatrix}$. This gate is also called the $\pi/8$ gate, as one can also write it in the form $\hat{T} = e^{i\pi/8} \begin{pmatrix} e^{-i\pi/8} & 0 \\ 0 & e^{i\pi/8} \end{pmatrix}$. Note that $\hat{T}^2 = \hat{S}$, $\hat{S}^2 = \hat{Z}$, $\hat{S}^4 = \hat{I}$.

10. A controlled-NOT $\hat{C}$ in the basis $|gg\rangle$, $|ge\rangle$, $|eg\rangle$, $|ee\rangle$, where the first state is the control and the second the target, is given by

$$\hat{C} = \begin{pmatrix} 1 & 0 & 0 & 0 \\ 0 & 1 & 0 & 0 \\ 0 & 0 & 0 & 1 \\ 0 & 0 & 1 & 0 \end{pmatrix}. \tag{A1}$$

Hence if the control is in $|g\rangle$ the CNOT is equivalent to $\hat{I}$ operating on the target, while if the control is in $|e\rangle$ the CNOT is equivalent to $\hat{X}$ operating on the target.

11. A Toffoli gate is a three-qubit, controlled-controlled NOT gate. In the basis $|ggg\rangle$, $|gge\rangle$, $|geg\rangle$, $|gee\rangle$, $|egg\rangle$, $|ege\rangle$, $|eeg\rangle$, $|eee\rangle$, where the first and second states are the two controls and the third state is the target, the Toffoli is represented by

$$\begin{pmatrix} 1 & 0 & 0 & 0 & 0 & 0 & 0 & 0 \\ 0 & 1 & 0 & 0 & 0 & 0 & 0 & 0 \\ 0 & 0 & 1 & 0 & 0 & 0 & 0 & 0 \\ 0 & 0 & 0 & 1 & 0 & 0 & 0 & 0 \\ 0 & 0 & 0 & 0 & 1 & 0 & 0 & 0 \\ 0 & 0 & 0 & 0 & 0 & 1 & 0 & 0 \\ 0 & 0 & 0 & 0 & 0 & 0 & 0 & 1 \\ 0 & 0 & 0 & 0 & 0 & 0 & 1 & 0 \end{pmatrix}. \quad (A2)$$

Hence if both controls are in $|e\rangle$ the Toffoli is equivalent to $\hat{X}$ operating on the target, while otherwise the Toffoli is equivalent to $\hat{I}$ operating on the target.

12. The $j$th Z (X) stabilizer operator $\hat{Z}_{sj}$ ($\hat{X}_{sj}$) is a product $\hat{Z}_{sj} = \hat{Z}_{j,a}\hat{Z}_{j,b}\hat{Z}_{j,c}\hat{Z}_{j,d}$ of its four neighboring $a$, $b$, $c$ and $d$ physical qubit $\hat{Z}$ operators (analogously for $\hat{X}$). Each cycle of the surface code yields a measurement of these operators, yielding the measurement outcome $Z_{sj}$ ($X_{sj}$).

We have chosen to modify the names we apply to many of the functional elements in the surface code, using terms we believe are more suggestive of each element's function. We have compiled these names changes as shown in Table V.

| Published terminology (Ref. [20]) | This paper |
|---|---|
| code state | quiescent state |
| X syndrome (light) | measure-X (orange/light) |
| Z syndrome (dark) | measure-Z (green/dark) |
| syndrome symbol ● | measure symbol ● |
| data symbol ○ | data symbol ● |
| $Z$, $X$, $H$, $I$ | $\hat{Z}$, $\hat{X}$, $\hat{H}$, $\hat{I}$ |
| smooth boundary | X boundary |
| rough boundary | Z boundary |
| rough/primal qubit | single X-cut qubit |
| smooth/dual qubit | single Z-cut qubit |
| double rough defect | (double) X-cut qubit |
| double smooth defect | (double) Z-cut qubit |
| | measurement $X_{abcd}$ |
| | measurement $Z_{abcd}$ |

TABLE V. Translation between the "traditional" language used in the published literature (e.g. from [20]) and the terminology used here.

# Appendix B: $\hat{Z}$ and $\hat{X}$ stabilizer circuits

In this Appendix, we explicitly work through the operation of the surface code $\hat{Z}$ and $\hat{X}$ stabilizer circuits shown in Fig. 1b and c of the main text, respectively. For simplicity, we will look at a system with just two data qubits, $a$ and $b$, stabilized by one measure-Z and one measure-X qubit; the extension to the full four-qubit stabilization is straightforward. The simplified layout and corresponding stabilizer circuits are shown in Fig. 34, which now involve two CNOTs per measure qubit instead of the four CNOTs in the full circuit.

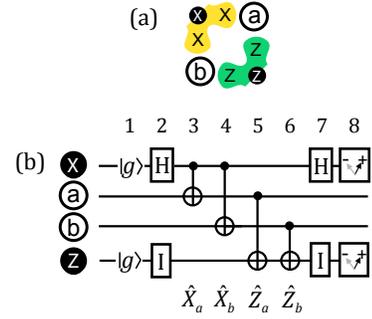

FIG. 34. (Color online) (a) Two data qubits $a$ and $b$ are stabilized by one measure-Z and one measure-X qubit, connected as shown. (b) Quantum circuit for the two measure qubits operating on the two data qubits. The CNOT order is critical: First, the measure-X qubit acts as the control of the CNOT on data qubits $a$ followed by that on $b$; the two CNOTs are preceded and followed by a Hadamard $\hat{H}$. The measure-Z qubit is then the target of a CNOT with $a$ as the control, followed by that with $b$ as a control. The two identity $\hat{I}$ operators for the measure-Z process, which are performed by simply waiting, ensure that the timing on the measure-Z qubit matches that of the measure-X qubit. The measurement operators that correspond to steps 3 through 6 of the control sequence, followed by the projective measurement at the end of the circuit, are indicated below the relevant CNOTs, with the measure-X qubit stabilizing the product $\hat{X}_a\hat{X}_b$ and the measure-Z qubit stabilizing $\hat{Z}_a\hat{Z}_b$.

The claim is that this circuit will stabilize the two data qubits $a$ and $b$ in a simultaneous eigenstate of $\hat{X}_a\hat{X}_b$ and $\hat{Z}_a\hat{Z}_b$, i.e. precisely the Bell states listed in Table II in the main text, with the measurement outcomes corresponding to the eigenvalues listed in that table. We show how this occurs, using an arbitrary entangled state of the two data qubits as an input to the circuits in Fig. 34. Note the circuit will entangle all the qubits together during the CNOT operations; we write the quantum states in the form $|\psi_X \psi_a \psi_b \psi_Z\rangle$, i.e. the first element is the state of the measure-X qubit, the second and third those of data qubits $a$ and $b$, respectively, and the fourth element that of the measure-Z qubit (this order makes the CNOTs easier to compute, with the measure-X controlling data qubits $a$ and $b$, and the data qubits controlling the measure-Z qubit).

The state after executing the $N$th step of the circuit is $|\psi_N\rangle$. We work through each numbered step in Fig. 34.

Step 1: The measure-X and measure-Z qubits are reset to their ground states. The data qubits can be entangled in a general two-qubit state; using the notation $|\psi_a\psi_b\rangle$ where $\psi_a$ represents the state of qubit $a$ and $\psi_b$ that of qubit $b$, we can write the data qubit state in the form

$$|\psi_{ab}\rangle = A|gg\rangle + B|ge\rangle + C|eg\rangle + D|ee\rangle, \quad \text{(B1)}$$

where the coefficients $A$, $B$, $C$ and $D$ are complex, and we ignore normalization throughout. The state of the full circuit is now

$$|\psi_1\rangle = |g\rangle \otimes (A|gg\rangle + B|ge\rangle + C|eg\rangle + D|ee\rangle) \otimes |g\rangle$$

$$= A|gggg\rangle + B|ggeg\rangle + C|gegg\rangle + D|geeg\rangle. \quad \text{(B2)}$$

Step 2: The identity $\hat{I}$ leaves the state of the measure-Z qubit unchanged. The measure-X qubit undergoes a Hadamard, which takes $|g\rangle \to |+\rangle = |g\rangle + |e\rangle$ and $|e\rangle \to |-\rangle = |g\rangle - |e\rangle$ (again, ignoring normalization). The state of the full circuit is now

$$\begin{aligned}|\psi_2\rangle =\ & A|gggg\rangle + A|eggg\rangle \\ & + B|ggeg\rangle + B|egeg\rangle \\ & + C|gegg\rangle + C|eegg\rangle \\ & + D|geeg\rangle + D|eeeg\rangle.\end{aligned} \quad \text{(B3)}$$

Step 3: The first CNOT transforms the states of a control and target qubit $|ct\rangle$ according to $|gg\rangle \to |gg\rangle$, $|ge\rangle \to |ge\rangle$, $|eg\rangle \to |ee\rangle$, and $|ee\rangle \to |eg\rangle$. By applying the CNOT between the control measure-X qubit and the target data qubit $a$, the 1st to 2nd element, we obtain

$$\begin{aligned}|\psi_3\rangle =\ & A|gggg\rangle + A|eegg\rangle \\ & + B|ggeg\rangle + B|eeeg\rangle \\ & + C|gegg\rangle + C|eggg\rangle \\ & + D|geeg\rangle + D|egeg\rangle.\end{aligned} \quad \text{(B4)}$$

Step 4: The second CNOT applies the control (the measure-X qubit) to the target (data qubit $b$), the 1st to 3rd element, yielding

$$\begin{aligned}|\psi_4\rangle =\ & A|gggg\rangle + A|eeeg\rangle \\ & + B|ggeg\rangle + B|eegg\rangle \\ & + C|gegg\rangle + C|egeg\rangle \\ & + D|geeg\rangle + D|eggg\rangle.\end{aligned} \quad \text{(B5)}$$

Step 5: The third CNOT applies the control (data qubit $a$) to the target (the measure-Z qubit), the 2nd to 4th elements,

$$\begin{aligned}|\psi_5\rangle =\ & A|gggg\rangle + A|eeee\rangle \\ & + B|ggeg\rangle + B|eege\rangle \\ & + C|gege\rangle + C|egeg\rangle \\ & + D|geeg\rangle + D|eggg\rangle.\end{aligned} \quad \text{(B6)}$$

Step 6: The fourth and last CNOT applies the control (data qubit $b$) to the target (the measure-Z qubit), the 3rd to 4th elements,

$$\begin{aligned}|\psi_6\rangle =\ & A|gggg\rangle + A|eeeg\rangle \\ & + B|ggee\rangle + B|eege\rangle \\ & + C|gege\rangle + C|egee\rangle \\ & + D|geeg\rangle + D|eggg\rangle.\end{aligned} \quad \text{(B7)}$$

Step 7: The measure-X qubit undergoes its second Hadamard, giving

$$\begin{aligned}|\psi_7\rangle =\ & A|{+}ggg\rangle + A|{-}eeg\rangle \\ & + B|{+}gee\rangle + B|{-}ege\rangle \\ & + C|{+}ege\rangle + C|{-}gee\rangle \\ & + D|{+}eeg\rangle + D|{-}ggg\rangle \\ =\ & (A+D)|g\rangle \otimes (|gg\rangle + |ee\rangle) \otimes |g\rangle \\ & + (A-D)|e\rangle \otimes (|gg\rangle - |ee\rangle) \otimes |g\rangle \\ & + (B+C)|g\rangle \otimes (|ge\rangle + |eg\rangle) \otimes |e\rangle \\ & + (B-C)|e\rangle \otimes (|ge\rangle - |eg\rangle) \otimes |e\rangle.\end{aligned} \quad \text{(B8)}$$

Step 8: The terminal $\hat{Z}$ measurements are performed on both the measure-Z and measure-X qubits. Each qubit has two possible outcomes, $+1$ for the $|g\rangle$ state and $-1$ for the $|e\rangle$ state. The four possible outcomes and the corresponding projection to the final state of the data qubits are

$$\begin{aligned}\{M_X, M_Z\} =\ & \{+1,+1\}; & |\psi_{ab}\rangle =\ & |gg\rangle + |ee\rangle, \\ & \{-1,+1\}; & & |gg\rangle - |ee\rangle, \\ & \{+1,-1\}; & & |ge\rangle + |eg\rangle, \\ & \{-1,-1\}; & & |ge\rangle - |eg\rangle.\end{aligned} \quad \text{(B9)}$$

The probabilities to project to each of these states is given by the modulus-squared of each corresponding amplitude in Eq. (B8).

Note that the measurement outcomes, and the corresponding states of the two data qubits, are precisely those appearing in Table II for the Bell states. Hence, given an arbitrary input state, at the end of the circuit we project the two data qubits onto one of the four eigenstates of $\hat{Z}_a\hat{Z}_b$ and $\hat{X}_a\hat{X}_b$.

If we take one of the output states in Eq. (B9), and use it as an input to the stabilizer circuit, you can easily check that the output will precisely reproduce the input, with the same measurement outcomes. For example, if we have the measurement outcome $\{M_X, M_Z\} = \{-1,-1\}$, the data qubits end up in the state $|ge\rangle - |eg\rangle$. This is the input state in Eq. (B1) with the coefficients $A = D = 0$ and $B = -C = 1$. Looking at Eq. (B8), the amplitude coefficients for $A + D$, $A - D$, and $B + C$ will all be zero, implying the output state will always project to the input state $|ge\rangle - |eg\rangle$. Hence, as promised, the stabilizer circuit returns the same measurement outcomes and the same data qubit state.

An alternative, and much more compact, way to see this result is to realize that we can perform the same calculation using stabilizers to identify the qubit states rather than using state notation itself (this assumes you have read the discussion of the Heisenberg representation of the CNOT transformation in Sect. XIV C.) We write the stabilizers in the form $\hat{Q}_X \otimes \hat{Q}_a \otimes \hat{Q}_b \otimes \hat{Q}_Z = \hat{Q}_X \hat{Q}_a \hat{Q}_b \hat{Q}_Z$ with subscripts corresponding to the qubit labeling in Fig. 34, and $\hat{Q}_j$ represents either $\hat{X}_j$ or $\hat{Z}_j$ on the $j$th qubit. The initial state of the system, after initialization of the $X$ and $Z$ qubits, is given by the two stabilizers $\hat{X}_X \hat{I}_a \hat{I}_b \hat{I}_Z$ and $\hat{I}_X \hat{I}_a \hat{I}_b \hat{Z}_Z$; this means that the system is in an eigenstate of these two operator products, and we actually know what the eigenvalue is for each stabilizer ($+1$ for each, as the $X$ qubit is initialized in $|+\rangle$ and the $Z$ qubit is initialized in $|g\rangle$). Now, the first CNOT, between $X$ and $a$, transforms the first stabilizer to $\hat{X}_X \hat{X}_a \hat{I}_b \hat{I}_Z$, using the Heisenberg transformation rule for the CNOT ($\hat{X}_X \otimes \hat{I}_a \rightarrow \hat{X}_X \otimes \hat{X}_a$). It does nothing to the second stabilizer, as $\hat{I}_X \otimes \hat{I}_a$ is trivially unchanged. The second CNOT, between $X$ and $b$, takes the first stabilizer to $\hat{X}_X \hat{X}_a \hat{X}_b \hat{I}_Z$, using the same rule, and again does nothing to the second stabilizer. The third CNOT, between $a$ and $Z$, takes the first stabilizer to $\hat{X}_X \hat{X}_a \hat{X}_b \hat{X}_Z$ and the second stabilizer to $\hat{I}_X \hat{Z}_a \hat{I}_b \hat{Z}_Z$, using the CNOT transformation $\hat{I}_a \otimes \hat{Z}_Z \rightarrow \hat{Z}_a \otimes \hat{Z}_Z$. Finally, the fourth CNOT between $b$ and $Z$, takes the first stabilizer to $\hat{X}_X \hat{X}_a \hat{X}_b \hat{I}_Z$, using $\hat{X}_b \otimes \hat{X}_Z \rightarrow \hat{X}_b \otimes \hat{I}_Z$, and the second stabilizer to $\hat{I}_X \hat{Z}_a \hat{Z}_b \hat{Z}_Z$. Note that the stabilizers commute as they each have operators on the two qubits $a$ and $b$. Now, the $\hat{X}$ measurement of the $X$ qubit at the end gives us the product $\hat{X}_a \hat{X}_b$, and does not interfere with the second stabilizer, as $\hat{X}_X$ commutes with that stabilizer. The $\hat{Z}$ measurement of $Z$ likewise commutes with the first stabilizer, and gives us the product $\hat{Z}_a \hat{Z}_b$. These two measurements thus completely identify the states of these two qubits.

We note that the sequence of CNOT operations in Fig. 34 is critical for properly producing the measurement stabilizers $\hat{X}_a \hat{X}_b$ and $\hat{Z}_a \hat{Z}_b$. If instead of using the sequence as given in Fig. 1b and c (top, left, right, bottom) the clockwise sequence is used (top, right, bottom, left), then you can work out the sequence of stabilizer transformations as was done in the preceding paragraph. The result is the pair of stabilizers $\hat{X}_X \hat{X}_a \hat{X}_b \hat{X}_Z$ and $\hat{Z}_X \hat{Z}_a \hat{Z}_b \hat{Z}_Z$. These two stabilizers no longer commute with the single qubit measurements of $\hat{X}_X$ and $\hat{Z}_Z$, so the final measurements will just give random results. Likewise, the sequence of CNOTs given in Fig. 1b and c produces the correct stabilizers for data-qubit pairs on both the up and down diagonals.

## Appendix C: X-cut qubit initialization in a $\hat{Z}$ eigenstate

In this appendix we give the detailed steps for the "difficult" initialization of an X-cut qubit in the $|g_L\rangle$ state, as illustrated in Fig. 14.

1. We start with a nominally infinite lattice with no qubit cuts (Fig. 14a).

2. A column of four measure-X qubits is turned off, opening a rectangular cut, and the six measure-Z qubits adjacent to the cut are switched to three-terminal stabilizer measurements. In order to maintain error tracking, we perform a $\hat{Z}$ measurement on the three data qubits 1, 2 and 3 inside the cut (Fig. 14b); this could be done by direct measurement of the data qubits, if the hardware implementation allows this, or by measuring with the idle measure-X qubits in the cut. The combination of the three-terminal measure-Z stabilizers and these three single-qubit $\hat{Z}$ measurements maintains error detection, as each three-terminal measure-Z result can be multiplied with the corresponding single data qubit $\hat{Z}$ measurement to compare to the prior four-terminal $\hat{Z}$ stabilizer measurements.

3. The data qubits inside the cut are reset to $|g\rangle$ (Fig. 14c). This operation will result in the logical qubit being initialized to $|g_L\rangle$.

4. The two measure-X qubits in the middle of the cut are turned back on, and all the measure-Z qubits are switched back to four-terminal measurements. The measure-Z outcomes can be compared with the previous cycle three-terminal measure-Z outcomes to maintain error detection, accounting of course for the fact that the internal data qubits were set to $|g\rangle$. As the internal data qubits were in $|g\rangle$ prior to the measure-X qubits being turned on, the two measure-X qubits will return random outcomes; however, the projective measurement of the measure-X qubits keeps the data qubits in an eigenstate of $\hat{Z}_L = \hat{Z}_1 \hat{Z}_2 \hat{Z}_3$, as $\hat{Z}_L$ commutes with all the stabilizers in the array. The ground-state initialization of the three data qubits thus ensures that the X-cut qubit is left in the $|g_L\rangle$ eigenstate of $\hat{Z}_L$.

The third step can be omitted if the measurements in step 2 are nondestructive and leave the data qubits in the measured eigenstates. In this case, the eigenvalue and corresponding eigenstate of $\hat{Z}_L$ will equal the product of the $\hat{Z}$ measurements for these qubits; if a different initial state is desired, an $\hat{X}_L$ bit-flip can be performed in software.

## Appendix D: Measuring an X-cut qubit in the $\hat{Z}_L$ basis

Here we give the procedure for the "difficult" measurement of an X-cut qubit in the $\hat{Z}_L$ basis, using Fig. 15 in the main text.

1. We start with the logical X-cut qubit in some state (Fig. 15a).

2. We turn off the two measure-X qubits between the two cuts, and also switch the neighboring measure-Z qubits from four-terminal to three-terminal measurements. We measure the three data qubits inside the rectangular cut in the $\hat{Z}$ basis (Fig. 15b); the product of these measurements is the value of $\hat{Z}_L$. By combining the three-terminal measure-Z outcomes with the single data qubit $\hat{Z}$ measurements, we can maintain the surface code error tracking.

3. We reset the three data qubits to their ground states $|g\rangle$. This step is not strictly necessary if the $\hat{Z}$ measurement in the prior step has a high fidelity for projecting to the $\hat{Z}$ eigenstates.

4. We "destroy" the logical qubit by resuming full four-terminal $\hat{Z}$ stabilization and turning all the measure-X qubits back on, leaving us with the original 2D array (Fig. 15d). Because the data qubits were set to $|g\rangle$, the measure-X qubits will report random outcomes on this step.

Note that the measurement $M_Z$ of the individual qubits in step 2 projects the data qubits onto a product eigenstate of $\hat{Z}_a$, $\hat{Z}_b$ and $\hat{Z}_c$ of the three individual qubits, which was not the state prior to this measurement; however, as $\hat{Z}_L$ commutes with the individual $\hat{Z}_j$ data qubit operators, this projection commutes with $\hat{Z}_L$, so the product of the individual measurement outcomes $Z_a Z_b Z_c$ is equal to the eigenvalue $Z_L$ of $\hat{Z}_L$.

## Appendix E: Making a larger qubit

Here we describe how to make a distance $d = 8$ logical qubit, twice the distance of the logical qubits we have been discussing, in which only one stabilizer was turned off per qubit hole. Figure 16, which accompanies this description, can be found in the main text.

We first define the the logical operator $\hat{Z}_L$ as the chain

$$\hat{Z}_L = \hat{Z}_1 \hat{Z}_2 \hat{Z}_3 \hat{Z}_4 \hat{Z}_5 \hat{Z}_6 \hat{Z}_7 \hat{Z}_8. \tag{E1}$$

This is shown outlined in red (gray) in Fig. 16; the analogous $\hat{X}_L$ chain, that links the two larger-distance qubits, is not shown.

1. In one surface code cycle, we stop measuring the four $\hat{Z}$ stabilizers $\hat{Z}_{s1}$, $\hat{Z}_{s2}$, $\hat{Z}_{s3}$ and $\hat{Z}_{s4}$ inside the $\hat{Z}_L$ loop. Note we are using a shorthand notation for the stabilizers, where $\hat{Z}_{s1}$ represents the four-element stabilizer $\hat{Z}_{s1} = \hat{Z}_{1a} \hat{Z}_{1b} \hat{Z}_{1c} \hat{Z}_{1d}$; $\hat{Z}_{1a}$ is the first $\hat{Z}$ physical qubit operator in the $\hat{Z}$ stabilizer loop, etc. The $\hat{Z}_L$ operator is the product of these four $\hat{Z}$ stabilizers, $\hat{Z}_L = \hat{Z}_{s1} \hat{Z}_{s2} \hat{Z}_{s3} \hat{Z}_{s4}$, so the initial qubit logical state will be either $|g_L\rangle$ or $|e_L\rangle$, as determined by the product of the stable outcomes $Z_{s1} Z_{s2} Z_{s3} Z_{s4} = \pm 1$.

2. In the same surface code cycle, we turn off the one $\hat{X}$ stabilizer $\hat{X}_{s1}$ inside the $\hat{Z}_L$ loop, and we also turn the $\hat{X}$ stabilizers bordering the qubit hole from four-terminal to three-terminal measurements. We perform $\hat{X}$ measurements of the four internal data qubits, projecting them onto a product of $\hat{X}$ eigenstates. This can be done either by direct measurements of these data qubits, if the hardware allows this, or by using the now-idle measure-Z qubits to perform the measurement. This measurement allows us to continue to track errors, by combining these measurements with the three-terminal $\hat{X}$ stabilizer measurements bordering the hole.

## Appendix F: One-cell qubit move

This appendix describes the details in moving a Z-cut logical qubit by one cell in the surface code array. The figure that relates to this process, Fig. 17, is in the main text. The move involves the following four steps:

1. We extend the logical operator $\hat{Z}_L$ by multiplying it by the set of single-qubit $\hat{Z}$ operators $\hat{Z}_6 \hat{Z}_7 \hat{Z}_8 \hat{Z}_9$ that comprise the $\hat{Z}$ stabilizer just below the lower Z-cut hole, as shown in Fig. 17a:

$$\begin{aligned} \hat{Z}_L^e &\equiv (\hat{Z}_6 \hat{Z}_7 \hat{Z}_8 \hat{Z}_9) \hat{Z}_L \\ &= \hat{Z}_3 \hat{Z}_4 \hat{Z}_5 \hat{Z}_7 \hat{Z}_8 \hat{Z}_9. \end{aligned} \tag{F1}$$

$\hat{Z}_L^e$ now encircles two cells, as illustrated in Fig. 17b. The result of operating with $\hat{Z}_L^e$ on a quiescent state $|\psi\rangle$ is the same as $\hat{Z}_L$, other than a possible sign change from the stabilizer outcome $Z_{6789} = \pm 1$.

2. We turn off the $\hat{Z}$ stabilizer $\hat{Z}_6 \hat{Z}_7 \hat{Z}_8 \hat{Z}_9$. We also switch the two $\hat{X}$ stabilizers that neighbor data qubit 6 from four-terminal to three-terminal measurements, so that they no longer measure data qubit 6. This data qubit is now not stabilized by any measurements; we therefore perform an $\hat{X}$ measurement of data qubit 6, projecting it to $|+\rangle$ or $|-\rangle$ with the eigenvalue outcome $X_6 = \pm 1$. This measurement is either performed directly on the data qubit, if allowed by the physical implementation, or by the idle measure-Z qubit just below data qubit 6 (see Fig. 17a). The product of this measurement outcome with the adjacent three-terminal $\hat{X}$ stabilizer measurements is compared with the prior four-terminal $\hat{X}$ stabilizer measurements to maintain error detection.

We extend $\hat{X}_L$ to $\hat{X}'_L$ by multiplying it by $\hat{X}_6$:

$$\hat{X}'_L = \hat{X}_6 \hat{X}_L = \hat{X}_1 \hat{X}_2 \hat{X}_3 \hat{X}_6. \tag{F2}$$

3. As the distance $d$ here is only three, we do not need to wait to stabilize the result. If we perform this process with a larger distance logical qubit, we would at this point have to wait $d/4$ (rounded up) surface code cycles to ensure no chain can wrap around a qubit hole.

4. We turn on the $\hat{Z}$ stabilizer $\hat{Z}_3 \hat{Z}_4 \hat{Z}_5 \hat{Z}_6$. We also switch the two $\hat{X}$ stabilizers that neighbor data qubit 6 back to four-terminal measurements, so data qubit 6 is now fully stabilized. A single measurement of each of these stabilizers is not sufficient to protect against errors, so we wait a minimum of $d$ cycles to properly establish their values.

We define the new logical operator $\hat{Z}'_L$ as the product of $\hat{Z}^e_L$ and the $\hat{Z}$ stabilizer we just turned on, $\hat{Z}_3 \hat{Z}_4 \hat{Z}_5 \hat{Z}_6$:

$$\begin{aligned} \hat{Z}'_L &= (\hat{Z}_3 \hat{Z}_4 \hat{Z}_5 \hat{Z}_6) \hat{Z}^e_L \\ &= \hat{Z}_6 \hat{Z}_7 \hat{Z}_8 \hat{Z}_9. \end{aligned} \tag{F3}$$

Clearly $\hat{Z}'_L$ is the loop shown in Fig. 17c, and $\hat{X}'_L$ is the extended chain shown in the same figure.

### Byproduct operators

One problem that is created in the move transformation is that the extension of $\hat{X}_L$ through multiplication by $\hat{X}_6$ can yield an extended $\hat{X}'_L$ that differs in sign from $\hat{X}_L$; this will occur if the measurement outcome $X_6 = -1$. Similarly the final $\hat{Z}'_L$ can differ in sign from $\hat{Z}_L$, by the product of the two $\hat{Z}$ stabilizer measurement outcomes involved in the move, $Z_{6789}$ and $Z_{3456}$. In order to prevent these unwanted sign changes from occurring, we could multiply $\hat{X}'_L$ by $X_6$ and $\hat{Z}'_L$ by $Z_{6789} Z_{3456}$, which would correct for any sign changes (note these multipliers would be stable measurement outcomes, not the operators). However, a more efficient way to account for this is to instead modify the quiescent state $|\psi\rangle$ by using logical bit- and phase-flips. It is convenient to define two parameters $p_X$ and $p_Z$, using

$$(-1)^{p_X} \equiv X_6, \tag{F4}$$

and

$$(-1)^{p_Z} \equiv Z_{3456} Z_{6789}. \tag{F5}$$

With these definitions, $p_X = 0(1)$ if the stable $X_6$ measurement is $+1(-1)$, that is, if $\hat{X}'_L$ does (does not) have the same sign as $\hat{X}_L$. Similarly, $p_Z = 0(1)$ if the product $Z_{3456} Z_{6789} = +1(-1)$, that is, if $\hat{Z}'_L$ does (does not) have the same sign as $\hat{Z}_L$.

Prior to performing any of the steps in the logical qubit move, the quiescent state is $|\psi\rangle$, which we will write as $|\psi^{\mathrm{pre}}\rangle$. The various operations and measurements in the move transform this to the post-move state, which in the ideal case would be $|\psi'\rangle$. Due to sign changes in the $\hat{X}_L$ and $\hat{Z}_L$ operators during the move, the transformation instead gives

$$|\psi\rangle \rightarrow \hat{X}'^{p_X}_L \hat{Z}'^{p_X}_L |\psi'\rangle. \tag{F6}$$

These additional operators are called "byproduct" operators. The expression Eq. (F6) looks complicated but is in fact quite simple: If $p_X = 0(1)$, then $\hat{X}'_L$ has the same (opposite) sign as $\hat{X}_L$, so $\hat{Z}'_L$ does not (does) appear in Eq. (F6). Similarly, if $p_Z = 0(1)$, then $\hat{Z}'_L$ has the same (opposite) sign as $\hat{Z}_L$, and $\hat{X}'_L$ does not (does) appear. The way the byproduct operators account for the sign change is through the anti-commutation of $\hat{X}_L$ and $\hat{Z}_L$, as can be seen by operating with $\hat{X}_L$ on the byproduct adjusted $|\psi'\rangle$:

$$\hat{X}'_L \left( \hat{X}'^{p_Z}_L \hat{Z}'^{p_X}_L |\psi'\rangle \right) = (-1)^{p_X} \hat{X}'^{p_Z}_L \hat{Z}'^{p_X}_L \hat{X}'_L |\psi'\rangle \tag{F7}$$

If $\hat{X}'_L$ has the same (opposite) sign as $\hat{X}_L$, then $p_X = 0(1)$ and the sign change is canceled by the factor $+1(-1)$. Similarly, the presence of the byproduct operator $\hat{X}'^{p_Z}_L$ will guarantee that $\hat{Z}'_L$ has the same sign when operating on $\hat{X}'^{p_Z}_L \hat{Z}'^{p_X}_L |\psi'\rangle$, through their anti-commutation.

The byproduct operators are just $\hat{X}_L$ and $\hat{Z}_L$, the bit- and phase-flip logical operators, and are handled by the software control system, by changing the signs of any logical measurements of that qubit in the appropriate fashion: If $p_X = 0(1)$, then a $\hat{Z}_L$ measurement will not have (will have) its sign reversed, and similarly for $p_Z$. The byproduct operators are never actually applied to the logical qubits directly.

A more complete description and formal theory for the byproduct (byproduct) operators can be found in Ref. [66].

### Appendix G: Multi-cell moves

A multi-cell qubit shift is performed using the following steps (Fig. 18 is in the main text):

1. We extend $\hat{Z}_L$ to $\hat{Z}^e_L$ by defining $\hat{Z}^e_L$ as

$$\hat{Z}^e_L = (\hat{Z}_{s2} \hat{Z}_{s3} \ldots \hat{Z}_{sn}) \hat{Z}_L. \tag{G1}$$

Here $\hat{Z}_{sj}$ is the $j$th $\hat{Z}$ stabilizer, i.e. $\hat{Z}_{sj} = \hat{Z}_{ja} \hat{Z}_{jb} \hat{Z}_{jc} \hat{Z}_{jd}$, the product of the four neighboring data qubit $\hat{Z}$ operators. By writing out each $\hat{Z}_{sj}$ in terms of its $\hat{Z}$ operators, you can easily convince yourself that $\hat{Z}_L^e$ is the chain of $\hat{Z}$ operators that forms the extended loop in Fig. 18b. Note that $\tilde{Z}_L^e$ can differ in sign from $\tilde{Z}_L$, depending on the value of the product of the pre-move stabilizer values $Z_{s2}^i Z_{s3}^i \ldots Z_{s,n}^i = \pm 1$; the byproduct operators will be corrected as described above.

2. We turn off the $\hat{Z}$ stabilizers $\hat{Z}_{sj}$ now enclosed by $\hat{Z}_L^e$, and also turn all the four-terminal $\hat{X}$ stabilizers that border $\hat{Z}_L^e$ to three-terminal stabilizers, leaving all the internal data qubits disconnected from the surface.

   In the same surface code cycle, we measure all the disconnected data qubits along $\hat{X}$, either directly or using the now-idle neighboring measure-Z qubits to perform this measurement. This yields the measurement outcomes $X_1, X_2 \ldots X_{n-1}$, each equal to $\pm 1$, which when multiplied by the appropriate three-terminal $\hat{X}$ stabilizers can be compared to the prior values of the corresponding four-terminal $\hat{X}$ stabilizers.

   We define the new $\hat{X}_L'$ operator as

   $$\hat{X}_L' = (\hat{X}_1 \ldots \hat{X}_{n-1})\hat{X}_L. \tag{G2}$$

   You can easily see that $\hat{X}_L'$ is the chain of $\hat{X}$ operators shown in Fig. 18c. Any sign change from the product of measurement outcomes $X_1 \ldots X_{n-1} = \pm 1$ will be accounted for with a byproduct operator, and corrected for in any subsequent measurement.

3. We define the new logical operator $\hat{Z}_L'$ by

   $$\begin{aligned}
   \hat{Z}_L' &= (\hat{Z}_{s1}\hat{Z}_{s2} \ldots \hat{Z}_{s,n-1})\hat{Z}_L^e \\
   &= (\hat{Z}_{s1} \ldots \hat{Z}_{s,n-1})(\hat{Z}_{s2} \ldots \hat{Z}_{s,n})\hat{Z}_L \\
   &= \hat{Z}_{s1}\hat{Z}_{s,n}\hat{Z}_L \\
   &= \hat{Z}_{s,n}.
   \end{aligned} \tag{G3}$$

   You can easily convince yourself that $\hat{Z}_L'$ is just the loop of physical $\hat{Z}$ operators surrounding the lower Z-cut hole shown in Fig. 18c, which is algebraically the same as the $n$th stabilizer $\hat{Z}_{s,n}$.

   We turn on the $\hat{Z}$ stabilizers $\hat{Z}_{s1}, \ldots, \hat{Z}_{s,n-1}$, and wait at least $d$ surface code cycles to establish the values of these stabilizers. The product of the stable post-move values, $Z_{s1}^f Z_{s2}^f \ldots Z_{s,n-1}^f = \pm 1$, will be used to determine the byproduct operators.

4. We correct the logical wavefunction for any byproduct operators. We define the byproduct powers in the same way as for the one-cell move,

   $$(-1)^{p_X} = X_1 X_2 \ldots X_n, \tag{G4}$$

and

$$\begin{aligned}
(-1)^{p_Z} &= (Z_{s1}^f Z_{s2}^f \ldots Z_{s,n-1}^f) \\
&\quad \times (Z_{s2}^i Z_{s3}^i \ldots Z_{s,n}^i).
\end{aligned} \tag{G5}$$

Using these, we multiply the (ideal) post-move state $|\psi'\rangle$ by byproduct operators to obtain the sign-corrected post-move logical state:

$$\hat{Z}_L'^{p_Z} \hat{X}_L'^{p_X} |\psi'\rangle. \tag{G6}$$

## Appendix H: Single qubit braid transformation

This appendix gives a more detailed explanation of the sign changes involved in the braid transformation of a single qubit. In Fig. 19a through f (Fig. 20a through e), shown in the main text, we show how the $\hat{X}_L$ ($\hat{Z}_L$) operator is transformed by the two moves in the braid. After the extension for the first move is closed up in Fig. 19c, $\hat{X}_L$ is extended to $\hat{X}_L'$ by multiplying $\hat{X}_L$ by all the data qubits isolated in the first move:

$$\hat{X}_L \rightarrow \hat{X}_L' = \left(\hat{X}_1 \ldots \hat{X}_8\right)\hat{X}_L. \tag{H1}$$

The second move extends $\hat{X}_L'$ to $\hat{X}_L''$ (Fig. 19e), again by multiplying $\hat{X}_L'$ by all the data qubits isolated in the second move:

$$\hat{X}_L' \rightarrow \hat{X}_L'' = \left(\hat{X}_9 \ldots \hat{X}_{12}\right)\hat{X}_L'. \tag{H2}$$

These transformations may involve a sign change from $\hat{X}_L$ to $\hat{X}_L''$, as with the one-cell and multi-cell moves, depending on the stable measurement outcomes of the data qubits used in the extension of $\hat{X}_L$, which were all measured along $\hat{X}$. The sign change is captured by the power $p_X$, $(-1)^{p_X} = (X_1^e \ldots X_8^e)(X_9^e \ldots X_{12}^e)$, with each parenthetical set the product of measurement outcomes from one of the two moves. The power $p_X$ is equal to 0(1) if $\hat{X}_L''$ has the same (different) sign from $\hat{X}_L$. Now, the loop of physical qubit operators encloses only fully-stabilized cells, as shown in Fig. 19e, and the surface code ensures that the wavefunction $|\psi\rangle$ is an eigenstate of this loop of operators. Hence the measurement product giving $p_X$ has to be equal to the product of all the $\hat{X}$ stabilizers enclosed by the loop:

$$\begin{aligned}
(-1)^{p_X} &= (X_1^e \ldots X_8^e)(X_9^e \ldots X_{12}^e) \\
&= X_{s1}^f X_{s2}^f \ldots X_{s9}^f,
\end{aligned} \tag{H3}$$

the product of the stable measure-X stabilizers $X_{s,j}^f = \pm 1$ enclosed by the loop. A way to see this is to write out each of the stabilizers $\hat{X}_{s1}$ through $\hat{X}_{s9}$ in terms of their four data qubit $\hat{X}_j$ operators, and then multiply these all together to get the product $\hat{X}_{s1}\hat{X}_{s2} \ldots \hat{X}_{s9}$. All the data qubits shared by any two stabilizers appear twice in this product, and thus cancel as $\hat{X}_j^2 = \hat{I}$, leaving only the data qubits on the periphery of this set of stabilizers. The product $(X_1^e \ldots X_8^e)(X_9^e \ldots X_{12}^e)$ that determines $p_X$ is precisely the measurement outcome of these peripheral data qubits.

The $\hat{Z}_L$ loop operator is similarly transformed by the braid to $\hat{Z}_L''$, using two sets of expansions followed by contractions, and any sign change is determined by the $\hat{Z}$ stabilizers that are enclosed in the two expansions. The sign change is equal to $(-1)^{p_Z}$, where $p_Z$ is equal to $0(1)$ if $\hat{Z}_L''$ has the same (opposite) sign as $\hat{Z}_L$. In terms of all the stabilizer values in the move, we have

$$
\begin{aligned}
(-1)^{p_Z} &= (Z_{s9}^f \ldots Z_{s12}^f)(Z_{s10}^i \ldots Z_{s13}^i) \\
&\quad \times (Z_{s1}^f \ldots Z_{s8}^f)(Z_{s2}^i \ldots Z_{s9}^i).
\end{aligned}
\tag{H4}
$$

The braid transforms the 2D array wavefunction $|\psi\rangle$ in the two move operations, ending up after the second move with the wavefunction $|\psi''\rangle$. Any sign changes indicated by the powers $p_X$ and $p_Z$ are accounted for by modifying the post-braid state $|\psi''\rangle$ using byproduct operators that are tracked, and corrected for, in the control software:

$$
\hat{Z}_L''^{p_Z} \hat{X}_L''^{p_Z} |\psi''\rangle.
\tag{H5}
$$

The notation reminds us that $|\psi''\rangle$ is the ideal result, with no sign changes, but that because of move-induced sign changes, we need to keep track of these sign changes with the additional byproduct operators appearing in Eq. (H5). Hence we see that other than sign changes and byproduct operators, determined by projective measurements during the move and thus accountable in software, the braid transformation around a fully stabilized part of the array leaves the two logical operators of the Z-cut qubit unchanged.

## Appendix I: Two-qubit braid transformation

Here we work through braiding a Z-cut qubit hole through an X-cut qubit, as shown in Figs. 21, 22 and 23 (main text). We want to show that the braid between these two qubit types is equivalent to a logical CNOT, with the Z-cut qubit as the control and the X-cut qubit the target. The same result holds when braiding an X-cut qubit through a Z-cut, with the X-cut qubit still the target of the CNOT with the Z-cut as the control, but braiding two Z-cut qubits or two X-cut qubits requires other logic circuitry, as discussed in Sect. XIV D.

As we discuss in the main text, it turns out we only need to show that the braid transforms a total of four of the sixteen two-qubit operator combinations correctly in order to prove a braid is a CNOT: $\hat{X}_L \otimes \hat{I}_L$, $\hat{I}_L \otimes \hat{X}_L$, $\hat{Z}_L \otimes \hat{I}_L$ and $\hat{I}_L \otimes \hat{Z}_L$. We consider each of these in turn, in addition to working out the transformation of $\hat{X}_L \otimes \hat{X}_L$, providing an example of how two sequential braids are equivalent to an identity operation.

$\hat{X}_L \otimes \hat{I}_L \to \hat{X}_L \otimes \hat{X}_L$: We first consider the effect of the braid on the two-qubit operator $\hat{X}_L \otimes \hat{I}_L = \hat{X}_{L1}\hat{I}_{L2}$, as shown in Fig. 21. The $\hat{X}_{L1}$ bit-flip is directed at the first, Z-cut qubit, and $\hat{I}_{L2}$ is directed at the second, X-cut qubit (the subscripts 1 and 2 indicate on which logical qubit the operator acts). In the first move for the braid, we have

$$
\begin{aligned}
\hat{X}_{L1}\hat{I}_{L2} &\to \left(\hat{X}_1 \ldots \hat{X}_8 \hat{X}_{L1}\right)\hat{I}_{L2} \\
&= \hat{X}_{L1}'\hat{I}_{L2}.
\end{aligned}
\tag{I1}
$$

Note we are mixing logical and physical qubit operators, but as the logical operators are constructed from products of physical qubit operators, this is formally acceptable.

In the second move, we have

$$
\begin{aligned}
\hat{X}_{L1}'\hat{I}_{L2} &\to \left(\hat{X}_9 \ldots \hat{X}_{12} \hat{X}_{L1}'\right)\hat{I}_{L2} \\
&= \hat{X}_{L1}''\hat{I}_{L2}.
\end{aligned}
\tag{I2}
$$

As shown in Fig. 21e, the combination of the two moves of $\hat{X}_{L1}$ generates an operator $\hat{X}_{L2}''$ that comprises the original $\hat{X}_{L1}$ chain along with a closed loop of data qubit operators, with the loop enclosing the upper hole of the X-cut qubit 2, as well as a number of fully stabilized cells. We can re-write this as

$$
\begin{aligned}
\hat{X}_{L1}''\hat{I}_{L2} &= \hat{X}_{\text{loop}}\hat{X}_{L1}'''\hat{I}_{L2} \\
&= \left(\hat{X}_{s1} \ldots \hat{X}_{s4}\hat{X}_{L2}\hat{X}_{s6} \ldots \hat{X}_{s9}\right)\hat{X}_{L1}''' \\
&\to \hat{X}_{L1}'''\hat{X}_{L2}.
\end{aligned}
\tag{I3}
$$

To obtain this result, we have deformed the closed loop $\hat{X}_{\text{loop}}$ from the transformed $\hat{X}_{L1}'$ through each of the enclosed $\hat{X}$ stabilizers $\hat{X}_{sj}$, including a loop of physical $\hat{X}$ bit-flips that wrap tightly around the upper hole of X-cut qubit 2, which is precisely an $\hat{X}_{L2}$ bit-flip. The $\hat{X}$ stabilizers resolve to measurement eigenvalues $\pm 1$, leaving the outer product of logical operators $\hat{X}_{L2}\hat{X}_{L1}''' = \hat{X}_{L1}'''\hat{X}_{L2}$, as in the last line of Eq. (I3).

There are a number of sign changes that can occur, depending on the stabilizer outcomes. We again use powers $p_X$ and $p_Z$ to capture these sign changes. The braid of the first logical qubit generates a sign change $(-1)^{p_{X1}} \equiv (X_1^e \ldots X_8^e)(X_9^e \ldots X_{12}^e) = \pm 1$, equal to the product of the stable single data qubit measurements in the two moves. This is the same sign change as for an empty-loop braid.

The braid also generates sign changes on the second qubit, as in the braid we have transformed the identity operator on that qubit to an $\hat{X}_{L2}$ logical operator. The sign change is very similar to that for the first qubit, where we define a power $p_{X2}$ given by $(-1)^{p_{X2}} \equiv X_{s1}^f \ldots X_{s4}^f X_{s6}^f \ldots X_{s9}^f = \pm 1$, the product of the $\hat{X}$ stabilizers involved in deforming the loop $\hat{X}_{\text{loop}}$ surrounding the second qubit's upper hole to the loop that is the $\hat{X}_{L2}$ operator.

To account for the sign changes from the two logical qubits, we modify the post-braid wavefunction $|\psi''\rangle$ with two byproduct operators $\hat{Z}_{L1}^{p_{X1}}$ and $\hat{Z}_{L2}^{p_{X2}}$, such that we end up with

$$\hat{Z}_{L1}^{p_{X1}} \hat{Z}_{L2}^{p_{X2}} |\psi''\rangle. \qquad (I4)$$

This is not yet the whole story, as we still need to determine the byproduct operators associated with any braid-induced sign changes of the $\hat{Z}_{L1}$ and $\hat{Z}_{L2}$ operators; see Eq. (I8) for the complete expression.

Ignoring any byproduct operators, the braid therefore generates the transformation

$$\hat{X}_L \otimes \hat{I}_L \rightarrow \hat{X}_L \otimes \hat{X}_L. \qquad (I5)$$

$\hat{X}_L \otimes \hat{X}_L \rightarrow \hat{X}_L \otimes \hat{I}_L$: Here we apply the braid transformation to the outer product of two $\hat{X}_L$ operators on the two logical qubits. This is thus a braid transformation of the result of the braid we just discussed, $\hat{X}_L \otimes \hat{I}_L \rightarrow \hat{X}_L \otimes \hat{X}_L$. A braid is not a reversible process, as there are projective measurements involved during the moves, but as we will see, the braid transformation of $\hat{X}_L \otimes \hat{X}_L$ reverses the braid transformation of $\hat{X}_L \otimes \hat{I}_L$. The simple argument is the following: When we start with $\hat{X}_L \otimes \hat{X}_L$, the braid process "wraps" the $\hat{X}_L$ chain from the Z-cut qubit 1 around the loop representing $\hat{X}_L$ for the X-cut qubit 2, just as with $\hat{X}_L \otimes \hat{I}_L$. We can again deform the loop in Z-cut qubit 1's transformed $\hat{X}_L$ through the enclosed $\hat{X}$ stabilizers, so that it wraps tightly around the X-cut qubit 2, giving us the equivalent of two $\hat{X}_L$ bit-flips on the X-cut qubit 2; these however cancel out ($\hat{X}_L^2 = \hat{I}_L$), leaving us with $\hat{I}_L$ as the resulting operation on qubit 2. Note there is no net modification of the Z-cut qubit 1 $\hat{X}_L$ operator during this process, other than byproduct operations on the wavefunction. Hence we find $\hat{X}_L \otimes \hat{X}_L \rightarrow \hat{X}_L \otimes \hat{I}_L$ under the braid transformation.

We write out the detailed process. For the first move, just as with $\hat{X}_{L1} \otimes \hat{I}_{L2}$, we have

$$\begin{aligned} \hat{X}_{L1} \hat{X}_{L2} &\rightarrow \left( \hat{X}_1 \ldots \hat{X}_8 \hat{X}_{L1} \right) \hat{X}_{L2} \\ &= \hat{X}'_{L1} \hat{X}_{L2}. \end{aligned} \qquad (I6)$$

In the second move, we have

$$\begin{aligned} \hat{X}'_{L1} \hat{X}_{L2} &\rightarrow \left( \hat{X}_9 \ldots \hat{X}_{12} \hat{X}'_{L1} \right) \hat{X}_{L2} \\ &\rightarrow \left( \hat{X}_{s1} \ldots \hat{X}_{s4} \hat{X}_{L2} \hat{X}_{s6} \ldots \hat{X}_{s9} \right) \hat{X}''_{L1} \hat{X}_{L2} \\ &\rightarrow \hat{X}''_{L1} \hat{X}''_{L2} \hat{X}_{L2} \\ &\rightarrow \hat{X}''_{L1} \hat{I}_{L2}, \end{aligned} \qquad (I7)$$

where we use the double primes on $\hat{X}''_{L2}$ to distinguish the second $\hat{X}_{L2}$ logical operator on qubit 2 from the original $\hat{X}_{L2}$. However the product of these $\hat{X}$ operators is the identity, as in the last line of this equation.

In general, performing the same braid operation twice is equivalent to an identity, other than the appearance of byproduct operators.

$\hat{I}_L \otimes \hat{X}_L \rightarrow \hat{I}_L \otimes \hat{X}_L$: Consider now a braid involving $\hat{I}_{L1}$ directed at the Z-cut qubit 1, and $\hat{X}_{L2}$ directed at the X-cut qubit 2; this is shown in Fig. 22a through d. The $\hat{X}_{L2}$ operator is a loop of $\hat{X}$ bit-flips on the data qubits surrounding the X-cut qubit's upper hole. The braid has no effect on the identity $\hat{I}_{L1}$, as the identity involves no operations, so dragging the upper qubit hole around a closed loop does not leave a trail of operators from the moving qubit that can interact with the X-cut qubit 2, and vice versa. Hence the braid leaves these two operators unchanged, as in Fig. 22d, and $\hat{I}_L \otimes \hat{X}_L \rightarrow \hat{I}_L \otimes \hat{X}_L$ (other than byproduct operators not shown here).

$\hat{I}_L \otimes \hat{Z}_L \rightarrow \hat{Z}_L \otimes \hat{Z}_L$: Consider now the braid involving $\hat{I}_{L1}$ directed at the Z-cut qubit 1, and $\hat{Z}_{L2}$ directed at the X-cut qubit 2; this is shown in Fig. 23a. This is completely analogous to $\hat{X}_L \otimes \hat{I}_L$ but with the roles of the Z- and X-cut qubits exchanged. We deform the X-cut qubit 2 $\hat{Z}_{L2}$ operator so that it has the form shown in Fig. 23b. The qubit 1 hole is moved through this path until it returns to its original location, shown in Fig. 23c. We can then multiply the $\hat{Z}_{L2}$ operator by all the $\hat{Z}$ stabilizers shown in the dashed boxes in Fig. 23d, $\hat{Z}_{s1}, \hat{Z}_{s2} \ldots \hat{Z}_{s7}$, whose stable measurement outcomes are all known, using the identity $\hat{Z}_{L2} = \hat{Z}_{L1} \hat{Z}_{s1} \ldots \hat{Z}_{s7} \hat{Z}'_{L2}$, where $\hat{Z}_{L1}$ is the loop of $\hat{Z}$ data qubit operators surrounding the lower qubit 1 hole, thus corresponding to $\hat{Z}_L$ on that qubit, and $\hat{Z}'_{L2}$ is the original $\hat{Z}_{L2}$ chain, as shown in Fig. 23d. We have thus generated a $\hat{Z}_{L1}$ phase-flip operation on Z-cut qubit 1. Other than possible byproduct operators, the $\hat{Z}_{L2}$ operation on the X-cut qubit 2 is not modified by the braid. Hence we see that $\hat{I}_L \otimes \hat{Z}_L$ transforms to $\hat{Z}_L \otimes \hat{Z}_L$. Performing this braid a second time, i.e. starting now with $\hat{Z}_L \otimes \hat{Z}_L$, will generate a second $\hat{Z}_{L1}$ operation on the Z-cut qubit 1, so this will transform back to the original pair of operators, i.e. to $(\hat{Z}_L \cdot \hat{Z}_L) \otimes \hat{Z}_L = \hat{I}_L \otimes \hat{Z}_L$, in a manner completely analogous to $\hat{X}_L \otimes \hat{X}_L \rightarrow \hat{X}_L \otimes \hat{I}_L$.

$\hat{Z}_L \otimes \hat{I}_L \rightarrow \hat{Z}_L \otimes \hat{I}_L$: Finally, consider the braid transformation involving a $\hat{Z}_{L1}$ operation on the Z-cut qubit 1, and $\hat{I}_{L2}$ directed at the X-cut qubit 2. This situation is completely analogous to $\hat{I}_L \otimes \hat{X}_L$: $\hat{Z}_{L1}$ is a loop of $\hat{Z}$ phase-flips of the data qubits surrounding the Z-cut hole, and during the braid the wavefunction acquires byproduct operators. The loop however does not enclose the X-cut qubit 2 in a way that interacts with that qubit, so the braid does nothing to X-cut qubit 2. Hence we see that other than byproduct operators, $\hat{Z}_L \otimes \hat{I}_L$ transforms to $\hat{Z}_L \otimes \hat{I}_L$.

**Byproduct operators for two-qubit transformations**

When discussing the $\hat{X}_L \otimes \hat{I}_L \rightarrow \hat{X}_L \otimes \hat{X}_L$ transformation, we gave expressions for the powers $p_{X1}$ and $p_{X2}$ that determine whether there is a sign change associated with the final $\hat{X}_{L1} \hat{X}_{L2}$ operators that must be corrected with phase-flips $\hat{Z}_{L1}$ and $\hat{Z}_{L2}$. Braid transformations that result in $\hat{Z}_{L1}$ or $\hat{Z}_{L2}$ operators must similarly be corrected for any unwanted sign changes in these operators, as determined in the usual way by the outcomes of the various stabilizer and single data qubit measurements involved in the braid transformation: If there is (is not) a sign change due to the braid transformation of the $\hat{Z}_{L1}$ operator, the power $p_{Z1} = 1(0)$, while if there is (is not) a sign change due to the transformation of the $\hat{Z}_{L2}$ operator, the power $p_{Z2} = 1(0)$. In either case, the transformed state wavefunction is modified by the byproduct operators $\hat{X}_{L1}^{p_{Z1}}$ and $\hat{X}_{L2}^{p_{Z2}}$ to indicate these sign changes.

Combined with the byproduct operators $\hat{Z}_{L1}$ and $\hat{Z}_{L2}$, the final wavefunction $|\psi''\rangle$ following a braid is given by

$$(\hat{Z}_{L1})^{p_{X1}}(\hat{X}_{L1})^{p_{Z1}}(\hat{Z}_{L2})^{p_{X2}}(\hat{X}_{L2})^{p_{Z2}}|\psi''\rangle. \quad \text{(I8)}$$

## Appendix J: Logical Hadamard process

The process for performing a logical Hadamard on a single logical qubit is detailed in Figs. 26, 27 and 28. The detailed sequence is as follows:

1. We turn off a ring of $\hat{X}$ stabilizers surrounding both Z-cut holes of the logical qubit to be Hadamard-transformed, and also reduce the $\hat{Z}$ stabilizers on either side of the ring from four- to three- and two-terminal measurements, effectively isolating the target qubit in a patch of the 2D array. The unstabilized data qubits in the gap between the two Z boundaries are measured along $\hat{Z}$, maintaining the surface code error tracking. This is shown in Fig. 27a. By comparing this figure to Fig. 26, it is apparent that the ring boundary comes very close to the logical qubits outside the ring, and it may seem that the proximity of these logical qubits reduces the array distance $d$ to a small value; however, the internal Z boundary of the moat does not allow short undetectable error chains to reach the internal X boundaries of the Z-cut logical qubits outside the ring, so actually the distance $d = 7$ is preserved. If any qubits exterior to the ring were X-cut qubits, however, these would have to be kept a distance $d$ from the ring. The same applies to the Z-cut qubit inside the ring; it is also protected by the internal Z boundary of the ring.

2. We deform the $\hat{Z}_L$ loop by multiplying it by all the $\hat{Z}_L^s$ stabilizers shown outline in Fig. 27b; this leaves us with the $\hat{Z}_L$ chain of operators going from the left to the right boundary of the isolated 2D patch, shown as the horizontal solid line in Fig. 27b.

3. We turn off, or reduce in terminal number, all the $\hat{X}$ and $\hat{Z}$ stabilizers inside the ring, creating a "moat", leaving those within the dashed box in Fig. 27c; this leaves us with the smaller 2D patch shown in Fig. 27d, eliminating both qubit holes, with $\hat{Z}_L$ still crossing from left to right, and $\hat{X}_L$ now crossing from top to bottom. Note this 2D patch is now just a larger ($d = 7$) version of the $d = 5$ array qubit we discussed earlier (Fig. 3); the two X boundaries on top and bottom, and Z boundaries on left and right, make this a logical qubit with two degrees of freedom. The data qubits just beyond the two internal X boundaries of the moat are measured in $\hat{X}$ to preserve error tracking with the adjacent three-terminal $\hat{X}$ stabilizers still active on the inside boundary of the moat.

4. We perform the key to this process, executing physical Hadamards on all the data qubits in the patch. As this changes the eigenbases from $\hat{Z}$ to $\hat{X}$ and vice versa, we change the identity of the measure qubits from $\hat{X}$ to $\hat{Z}$ stabilization and vice versa, as shown in Fig. 27e. Also, the $\hat{X}_L$ and $\hat{Z}_L$ logical operators swap their identities. This step, and the two that follow, are done in between two cycles of the surface code (so the measure-X and measure-Z qubits do not perform any stabilization on the isolated patch during these three steps).

5. The $\hat{X}$ and $\hat{Z}$ stabilizers in the patch are now misaligned with those in the larger 2D array. To correct this, we perform a data qubit-measure qubit swap, between each patch data qubit and the measure qubit directly above it.

6. We perform a second swap, between each measure qubit and the patch data qubit to its left, as shown in Fig. 27f. The data qubits now hold the Hadamard-transformed logical state of the patch. The surface code cycle is re-started, with the measure qubits continuing to measure in X and Z as before the physical Hadamard, in alignment with the larger 2D array; the two swaps ensure that the Hadamard-transformed state in the 2D patch is consistent with this stabilization.

7. Most of the $\hat{X}$ and $\hat{Z}$ stabilizers isolating the 2D patch from the 2D array are turned back on, reducing the width of the moat to a ring one data qubit wide, separating the patch from the 2D array, as we had in Fig. 27a. This is done in a way that creates two logical qubit holes in the patch as shown in Fig. 28g, with the two qubit holes positioned so that the $\hat{X}_L$ operator chain terminates on the internal X boundaries of the two qubit holes.

8. The $\hat{Z}_L$ chain is deformed so it wraps tightly around the right qubit Z-cut hole, as shown in Fig. 28g. This is done by multiplying $\hat{Z}_L$ by all the outlined $\hat{Z}$ stabilizers.

9. The left and right qubit holes are moved in the usual way around the patch, as shown in Fig. 28h partway through the move transformation, and in Fig. 28i, after the open cells have been re-stabilized, closing the move cut. The holes move must be split into two steps to avoid reducing the distance $d$ of logical qubit. We wait $d$ surface code cycles to establish all stabilizer values in time.

10. Each of the two Z-cut holes is moved one cell to return each to its original start point, as shown in Fig. 28j.

11. The stabilizers isolating the patch from the 2D array are all turned back on, reunifying the patch with the array and completing the Hadamard transformation, after waiting $d$ surface code cycles to establish the stabilizer values in time.

## Appendix K: Short qubits

Here we outline how to create and inject a state into a short X-cut qubit (the process for a short Z-cut qubit is completely analogous, simply exchanging the roles of X and Z). Figure 31 is in the main text.

1. We start with a completely stabilized section of the 2D array, as in Fig. 31(a).

2. We turn off the two $\hat{X}$ stabilizers $\hat{X}_1\hat{X}_2\hat{X}_3\hat{X}_5$ and $\hat{X}_5\hat{X}_7\hat{X}_8\hat{X}_9$, creating a pair of X-cut holes. These holes are separated by just one data qubit, qubit 5; this is the "short qubit".

   During the same surface code cycle, we perform measurements of the two $\hat{Z}$ stabilizers $\hat{Z}_{s1} = \hat{Z}_2\hat{Z}_4\hat{Z}_5\hat{Z}_7$ and $\hat{Z}_{s2} = \hat{Z}_3\hat{Z}_5\hat{Z}_6\hat{Z}_7$, and at the same time measure data qubit 5 along $\hat{X}$, yielding the eigenvalue $X_5$. Note that $\hat{X}_5$ anti-commutes with the two $\hat{Z}$ stabilizers $\hat{Z}_{s1}$ and $\hat{Z}_{s2}$, so these three measurements cannot result in a simultaneous eigenstate of all three stabilizers. However, the operator formed by the product $\hat{Z}_{s1}\hat{Z}_{s2} = \hat{Z}_2\hat{Z}_3\hat{Z}_4\hat{Z}_6\hat{Z}_7\hat{Z}_8$ commutes with $\hat{Z}_{s1}$, $\hat{Z}_{s2}$ and with $\hat{X}_5$, so the simultaneous measurements will select states that are eigenstates of $\hat{Z}_{s1}\hat{Z}_{s2}$ and $\hat{X}_5$. The measurement outcome $Z_{s1}Z_{s2}$ will be equal to the product of $Z_{s1}$ and $Z_{s2}$ from the previous surface code cycle, with $X_5 = \pm 1$.

3. We have now created a short qubit that we can manipulate. We prepare the qubit, presently in a known eigenstate of $\hat{X}_5$, by applying the appropriate rotation about the Z axis, using $\hat{R}_Z(\theta)$ (see Eq. 52)) to put the qubit in $|g_L\rangle + e^{i\theta}|e_L\rangle$; the specific rotation will depend on whether the logical qubit was initialized in $|+_L\rangle$ ($X_5 = +1$) or $|-_L\rangle$ ($X_5 = -1$). If we are interested in injecting the $|Y_L\rangle$ state, the angle $\theta = \pi/2$, while for the $|A_L\rangle$ state we would use $\theta = \pi/4$.

4. The surface code stabilization sequence is now restarted, with the two $\hat{X}$ stabilizers left idle. The two holes of the qubit can now be separated and enlarged to better protect against errors. The state in the qubit can now be purified in a distillation process, prior to its use in an $\hat{S}_L$ or $\hat{T}_L$ gate.

## Appendix L: $\hat{S}_L$ and $\hat{T}_L$ distillation sub-circuits

In this Appendix, we look explicitly at some of the operations used in the $|Y_L\rangle$ and $|A_L\rangle$ distillation circuits, focussing on the terminal $\hat{S}_L$ and $\hat{T}_L^\dagger$ gates respectively, followed the measurement $M_X$, i.e. the last two steps for each ancilla qubit's distillation in Figs. 32 and 33. The $\hat{S}_L$ gate as shown in Fig. 29 is deterministic, but to provide a parallel discussion to the analysis of the $|A_L\rangle$ distillation, we instead use the circuit in Fig. 30, replacing the $|A_L\rangle$ ancilla state in that circuit with $|Y_L\rangle$. Using $|Y_L\rangle$, this circuit will execute $\hat{S}_L$, although only about half the time, as using this non-deterministic circuit the output states need to be corrected the other half of the time, just as for the $\hat{T}_L$ gate.

Figure 35 shows the corresponding sub-circuits, along with equivalent circuits that include the conditional measurements. As all states and operations will be logical in this appendix, we will drop the $L$ subscript for compactness.

We first consider Fig. 35a, which implements the $\hat{S}$ gate. For the state going into the $M'_X$ measurement, the CNOT followed by the $M'_Z$ measurement gives $|\psi'\rangle = \hat{S}|\psi\rangle$ for $M'_Z = +1$, and $|\psi'\rangle = \hat{X}\hat{Z}\hat{S}|\psi\rangle$ for $M'_Z = -1$.

The result of the subsequent $M'_X$ measurement is then given by

$$
\begin{aligned}
M'_X &= M_X[\hat{S}|\psi\rangle] && \text{(for } M'_Z = +1\text{), and} \\
M'_X &= M_X[\hat{X}\hat{Z}\hat{S}|\psi\rangle] && \text{(for } M'_Z = -1) \\
&= M_X[\hat{Z}\hat{S}|\psi\rangle] \\
&= -M_X[\hat{S}|\psi\rangle],
\end{aligned}
\tag{L1}
$$

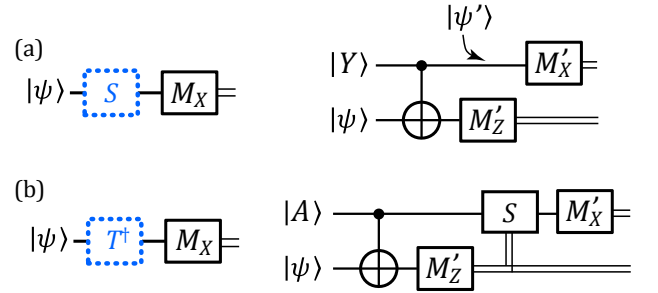

FIG. 35. (Color online) (a) Left panel shows original portion of the $|\hat{Y}\rangle$ distillation circuit, involving the $\hat{S}$ gate followed by measurement $M_X$, whereas right panel shows the expanded $\hat{S}$ circuit displaying the logical CNOT and measurements $M'_Z$ and $M'_X$. (b) Same as (a) but for a portion of the $|A\rangle$ distillation circuit, involving the $\hat{T}^\dagger$ gate. A conditional $\hat{S}$ gate is applied to the upper qubit if $M'_Z = +1$.

| Input | $|Y\rangle$ | $\hat{Z}|Y\rangle = |Y^\star\rangle$ | $\hat{X}|Y\rangle = i|Y^\star\rangle$ | $\hat{Y}|Y\rangle = -i|Y\rangle$ |
|---|---|---|---|---|
| $\theta$ | $\pi/2$ | $-\pi/2$ | $-\pi/2$ | $\pi/2$ |
| $M'_Z =$ | $M'_X[\hat{S}]$ | $M'_X[\hat{Z}\hat{S}]$ | $M'_X[\hat{Z}\hat{S}]$ | $M'_X[\hat{S}]$ |
| $+1$ | | $= -M'_X[\hat{S}]$ | $= -M'_X[\hat{S}]$ | |
| $M'_Z =$ | $M'_X[\hat{X}\hat{Z}\hat{S}]$ | $M'_X[\hat{X}\hat{S}]$ | $M'_X[\hat{X}\hat{S}]$ | $M'_X[\hat{X}\hat{Z}\hat{S}]$ |
| $-1$ | $= -M'_X[\hat{S}]$ | $= +M'_X[\hat{S}]$ | $= +M'_X[\hat{S}]$ | $= -M'_X[\hat{S}]$ |

TABLE VI. Table of measurement outcomes for the $\hat{S}$ operation and $M'_Z$ and $M'_X$ measurement. For compactness, the state $|\psi\rangle$ has been dropped from inside the $M'_X$ measurement brackets. The second row shows the angle $\theta$ for the input state $|g\rangle + \exp(i\theta)|e\rangle$ after application of $\hat{Z}$, $\hat{X}$, or $\hat{Y}$ errors to the state $|Y\rangle$.

using $M_X[\hat{X}|\psi\rangle] = M_X[|\psi\rangle]$ and $M_X[\hat{Z}|\psi\rangle] = -M_X[|\psi\rangle]$. Hence when the circuit fails to generate $\hat{S}$ and instead generates $\hat{X}\hat{Z}\hat{S}$, the final measurement acquires a minus sign. This result is summarized in the $|Y\rangle$ column of Table VI. From these equations, it is easy to see that the net result for the desired $M_X$ measurement is equal to the product of the two sub-measurements

$$M_X[\hat{S}|\psi\rangle] = M'_X M'_Z . \tag{L2}$$

As the goal of distillation is to reduce errors in $|Y\rangle$, we next consider the effect of errors on the input $|Y\rangle$ ancilla state, and how these affect the output measurement. These errors can be modeled as probabilistic applications of $\hat{X}$, $\hat{Y}$, or $\hat{Z}$ to the $|Y\rangle$ state, which give $\hat{Z}|Y\rangle = |Y^\star\rangle = |g\rangle - i|e\rangle$, $\hat{X}|Y\rangle = i|Y^\star\rangle$ and $\hat{Y}|Y\rangle = -i|Y\rangle$, where $\pm i$ represents an overall phase factor that can be dropped. With a general input state given by $|g\rangle + \exp(i\theta)|e\rangle$, the $|Y\rangle$ ($|Y^\star\rangle$) state corresponds to $\theta = \pi/2$ ($\theta = -\pi/2$). When we have the erroneous input $|Y^\star\rangle$ state, i.e. $\theta = -\pi/2$, the $M'_Z = +1$ measurement outcome feeds the state $\hat{S}^\dagger|\psi\rangle = \hat{Z}\hat{S}|\psi\rangle$ to the measurement $M'_X$, while $M'_Z = -1$ feeds the state $\hat{X}\hat{S}|\psi\rangle$ to the measurement. Table VI lists the $M'_X$ measurement results for both $M'_Z$ outcomes and all error types, and shows that the measurement result $M_X$ is reversed in sign for $\hat{Z}$ and $\hat{X}$ errors, signalling the error in the distillation process. Although $\hat{Y}$ errors are undetectable (the overall phase factor cannot be measured), the distillation is not affected since this error produces no significant change to the input state $\hat{Y}|Y\rangle = -i|Y\rangle$. All errors are thus described by $\hat{X}$ or $\hat{Z}$ operations and are successfully and directly detected by the circuit.

For the $\hat{T}^\dagger$ circuit in Fig. 35b, the measurement outcome $M'_Z = +1$ indicates the circuit produced $\hat{T}|\psi\rangle$, which is corrected by applying $\hat{S}$, using the identity $\hat{S}\hat{T} = \hat{Z}\hat{T}^\dagger$, the desired result accompanied by a byproduct $\hat{Z}_L$ operator. For $M'_Z = -1$, the circuit produces

$\hat{X}\hat{T}^\dagger|\psi\rangle$). The result of the $M'_X$ measurement is thus

$$
\begin{aligned}
M'_X &= M_X[\hat{S}\hat{T}|\psi\rangle] \qquad \text{(for } M'_Z = +1\text{)} \\
&= M_X[\hat{Z}\hat{T}^\dagger|\psi\rangle] \\
&= -M_X[\hat{T}^\dagger|\psi\rangle], \text{ and} \\
M'_X &= M_X[\hat{X}\hat{T}^\dagger|\psi\rangle] \qquad \text{(for } M'_Z = -1\text{)} \\
&= M_X[\hat{T}^\dagger|\psi\rangle].
\end{aligned} \tag{L3}
$$

Hence we find for the $\hat{T}^\dagger$ circuit

$$M_X[\hat{T}^\dagger|\psi\rangle] = -M'_X M'_Z. \tag{L4}$$

For a general ancilla state $|g\rangle + e^{i\theta}|e\rangle$, passing the input state $|\psi\rangle = \alpha|g\rangle + \beta|e\rangle$ to the $\hat{T}^\dagger$ circuit generates two outcomes depending on the $M'_Z$ measurement. For $M'_Z = +1$, $\hat{S}$ is applied and the result $\alpha|g\rangle + \beta e^{i(\theta - \pi/2)}|e\rangle$ is passed to the $M'_X$ measurement. For the $M'_Z = -1$ outcome, the conditional $\hat{S}$ is not applied, and instead the result $\beta|g\rangle + \alpha e^{i\theta}|e\rangle = e^{i\theta}\hat{X}(\alpha|g\rangle + \beta e^{-i\theta}|e\rangle)$ is passed to the $M'_X$ measurement. In the latter case the phase factor and the $\hat{X}$ do not affect the measurement, so the $M'_X$ measurement is performed on $\alpha|g\rangle + \beta e^{-i\theta}|e\rangle$. For the $|A\rangle$ ancilla with $\theta = \pi/4$, the two states sent to the measurement are identical, as per (L3). Errors on the $|A\rangle$ ancilla however change the angle. For a $\hat{Z}$ error the ancilla angle changes to $\theta = -3\pi/4$, for $\hat{X}$ it becomes $e^{i\pi/4}|A^\star\rangle$, equivalent to $\theta = -\pi/4$, and for $\hat{Y}$ the angle is $\theta = 3\pi/4$.

The results for the correct $|A\rangle$, as well as for $\hat{Z}$, $\hat{X}$, and $\hat{Y}$ errors to the ancilla, are listed in Table VII. To calculate some of these entries, we used the identity

$$
\begin{aligned}
\hat{T} &= \hat{S}\hat{T}^\dagger \\
&= \left(\frac{1+i}{\sqrt{2}}\hat{I} + \frac{1-i}{\sqrt{2}}\hat{Z}\right)\hat{T}^\dagger \\
&= e^{i\pi/4}\hat{T}^\dagger + e^{-i\pi/4}\hat{Z}\hat{T}^\dagger.
\end{aligned} \tag{L5}
$$

A measurement $M'_X$ of $\hat{T}|\psi\rangle$ thus yields the same as $M'_X$ on $\hat{T}^\dagger|\psi\rangle$ about half the time (meaning things worked out), while it yields $-M'_X$ the other half of the time (meaning we throw the result away). We see from Table VII that whenever an error in the $|A\rangle$ state generates the incorrect output, it is signalled by a sign change in the measurement $M'_X$, so that state would be thrown away.

When computing the effect of errors on the input state $|A\rangle$, simulations of the (error-free) logic circuit show that the $\hat{T}^\dagger$ circuit with the expected Reed code stabilizer outcomes produces a perfect output when zero, one or two ancilla states have $\hat{X}$, $\hat{Z}$ or $\hat{Y}$ errors, but when three or more ancilla states have errors, the output is purified with a probability scaling as the error rate $p^3$; statistical arguments indicate that the error rate should be $35p^3$ as above. The rate at which the distillation circuit fails is $1 - 15p$.

## Appendix M: Estimating the time and size of a factoring circuit

We provide here a few more details on the surface code quantum computer size and execution time that are needed to factor a $N = 2000$ bit number into its primes.

We use the general Shor circuit as described in Ref. [35], using the addition circuit described in Ref. [37]. The most resource-intensive part of Shor's algorithm is the modular exponentiation circuit, which is the backbone of the factoring algorithm; in this implementation, modular exponentiation involves $40N^3$ sequential Toffoli gates, where each Toffoli gate uses a total of seven $\hat{T}_L$ non-Clifford gates, using three in parallel, then one, then three in parallel again. The $\hat{T}_L$ gate sequence is essentially incompressible, and is the most resource-intensive part of the modular exponentiation circuit, because each $\hat{T}_L$ gate uses up one highly-distilled $|A_L\rangle$ ancilla state.

A highly time-optimized version of the $\hat{T}_L$ circuit from Ref. [41] indicates that each $\hat{T}_L$ gate can be completed in one measurement time $t_M$, a fraction of the surface code cycle time, so each Toffoli gate requires a time $3t_M$ to complete. The total time for the exponentiation circuit is thus $120N^3 t_M$. With a target measurement time of $t_M = 100$ ns (see Sect. XVII), this implies 26.7 hours to factor a $N = 2000$ bit number. The remainder of the surface code must be designed to supply this circuit with sufficient resources to allow execution in this time.

Each of the $\hat{T}_L$ gates consumes one ancilla $|A_L\rangle$ state, so in total the exponentiation circuit consumes $7 \times 40N^3 = 280N^3 \approx 2 \times 10^{12}$ $|A_L\rangle$ states. These states must be generated at a rate sufficient to keep pace with the exponentiation circuit, and generated with a logical error rate sufficiently small that the exponentiation circuit makes a negligible number of errors; we would like the rate of errors in the final $|A_L\rangle$ states to be much less than $P_A = (280N^3)^{-1} \approx 4 \times 10^{-13}$ to ensure good fidelity.

With a state injection error rate $p_I = 0.005$ (meaning a 0.5% error rate in the $|A_L\rangle$ states injected in short qubits), the first stage of distillation will yield $|A_L\rangle$ states with an error rate $p_1 = 35p_I^3 \approx 4 \times 10^{-6}$, assuming the distillation circuit is error-free; the second stage will yield states with an error rate $p_2 = 35p_1^3 \approx 3 \times 10^{-15}$. As this error rate is below the target rate $P_A$ for errors in the $|A_L\rangle$ state, two stages of distillation are sufficient with this injection error rate.

The distillation circuits are of course not flawless; given a per-surface code step physical qubit error rate of $p = 10^{-3}$, we must find the surface code distance $d$ that yields a sufficiently low logical error rate. The second and last stage of distillation for an $|A_L\rangle$ state takes 16 logical qubits (15 ancillae plus one logical qubit for the Bell pair) and is executed in $8 \times 1.25 \times d_2$ surface code cycles, including the various CNOTs in a compressed format and sufficient distance $d_2$ in time, as we perform this distillation with a surface code distance $d_2$. The prior (first) stage of distillation requires more logical qubits, using $15 \times 16$ logical qubits (15 sets of distillation circuits, each with 16 logical qubits), operated with a surface code distance $d_1$ in $10d_1$ surface code cycles. We want to reduce the distance $d$ in the first stage of distillation over that in the second stage, as this reduces the required surface code footprint; the distance reduction will concomitantly increase the logical error rate over that of the following stage, but these errors will be distilled out.

For the second and final stage of distillation, an error rate $P_{L2}$ per surface code cycle with a distance $d_2$ surface code will yield an $|A_L\rangle$ error rate of about $16 \times 2 \times 3 \times 1.25d_2 \times P_{L2} = 120d_2P_{L2}$ (16 logical qubits, 2 types of logical qubits, multiplier of 3 kinds of error chains, and $5d_2/4$ surface code cycles). We need $d_2$ sufficient to keep this error rate below $P_A$. With an error rate $p = 10^{-3}$, a distance $d_2 = 34$ code yields $P_{L2} \approx 3 \times 10^{-19}$, with $120d_2P_L \approx 1 \times 10^{-15} < P_A$, just below the target error rate. For the first stage of distillation, a logical error rate $P_{L1}$ with a distance $d_1$ surface code will generate states with an error rate $15 \times 16 \times 2 \times 3 \times 1.25d_1 \times P_{L1} = 1800d_1P_{L1}$. This rate can be higher than $P_A$ as we distill the output, so we actually only need $35(1800d_1P_{L1})^3 < P_A$. A distance $d_1 = 17$ code will give $P_{L1} \approx 1 \times 10^{-10}$, so $|A_L\rangle$ states will be output with an error rate of about $3 \times 10^{-6}$. The following distillation stage will reduce this to $1 \times 10^{-15}$, again just below the target rate.

The first stage of distillation occupies the largest footprint in the surface code, since to generate one final purified $|A_L\rangle$ state we need $16 \times 15 = 240$ logical qubits. In a distance $d_1 = 17$ code, a logical qubit takes $2.5 \times 1.25 \times (2d_1)^2 \approx 3600$ physical qubits. Hence the first stage of distillation will take $240 \times 3600 \approx 8 \times 10^5$ physical qubits, with the distillation taking $10d_1 = 170$ surface code cycles. The second and final stage uses a $d_2 = 34$ code, so one logical qubit takes $2.5 \times 1.25 \times (2d_2)^2 \approx 14500$ physical qubits. The distillation then takes $16 \times 14500 \approx 2.4 \times 10^5$ physical qubits and about 340 surface code cycles. The surface code footprint occupied by the first stage can be re-used in the second stages, so in total about 800,000 physical qubits are required to generate a sufficiently purified $|A_L\rangle$ state, which takes about 500 surface code cycles. The surface code footprint used to generate this one state is active in a volume comprising a square pyramid in space-time, with a large base for the first stage of distillation, reduced by a factor of about three in the subsequent stage. The total space-time volume of the two stages is $170 \times 8 \times 10^5 + 340 \times 2.4 \times 10^5 \sim 2.2 \times 10^8$. The space-time volume available when using the full footprint for the full duration is $500 \times 8 \times 10^5 \sim 4 \times 10^8$. The full footprint can thus be used to generate approximately two $|A_L\rangle$ states in the same number of surface code cycles, forming what we call the "AA factory".

TABLE VII. Table of measurement outcomes for the $\hat{T}$ operation and $M'_Z$ and $M'_X$ measurements. Columns show ideal input state $|A\rangle$ and after application of $\hat{Z}$, $\hat{X}$, or $\hat{Y}$ errors. Entries with $\vee$ ("or") yield one entry or the other, so approximately half the time succeed and the other half fail.

| Input | $|A\rangle$ | $\hat{Z}|A\rangle$ | $\hat{X}|A\rangle$ | $\hat{Y}|A\rangle$ |
|---|---|---|---|---|
| $\theta$ | $\pi/4$ | $-3\pi/4$ | $-\pi/4$ | $3\pi/4$ |
| $M'_Z =$ +1 | $M'_X[\hat{S}\hat{T}]$ $= -M'_X[\hat{T}^\dagger]$ | $M'_X[\hat{S}\hat{S}^\dagger\hat{T}^\dagger]$ $= +M'_X[\hat{T}^\dagger]$ | $M'_X[\hat{S}\hat{T}^\dagger]$ $= M'_X[\hat{T}]$ $= -M'_X[\hat{T}^\dagger]$ $\vee +M'_X[\hat{T}^\dagger]$ | $M'_X[\hat{S}\hat{S}\hat{T}]$ $= -M'_X[\hat{T}]$ $= +M'_X[\hat{T}^\dagger]$ $\vee -M'_X[\hat{T}^\dagger]$ |
| $M'_Z =$ -1 | $M'_X[\hat{X}\hat{T}^\dagger]$ $= +M'_X[\hat{T}^\dagger]$ | $M'_X[\hat{X}\hat{S}\hat{T}]$ $= -M'_X[\hat{T}^\dagger]$ | $M'_X[\hat{X}\hat{T}]$ $= M'_X[\hat{T}]$ $= +M'_X[\hat{T}^\dagger]$ $\vee -M'_X[\hat{T}^\dagger]$ | $M'_X[\hat{X}\hat{S}^\dagger\hat{T}^\dagger]$ $= -M'_X[\hat{T}]$ $= -M'_X[\hat{T}^\dagger]$ $\vee +M'_X[\hat{T}^\dagger]$ |

Each surface code cycle involves single physical qubit resets and gates, physical qubit CNOTs, and readout of the measure qubits; as discussed in Sect. XVII, we believe a 200 ns time for this cycle is not unreasonable, limited mainly by the measurement time $t_M$, which we take as 100 ns, but also in part by microwave technology as well as by classical processing speeds. Hence an AA factory can produce two $|A_L\rangle$ states every $500 \times 200$ ns $\approx 100 \, \mu$s, and can generate $2 \times 10^9$ states in the 26.7 hour factoring time. The $280N^3 \approx 2.2 \times 10^{12}$ required $|A_L\rangle$ states translates to then needing about 1200 AA factories working in parallel, and thus $1200 \times 8 \times 10^5 \approx 10^9$ physical qubits. The remainder of Shor's algorithm requires about $2N = 4000$ logical qubits, which in a $d_2 = 34$ distance surface code takes about $4000 \times 14500 \approx 5.6 \times 10^7$ additional physical qubits, adding a fairly negligible 6% to the AA factory footprint, for a total of about a billion physical qubits.

Improving the performance of the surface code can reduce these numbers somewhat. Eliminating the first stage of $|A_L\rangle$ distillation (so we only did one round of distillation) would require a state injection error rate $p_I \lesssim (P_A/35)^{1/3} \approx 2 \times 10^{-5}$, which is probably not realizable. Improving the physical qubit error rate by a factor of ten, to $p = 10^{-4}$, would reduce the distance $d$ at the top level, allowing a $d_2 = 16$ top-level surface code, with $P_{L2} \approx 3 \times 10^{-19}$. The first stage of distillation with 240 logical qubits could be run using a $d_1 = 8$ code with $P_{L1} \approx 3 \times 10^{-15}$, with an output error rate of $\approx 4 \times 10^{-11}$. The total footprint for a AA factory would then be $240 \times 2.5 \times 1.25 \times (2d_1)^2 \approx 2 \times 10^5$ physical qubits, and would produce $|A_L\rangle$ states at twice the rate due to the reduction in the code distance, so only half the number of factories would be needed, for a total of about 120 million physical qubits. The $2N = 4000$ computational qubits in a $d_2 = 16$ surface code would take $4000 \times 3200 \approx 12$ million qubits. The overall surface code could thus be implemented with about 130 million physical qubits, albeit with no change in the overall execution time.

[1] N. D. Mermin, *Quantum Computer Science: An Introduction* (Cambridge University Press, 2007).

[2] M. A. Nielsen and I. L. Chuang, *Quantum Computation and Quantum Information* (Cambridge University Press, 2000).

[3] P. W. Shor, in *Algorithmic Number Theory: First International Symposium ANTS-I* (1994) p. 289.

[4] L. K. Grover, Phys. Rev. Lett. **79**, 325 (1997).

[5] L. K. Grover, in *Proc. 28th Annual ACM Symposium on the Theory of Computing (STOC)* (1996) pp. 212–219.

[6] T. D. Ladd, F. Jelezko, R. Laflamme, Y. Nakamura, C. Monroe, and J. L. O'Brien, Nature **464**, 45 (2010), arXiv:1009.2267.

[7] I. Buluta, S. Ashhab, and F. Nori, Rep. Prog. Phys. **74**, 104401 (2011), arXiv:1002.1871.

[8] S. B. Bravyi and A. Y. Kitaev, "Quantum codes on a lattice with boundary," (1998), arXiv:quant-ph/9811052, http://arxiv.org/abs/quant-ph/9811052.

[9] E. Dennis, A. Y. Kitaev, A. Landahl, and J. Preskill, J. Math. Phys. **43**, 4452 (2002), http://arxiv.org/abs/quant-ph/0110143.

[10] D. Gottesman, *Stabilizer Codes and Quantum Error Correction*, Ph.D. thesis, Caltech (May 1997), arXiv:quant-ph/9705052.

[11] A. Y. Kitaev, "Quantum communication, computing and measurement," (Plenum Publishing Corporation, 1997) Chap. Quantum error correction with imperfect gates.

[12] A. Y. Kitaev, Russian Math Surveys **52**, 1191 (1997).

[13] A. Y. Kitaev, "Fault-tolerant quantum computation by anyons," (1997), arXiv:quant-ph/9707021.

[14] A. Y. Kitaev, Annals of Physics **303**, 2 (2003), http://arxiv.org/abs/quant-ph/9707021.

[15] M. H. Freedman and D. A. Meyer, Foundations of Computational Mathematics **1**, 325 (2001), http://arxiv.org/abs/quant-ph/9810055.

[16] C. Wang, J. Harrington, and J. Preskill, Annals of Physics **303**, 31 (2003), http://arxiv.org/abs/quant-ph/0207088.

[17] R. Raussendorf, J. Harrington, and K. Goyal, Annals of Physics **321**, 2242 (2006), http://arxiv.org/abs/quant-ph/0510135.

[18] R. Raussendorf and J. Harrington, Phys. Rev. Lett. **98**, 190504 (2007), http://arxiv.org/abs/quant-ph/0610082.

[19] R. Raussendorf, J. Harrington, and K. Goyal, New Journal of Physics **9**, 199 (2007), http://arxiv.org/abs/quant-ph/0703143.

[20] A. G. Fowler, A. M. Stephens, and P. Groszkowski, Phys. Rev. A **80**, 052312 (2009), http://arxiv.org/abs/0803.0272.

[21] A. G. Fowler, D. S. Wang, and L. C. L. Hollenberg, Quant. Info. Comput. **11**, 8 (2011), arXiv:quant-ph/1004.0255.

[22] D. S. Wang, A. G. Fowler, and L. C. L. Hollenberg, Phys. Rev. A **83**, 020302(R) (2011), arXiv:quant-ph/1009.3686.

[23] A. G. Fowler, A. C. Whiteside, and L. C. L. Hollenberg, Phys. Rev. Lett. **108**, 180501 (2012), arXiv:1110.5133.

[24] G. Duclos-Cianci and D. Poulin, Physical Review Letters **104**, 050504 (Feb. 2010), arXiv:quant-ph/0911.0581.

[25] G. Duclos-Cianci and D. Poulin, "A renormalization

group decoding algorithm for topological quantum codes," (Jun. 2010), arXiv:quant-ph/1006.1362.

[26] H. Bombin, R. S. Andrist, M. Ohzeki, H. G. Katzgraber, and M. A. Martin-Delgado, "Strong resilience of topological codes to depolarization," (Feb. 2012), arXiv:quant-ph/1202.1852.

[27] J. R. Wootton and D. Loss, "High threshold error correction for the surface code," (Feb. 2012), arXiv:quant-ph/1202.4316.

[28] A. G. Fowler, A. C. Whiteside, A. L. McInnes, and A. Rabbani, "Topological code Autotune," (Feb. 2012), arXiv:quant-ph/1202.6111.

[29] D. S. Wang, A. G. Fowler, C. D. Hill, and L. C. L. Hollenberg, Quant. Info. Comp. **10**, 780 (2010), arXiv:0907.1708.

[30] H. Bombin, G. Duclos-Cianci, and D. Poulin, "Universal topological phase of 2D stabilizer codes," (Mar. 2011), arXiv:quant-ph/1103.4606.

[31] A. J. Landahl, J. T. Anderson, and P. R. Rice, "Fault-tolerant quantum computing with color codes," (Aug. 2011), arXiv:quant-ph/1108.5738.

[32] P. Sarvepalli and R. Raussendorf, "Efficient Decoding of Topological Color Codes," (Nov. 2011), arXiv:quant-ph/1111.0831.

[33] K. M. Svore, D. P. DiVincenzo, and B. M. Terhal, Quantum Inf. Comput. **7**, 297 (2007).

[34] F. M. Spedalieri and V. P. Roychowdhury, Quantum Information & Computation **9**, 666 (2009).

[35] V. Vedral, A. Barenco, and A. Ekert, Phys. Rev. A **54**, 147 (Jul 1996), http://link.aps.org/doi/10.1103/PhysRevA.54.147.

[36] S. Beauregard, "Circuit for Shor's algorithm using 2n+3 qubits," (May 2002), arXiv:quant-ph/0205095.

[37] S. A. Cuccaro, T. G. Draper, S. A. Kutin, and D. P. Moulton, "A new quantum ripply-carry addition circuit," (2004), arXiv:quant-ph/0410184.

[38] C. Zalka, "Fast versions of Shor's quantum factoring algorithm," (Jun. 1998), arXiv:quant-ph/9806084.

[39] R. Van Meter, K. M. Itoh, and T. D. Ladd, "Architecture-Dependent Execution Time of Shor's Algorithm," (Jul. 2005), arXiv:quant-ph/0507023.

[40] A. Y. Kitaev, A. H. Shen, and M. N. Vyalyi, *Classical and Quantum Computation* (American Mathematical Society, Providence, RI, 2002).

[41] A. G. Fowler(2012), unpublished.

[42] P. W. Shor, Phys. Rev. A **52**, 2493 (1995).

[43] J. Edmonds, Canad. J. Math. **17**, 449 (1965).

[44] J. Edmonds, J. Res. Nat. Bur. Standards **69B**, 125 (1965).

[45] A. G. Fowler, "Analytic asymptotic performance of topological codes," (2012), arXiv:quant-ph/1208.1334.

[46] A. G. Fowler and S. J. Devitt, arXiv:1209.0510(2012).

[47] D. Gottesman, "The Heisenberg representation of quantum computers," (2008), arXiv:quant-ph/9807006.

[48] J. J. Sakurai, *Modern Quantum Mechanics* (Addison Wesley, 1994).

[49] A. G. Fowler, "Low-overhead surface code logical H," (2012), arXiv:quant-ph/1202.2369.

[50] A. M. Steane, Proc. R. Soc. Lond. A **452**, 2551 (1996), quant-ph/9601029.

[51] S. Bravyi and A. Kitaev, Phys. Rev. A **71**, 022316 (Feb 2005), http://link.aps.org/doi/10.1103/PhysRevA.71.022316.

[52] I. Bloch, J. Dalibard, and W. Zwerger, Rev. Mod. Phys. **80**, 885 (2008).

[53] I. Bloch, J. Dalibard, and S. Nascimbene, Nature Physics **8**, 267 (2012).

[54] D. Leibfried, R. Blatt, C. Monroe, and D. J. Wineland, Rev. Mod. Phys. **75**, 281 (2003).

[55] R. Blatt and D. Wineland, Nature **453**, 1008 (2008).

[56] D. Kielpinski, C. R. Monroe, and D. J. Wineland, Nature **417**, 709 (2002).

[57] M. A. Eriksson, M. Friesen, S. N. Coppersmith, R. Joynt, L. J. Klein, K. Slinker, C. Tahan, P. M. Mooney, J. O. Chu, and S. J. Koester, Quantum Information Processing **3**, 133 (2004).

[58] Y. Makhlin, G. Schön, and A. Shnirman, Rev. Mod. Phys. **73**, 357 (2001).

[59] M. H. Devoret and J. M. Martinis, Quantum Inf. Process. **3**, 163 (2004).

[60] J. Q. You and F. Nori, Physics Today **58**, 42 (2005).

[61] J. Clarke and F. K. Wilhelm, Nature **453**, 1031 (2008).

[62] R. J. Schoelkopf and S. M. Girvin, Nature **451**, 664 (2008).

[63] D. P. DiVincenzo, Phys. Scr. **T137**, 014020 (2009).

[64] H. Paik, D. I. Schuster, L. S. Bishop, G. Kirchmair, G. Catelani, A. P. Sears, B. R. Johnson, M. J. Reagor, L. Frunzio, L. I. Glazman, S. M. Girvin, M. H. Devoret, and R. J. Schoelkopf, Phys. Rev. Lett. **107**, 240501 (Dec 2011), http://link.aps.org/doi/10.1103/PhysRevLett.107.240501.

[65] M. Mariantoni, H. Wang, T. Yamamoto, M. Neeley, R. C. Bialczak, Y. Chen, M. Lenander, E. Lucero, A. D. O'Connell, D. Sank, M. Weides, J. Wenner, Y. Yin, J. Zhao, A. N. Korotkov, A. N. Cleland, and J. M. Martinis, Science **334**, 61 (2011).

[66] R. Raussendorf, D. E. Browne, and H. J. Briegel, Phys. Rev. A **68**, 022312 (Aug 2003), http://link.aps.org/doi/10.1103/PhysRevA.68.022312.